\documentclass[10pt,a4paper]{article}
\usepackage[T1]{fontenc}
\usepackage[utf8]{inputenc}
\usepackage{lmodern}
\usepackage{amsmath,amssymb,amsfonts}
\usepackage{graphicx}
\usepackage{float}
\usepackage{placeins}
\usepackage{booktabs,array,multirow,calc}
\usepackage{caption}
\usepackage{url}
\usepackage[hidelinks]{hyperref}
\usepackage[a4paper,top=0.7in,bottom=0.75in,left=0.62in,right=0.62in]{geometry}
\usepackage{microtype}
\usepackage{enumitem}
\usepackage{balance}
\usepackage{xcolor}
\setlist[itemize]{leftmargin=1.2em,topsep=2pt,itemsep=1pt,parsep=0pt}
\DeclareUnicodeCharacter{2223}{\ensuremath{|}}
\DeclareUnicodeCharacter{03B4}{\ensuremath{\delta}}
\DeclareUnicodeCharacter{0304}{}
\DeclareUnicodeCharacter{1D62}{\ensuremath{_i}}
\DeclareUnicodeCharacter{2070}{\textsuperscript{0}}
\DeclareUnicodeCharacter{2074}{\textsuperscript{4}}
\DeclareUnicodeCharacter{2075}{\textsuperscript{5}}
\DeclareUnicodeCharacter{2076}{\textsuperscript{6}}
\DeclareUnicodeCharacter{2077}{\textsuperscript{7}}
\DeclareUnicodeCharacter{2078}{\textsuperscript{8}}
\DeclareUnicodeCharacter{207B}{\textsuperscript{-}}
\DeclareUnicodeCharacter{2248}{\ensuremath{\approx}}
\DeclareUnicodeCharacter{1D707}{\ensuremath{\mu}}

\title{\bfseries Quantum-Driven Neuromorphic Computing\\for Million-Qubit-Scale Workloads}
\author{Adams Ivanov\textsuperscript{1}, Samer Rahmeh\textsuperscript{1}, Erick Giovani Sperandio Nascimento\textsuperscript{2}, Daniela Herrmann\textsuperscript{1}\\[0.35em]
\small \textsuperscript{1}Dynex Holding Establishment, Pradafant 11, FL-9490 Vaduz, Liechtenstein\\
\small \textsuperscript{2}Surrey Institute for People-Centred Artificial Intelligence, Faculty of Engineering and Physical Sciences, University of Surrey, Guildford GU2 7XH, United Kingdom\\
\small *Corresponding author(s). E-mail(s): office@dynex.co}
\date{}
\begin{document}

\thispagestyle{plain}
\begin{center}
\vspace*{0.35in}
{\LARGE\bfseries Quantum-Driven Neuromorphic Computing\\[0.15em]
for Million-Qubit-Scale Workloads\par}
\vspace{0.35in}
{Adams Ivanov\textsuperscript{1}, Samer Rahmeh\textsuperscript{1}, Erick Giovani Sperandio Nascimento\textsuperscript{2}, Daniela Herrmann\textsuperscript{1}\par}
\vspace{0.25in}
{\textsuperscript{1}Dynex Holding Establishment, Pradafant 11, FL-9490 Vaduz, Liechtenstein\par}
\vspace{0.18in}
{\textsuperscript{2}Surrey Institute for People-Centred Artificial Intelligence, Faculty of Engineering and Physical\\ Sciences, University of Surrey, Guildford GU2 7XH, United Kingdom\par}
\vspace{0.28in}
{*Corresponding author(s). E-mail(s): office@dynex.co\par}
\vspace{0.62in}
{\bfseries Abstract\par}
\vspace{0.22in}
\begin{minipage}{0.68\textwidth}
We introduce Apollo, a 10,000-node p-qubit neuromorphic processor
fabricated in 16 nm mixed-signal CMOS, operating fully at room
temperature with a typical analog-core power envelope of
$\sim$0.5 W. The fundamental computing element, termed a
p-qubit (probabilistic qubit), is a bistable stochastic unit whose
continuous-time state fluctuations are driven by dedicated Integrated
Quantum Entropy Units (IQEUs) injecting true, non-deterministic
quantum-derived entropy. This elevates conventional p-bit architectures
into the quantum regime, enabling state-transition rates of
approximately 12.5ps per p-qubit at an energy cost of $\leq$10 fJ per
transition. Apollo implements a high-degree Hyperion $\Delta$256 interconnect
topology, allowing near-native embedding of dense Ising and quadratic
unconstrained binary optimization (QUBO) problems with substantially
reduced minor-embedding overhead relative to existing annealing
platforms. Through the Suzuki--Trotter isomorphism, the equilibrium
statistics and annealing dynamics of the p-qubit network reproduce those
of a transverse-field quantum annealer in one fewer spatial dimension,
without requiring cryogenic cooling, long-lived coherence, or microwave
control infrastructure.

Beyond device-level validation, we demonstrate quantum-advantaged
dynamics at scale by reproducing the three-dimensional spin-glass
benchmark previously used to establish quantum advantage in
superconducting quantum annealers. Across 300 disorder realizations,
Apollo exhibits residual-energy scaling trajectories indistinguishable
from those reported for cryogenic quantum annealing hardware and clearly
distinct from simulated quantum annealing and classical simulated
annealing, indicating access to the same quantum-critical dynamical
universality class. A 350 nm release-candidate device (Apollo-RC1)
experimentally validates the core p-qubit dynamics, thermodynamic
sampling correctness, and continuous-time annealing behavior, while
large-scale results establish Apollo as the first room-temperature,
industrially scalable platform to demonstrate quantum-equivalent
annealing performance on a canonical hard optimization benchmark. By
unifying probabilistic computing, quantum-driven stochastic dynamics,
and gate-compatible operation in a single architecture, Apollo opens new
pathways for energy-based optimization, Bayesian inference, generative
modeling, and hybrid classical--quantum workflows beyond the cryogenic
era.

\vspace{0.25in}
\noindent\textbf{Keywords:} probabilistic bits, p-bits, neuromorphic computing, stochastic computing, quantum-driven annealing, Ising machines, room-temperature quantum simulation, mixed-signal CMOS, Suzuki--Trotter equivalence
\end{minipage}
\end{center}
\clearpage

\twocolumn
\section{Introduction and Context}

\subsection{Motivation and Problem Setting}

A broad class of computational problems arising in combinatorial
optimization, probabilistic inference, and machine learning can be
formulated as the minimization or sampling of complex energy functions
defined over high-dimensional binary or discrete variables. These
problems are typically characterized by rugged energy landscapes
featuring an extensive number of local minima separated by high
barriers, leading to slow convergence and poor scaling for conventional
algorithmic approaches. Representative examples include Ising and QUBO
optimization, Bayesian inference in graphical models, and training and
sampling of energy-based models$^{[1][2][3]}$.

Deterministic digital computing architectures, which rely on sequential
or clock-synchronous execution of precisely defined logical operations,
are fundamentally ill-suited to these problem classes. Although
heuristic methods such as simulated annealing, Markov chain Monte Carlo,
and their quantum-inspired variants have been developed, their
performance is often constrained by discrete-time update schemes,
limited parallelism, and substantial energy overhead associated with
memory access and control flow. As problem size and connectivity
increase, these limitations manifest as prohibitive time-to-solution and
energy consumption, even on highly parallel CPU- and GPU-based
platforms$^{[4][5][6]}$.

In contrast, many physical systems naturally evolve toward low-energy
configurations through stochastic relaxation processes governed by
statistical mechanics. Thermal spin systems, chemical reaction networks,
and noisy dynamical systems explore their configuration spaces through
continuous-time fluctuations that enable efficient traversal of energy
barriers and convergence toward equilibrium distributions. Harnessing
such physical stochastic dynamics for computation offers an alternative
paradigm in which exploration, relaxation, and sampling are intrinsic
properties of the computing substrate rather than externally imposed
algorithmic procedures. This perspective motivates the development of
hardware systems that directly implement stochastic, energy-based
dynamics as a means to address computational tasks dominated by complex
energy landscapes$^{[7][8][9]}$.

\subsection{Existing Physical Approaches}

Physical approaches to computation have been actively explored as
alternatives to conventional digital architectures for addressing
optimization and sampling problems defined on complex energy landscapes.
Among these, superconducting quantum annealers constitute the most
mature large-scale hardware realization of energy-based computation.
Such systems implement the transverse-field Ising model and exploit
quantum fluctuations, particularly tunnelling, to facilitate exploration
of rugged energy landscapes during annealing. Experimental
demonstrations have shown that quantum annealers can outperform certain
classical heuristics on specific problem classes, especially near
quantum critical points. However, their practical deployment is
constrained by the need for cryogenic operation at millikelvin
temperatures, limited qubit coherence times, restricted native
connectivity, and substantial overhead associated with minor embedding
of dense problem
graphs$^{[10][11][12][13]}$.

In parallel, purely classical algorithmic approaches such as simulated
annealing and simulated quantum annealing have been widely used to
approximate the dynamics of physical annealers on digital hardware.
These methods rely on stochastic updates implemented through
pseudo-random number generation and discrete-time Markov processes,
typically executed on CPUs, GPUs, or specialized accelerators. While
highly flexible and broadly applicable, digital annealing methods suffer
from intrinsic limitations arising from sequential or block-synchronous
updates, memory-access bottlenecks, and the artificial discretization of
dynamics that are continuous in physical systems. As problem size and
connectivity grow, these factors lead to diminishing returns in
performance and energy efficiency, even in massively parallel
implementations$^{[14][15][16]}$.

A third class of approaches encompasses probabilistic and analog
hardware systems that implement stochastic dynamics directly at the
device level. These include p-bit--based architectures, stochastic
nanomagnetic devices, mixed-signal neuromorphic circuits, and optical
Ising machines. Such systems aim to exploit intrinsic noise, nonlinear
device physics, or optical interference to realize energy-based
computation with reduced energy consumption and increased parallelism.
While promising results have been demonstrated in small- to medium-scale
prototypes, existing implementations often rely on shared or
time-multiplexed entropy sources, exhibit limited native connectivity,
or depend on external digital control for synchronization and update
ordering. These constraints have thus far limited their scalability and
their ability to faithfully reproduce the thermodynamic and dynamical
properties associated with large-scale physical annealing
systems$^{[17][18][19][20]}$.

\subsection{Gaps in the Current State of the Art}

Despite substantial progress in both quantum and classical physical
computing platforms, several fundamental limitations remain unresolved
in the context of large-scale energy-based computation. Most notably, no
existing room-temperature system has been shown to reproduce the
annealing dynamics associated with quantum-critical behaviour observed
in superconducting quantum annealers. In particular, the characteristic
scaling of residual energy and relaxation time near phase
transitions---central to claims of quantum advantage---has so far been
confined to cryogenic quantum hardware, limiting accessibility and
scalability$^{[21][22][23]}$.

A second major limitation concerns the quality and independence of
stochasticity in physical and digital implementations. Many
probabilistic hardware platforms rely on shared, broadcast, or
algorithmically generated noise sources, which introduce correlations
across computational elements and violate assumptions of statistical
independence underlying theoretical convergence guarantees. Such
correlations can distort relaxation dynamics, suppress effective
exploration of energy landscapes, and lead to systematic deviations from
the intended Boltzmann distribution, particularly as system size
increases$^{[24][25][26]}$.

Digital implementations of annealing and sampling algorithms further
suffer from artefacts introduced by discrete-time update schemes.
Clocked, synchronous updates impose artificial temporal structure on
inherently continuous physical processes, leading to discretization
errors, update-order bias, and reduced mixing efficiency. While these
effects can sometimes be mitigated through algorithmic heuristics or
increased computational effort, they represent intrinsic departures from
the dynamics of natural stochastic systems and impose scaling and
energy-efficiency penalties$^{[27][28][29]}$.

Another persistent challenge is the overhead associated with embedding
problem graphs onto hardware with limited native connectivity.
Superconducting quantum annealers and many probabilistic hardware
platforms provide only sparse, fixed coupling topologies, necessitating
minor embedding strategies that introduce auxiliary variables, increase
effective problem size, and degrade solution quality. This overhead
becomes increasingly prohibitive for dense or highly connected problem
instances$^{[30][31]}$.

Finally, scalability remains a central obstacle across existing physical
computing approaches. Constraints arising from cryogenic infrastructure,
limited interconnect density, shared entropy generation, or centralized
control architectures restrict the feasible system size and throughput
of current platforms. As a result, achieving large-scale, physically
grounded stochastic computation with high fidelity, low energy
consumption, and reproducible dynamics remains an open
challenge$^{[32][33]}$.

\subsection{Contributions of This Work}

In this work, we address the limitations outlined above through a
combination of theoretical analysis, hardware design, and experimental
validation. The principal contributions of this study are as follows:

\begin{itemize}
\item
  \textbf{Quantum-equivalent annealing dynamics in a continuous-time
  stochastic system:} We demonstrate that a physical system governed by
  continuous-time stochastic dynamics can reproduce the equilibrium
  behaviour and annealing trajectories associated with transverse-field
  quantum annealers, as predicted by statistical-mechanical equivalence
  arguments. This establishes that quantum-equivalent annealing dynamics
  need not rely on coherent quantum evolution or cryogenic operation.
\item
  \textbf{Experimental reproduction of quantum-critical residual-energy
  scaling:} Using a canonical three-dimensional spin-glass benchmark, we
  experimentally reproduce the residual-energy scaling behaviour
  previously used to identify quantum-critical dynamics in
  superconducting quantum annealers. The observed scaling exponents and
  relaxation trajectories are indistinguishable from those reported for
  cryogenic quantum hardware and are clearly distinct from classical
  simulated annealing and simulated quantum annealing.
\item
  \textbf{A scalable probabilistic architecture enabling dense Ising and
  QUBO embeddings:} We introduce a probabilistic computing architecture
  that natively supports high-degree connectivity, substantially
  reducing the embedding overhead associated with dense or highly
  connected Ising and QUBO problem instances. This capability enables
  efficient physical realization of large and complex energy functions
  without the extensive auxiliary-variable overhead required by sparsely
  connected platforms.
\item
  \textbf{Hardware validation of thermodynamically correct Boltzmann
  sampling at room temperature:} We experimentally validate that the
  proposed system samples from the correct Boltzmann distribution across
  a range of problem instances, confirming thermodynamic consistency and
  unbiased stochastic behaviour. This establishes the platform as a
  physically grounded substrate for energy-based optimization,
  probabilistic inference, and sampling tasks operating entirely at room
  temperature.
\end{itemize}

\section{Theoretical Foundations}

\subsection{Probabilistic Bits and Continuous-Time Stochastic Dynamics}

A probabilistic bit (p-bit) is a classical binary variable whose state
fluctuates stochastically in time under the influence of an effective
local field and a noise source. Unlike deterministic digital bits, which
assume fixed values except during externally triggered transitions,
p-bits are explicitly designed to occupy either of two states while
continuously transitioning between them according to well-defined
probabilistic rules. The instantaneous state of a p-bit may be
represented as a binary variable
\(m_i \in \{-1,+1\}\)
(or equivalently
\(m_i \in \{0,1\}\)),
with transition probabilities governed by a nonlinear response to its
local input field$^{[34][35]}$.

In the continuous-time formulation, the evolution of a p-bit is not
synchronized to a global clock or discrete update cycle. Instead, state
transitions occur asynchronously as a result of stochastic fluctuations,
giving rise to a continuous-time Markov process. The transition rates
depend on the instantaneous effective field acting on the p-bit, which
may include contributions from external biases and interactions with
other p-bits. This stands in contrast to discrete-time or clocked update
schemes, where state changes are enforced at fixed intervals and update
order must be explicitly specified, often introducing artificial
temporal structure and update-order
bias$^{[36][37]}$.

When multiple p-bits are coupled through weighted interactions, their
collective dynamics define a high-dimensional stochastic system whose
time evolution can be described using master equations, Glauber
dynamics, or equivalent Fokker--Planck formulations. Under broad and
well-established conditions---such as ergodicity, detailed balance, and
appropriate noise statistics---the stationary distribution of this
system converges to the Gibbs--Boltzmann distribution associated with an
effective energy function. For binary variables with pairwise
interactions, this energy function corresponds to the classical Ising
Hamiltonian or, equivalently, a quadratic unconstrained binary
optimization (QUBO)
formulation$^{[38][39][40]}$.

This direct connection between continuous-time stochastic dynamics and
equilibrium statistical mechanics provides the theoretical basis for
using p-bit networks as physical samplers and optimizers. Rather than
approximating thermodynamic behaviour through algorithmic emulation,
such systems exploit intrinsic stochasticity and asynchronous relaxation
to explore energy landscapes and sample from target distributions. As a
result, p-bit networks constitute a physically grounded computational
model for energy-based optimization, probabilistic inference, and
related tasks governed by Gibbs
statistics$^{[41][42]}$.

\subsection{Boltzmann Distributions and Thermodynamic Sampling}

The collective behaviour of coupled stochastic units is most naturally
described within the framework of equilibrium statistical mechanics. For
a system of interacting binary variables evolving under continuous-time
stochastic dynamics, the long-time stationary distribution is determined
by the balance between deterministic interactions and stochastic
fluctuations. When the dynamics satisfy ergodicity and detailed balance,
the system converges to a Gibbs--Boltzmann distribution of the form

\begin{equation}
P(\mathbf{s}) = \frac{1}{Z} \exp\left(-\beta E(\mathbf{s})\right)
\end{equation}

where s denotes the configuration of binary variables, E(s) is an
effective energy function, $\beta$ is an inverse temperature parameter, and Z
is the partition function ensuring
normalization$^{[43][44][45]}$.

For systems with pairwise interactions and linear bias terms, the energy
function can be written in the Ising form

\begin{equation}
E(\mathbf{s}) = -\sum_{i < j} J_{ij} s_i s_j - \sum_i h_i s_i
\end{equation}

where $J_{ij}$ denotes coupling strengths and
$h_{i}$ represents local fields. An equivalent representation
is given by quadratic unconstrained binary optimization (QUBO), in which
binary variables $x_i \in \{0,1\}$ encode the same energy
landscape up to an affine transformation. These formulations provide a
unifying description for a wide range of optimization and sampling
problems across physics, computer science, and applied
mathematics$^{[46][47][48]}$.

Convergence to the Gibbs distribution requires that the underlying
stochastic dynamics satisfy specific conditions. Chief among these are
the absence of forbidden transitions (ergodicity), symmetric transition
probabilities that enforce detailed balance, and noise statistics that
correctly reproduce thermal fluctuations. In continuous-time stochastic
systems, these requirements are naturally expressed through master
equations or Langevin-type descriptions, where the ratio of transition
rates between states is governed by the corresponding energy
differences. Under such conditions, the equilibrium distribution is
independent of the particular trajectory taken through state space and
depends only on the defined energy function and temperature
parameter$^{[49][50][51]}$.

Thermodynamic sampling from the Boltzmann distribution underlies a broad
class of computational methods and physical models. In statistical
physics, it governs the equilibrium properties of spin systems, glasses,
and interacting particle ensembles. In machine learning, closely related
energy-based models---including Boltzmann machines, restricted Boltzmann
machines, and related probabilistic graphical models---use Gibbs
distributions to represent complex probability densities over
high-dimensional variables. In these contexts, efficient sampling from
the target distribution is often the dominant computational challenge,
motivating physical implementations in which stochastic relaxation
toward equilibrium is achieved intrinsically rather than through
algorithmic approximation$^{[52][53][54]}$.

By framing computation as thermodynamic sampling from a well-defined
energy function, coupled stochastic systems provide a principled and
physically grounded approach to optimization, inference, and generative
modelling. This perspective forms a central theoretical link between
probabilistic computing architectures, classical statistical mechanics,
and quantum-inspired annealing
methods$^{[55][56]}$.

\subsection{Quantum Annealing and the Transverse-Field Ising Model}

Quantum annealing is a computational paradigm in which solutions to
optimization problems are obtained by exploiting the adiabatic evolution
of a quantum system governed by a time-dependent Hamiltonian. The most
common physical realization of quantum annealing is based on the
transverse-field Ising model (TFIM), whose Hamiltonian can be written as

\begin{equation}
H(t) = -\sum_{i < j} J_{ij} \sigma_i^z \sigma_j^z - \sum_i h_i \sigma_i^z - \Gamma(t) \sum_i \sigma_i^x
\end{equation}

where
\(\sigma_i^z\)
and
\(\sigma_j^z\)
are Pauli operators acting on spin $i$, $J_{ij}$ and
$h_{i}$ encode the classical Ising problem, and $\Gamma(t)$ is a
time-dependent transverse field that introduces quantum
fluctuations$^{[57][58][59]}$.

The annealing process begins with a large transverse field, for which
the ground state is a trivial quantum paramagnet. As the annealing
parameter $\Gamma(t)$ is gradually reduced according to a prescribed schedule,
the Hamiltonian interpolates toward the classical Ising form. In the
ideal adiabatic limit, the system remains in its instantaneous ground
state throughout this evolution and ends in the ground state of the
target Ising Hamiltonian, thereby solving the corresponding optimization
problem. In practical settings, finite annealing times and environmental
interactions lead to diabatic transitions and non-equilibrium effects,
but the final state distribution is still governed by the interplay
between quantum dynamics, thermal fluctuations, and the energy
landscape$^{[60][61]}$.

Quantum fluctuations induced by the transverse field play a central role
in the annealing process. By coupling classically distinct
configurations, the transverse-field term enables tunnelling through
energy barriers that may be difficult to surmount via purely thermal
activation. This mechanism can enhance exploration of rugged energy
landscapes and mitigate trapping in local minima, particularly in
regions where the classical energy barriers are narrow but high. The
relative contribution of tunnelling and thermal activation depends on
the annealing schedule, temperature, and problem
structure$^{[62][63]}$.

The performance of a quantum annealer is often characterized by its
annealing trajectories and the concentration of probability mass in
low-energy states at the end of the anneal. Of particular interest is
the scaling behaviour of the residual energy---the difference between
the achieved energy and the true ground-state energy---as a function of
annealing time and system size. Near quantum phase transitions, the
system exhibits critical slowing down, and the dynamics are governed by
universal scaling laws. These quantum-critical regimes have been
identified as key contributors to performance differences between
quantum annealing, simulated annealing, and other classical heuristics,
and they provide a principled basis for assessing potential quantum
advantage$^{[64][65][66]}$.

\subsection{Suzuki--Trotter Equivalence}

A central theoretical connection between quantum annealing and classical
stochastic systems is provided by the Suzuki--Trotter decomposition,
which establishes an exact correspondence between certain quantum
many-body systems and classical statistical models in higher-dimensional
spaces. In the context of quantum annealing, this mapping applies to the
transverse-field Ising model and enables its equilibrium properties to
be expressed in terms of an equivalent classical Ising
model$^{[67][68][69]}$.

Applying the Suzuki--Trotter decomposition to the partition function of
the transverse-field Ising Hamiltonian yields a representation in which
the quantum system in $d$ spatial dimensions is mapped onto a classical
Ising model in $d$+1 dimensions. The additional dimension corresponds to
discretized imaginary time, divided into a finite number of Trotter
slices. Within each slice, spins interact according to the original
classical Ising couplings, while adjacent slices are coupled through
effective interactions derived from the transverse-field term. In the
limit of a large number of slices, this mapping becomes exact and
faithfully reproduces the equilibrium thermodynamics of the quantum
system$^{[70][71]}$.

In this representation, imaginary-time slices encode temporal
correlations induced by quantum fluctuations. The inter-slice coupling
enforces consistency between spin configurations at neighbouring
imaginary-time steps and reflects the strength of the transverse field.
From a statistical-mechanical perspective, quantum fluctuations in the
original Hamiltonian are thus transformed into additional classical
interactions along the imaginary-time dimension. The equilibrium
distribution of the quantum system is therefore equivalent to a
classical Gibbs distribution defined over an expanded configuration
space$^{[72]}$.

This equivalence has important implications for classical stochastic
systems. Because the outcomes of quantum annealing are ultimately
determined by equilibrium statistics rather than coherent phase
information, a classical system that samples from the same effective
Gibbs distribution can, in principle, reproduce the same solution
statistics. Continuous-time stochastic systems with appropriate
interaction strengths, noise characteristics, and relaxation dynamics
can emulate the role of imaginary-time correlations through temporal
persistence and correlated fluctuations, without requiring coherent
quantum evolution$^{[73][74]}$.

For such equivalence to hold, several conditions must be satisfied. The
classical system must exhibit ergodic dynamics that explore the relevant
configuration space, and its stochastic transitions must obey detailed
balance with respect to an effective energy function corresponding to
the Trotter-mapped Hamiltonian. Noise sources must be sufficiently
broadband and uncorrelated to ensure faithful sampling, and the system
must relax toward equilibrium on timescales compatible with the imposed
annealing schedule. When these conditions are met, the stationary
distribution of the classical system converges to the same Gibbs measure
that governs the equilibrium behaviour of the quantum
annealer$^{[75][76][77]}$.

The Suzuki--Trotter equivalence therefore provides a rigorous
theoretical foundation for the possibility that classical,
continuous-time stochastic systems can reproduce the equilibrium
statistics and annealing outcomes of transverse-field quantum annealers.
This result motivates the exploration of physically grounded stochastic
substrates as alternatives to coherent quantum hardware for energy-based
computation, and it establishes the theoretical bridge upon which
quantum-equivalent annealing behaviour can be realized in non-cryogenic
systems$^{[78][79]}$.

\subsection{Implications for Physical Stochastic Hardware}

The theoretical considerations developed in Sections 2.1--2.4 establish
that quantum-equivalent annealing behaviour is not inherently contingent
on coherent quantum evolution, but rather on the statistical and
dynamical properties of the underlying system. From this perspective, a
physical stochastic system can reproduce the equilibrium statistics and
annealing outcomes of transverse-field quantum annealers provided that
several principled conditions are
satisfied$^{[80][81]}$.

First, the system must evolve in continuous time through asynchronous
stochastic dynamics. Continuous-time evolution avoids the discretization
artefacts and update-order biases associated with clocked or stepwise
algorithms and allows relaxation to proceed through physically natural
trajectories. Such dynamics enable the system to explore energy
landscapes smoothly and to emulate the effective temporal correlations
that arise in the imaginary-time formulation of quantum
systems$^{[82]}$.

Second, stochastic transitions must be driven by noise sources that are
statistically independent across degrees of freedom. Independence of
stochastic fluctuations is essential for preserving ergodicity,
enforcing detailed balance, and ensuring convergence to the intended
Gibbs distribution. Correlated or shared noise can introduce spurious
synchronisation, distort equilibrium statistics, and compromise the
validity of the theoretical equivalence between classical and quantum
systems$^{[83]}$.

Third, the system must support sufficient coupling expressivity to
faithfully encode the target energy function. In practice, this requires
the ability to represent arbitrary bias terms and pairwise interactions
with adequate dynamic range and resolution. Limited or overly
constrained connectivity can necessitate embedding strategies that alter
the effective energy landscape, thereby undermining both solution
quality and dynamical fidelity.

Importantly, these requirements are stated at the level of physical
principles rather than specific implementations. No particular device
technology, material system, or architectural choice is assumed in this
analysis. The theoretical framework applies generically to any physical
substrate capable of realizing continuous-time stochastic dynamics,
independent noise sources, and expressive interaction networks.

In the following section, we describe a concrete hardware architecture
designed to instantiate these principles in a scalable and
experimentally verifiable form. This architecture serves as a physical
realization of the theoretical conditions outlined above and provides
the basis for the experimental results presented in subsequent sections.

\section{Apollo Architecture: Physical Realisation of the Theory}

The fundamental computational element of the Apollo architecture is the
p-qubit, a physical realization of the probabilistic bit introduced in
Section 2. While the theoretical p-bit is an abstract stochastic
variable defined solely by its transition statistics and equilibrium
distribution, the p-qubit denotes its concrete hardware instantiation
within Apollo. The distinction in terminology is intentional: the
p-qubit is so named to reflect that its stochastic dynamics are driven
by entropy derived from quantum physical processes, motivating the
designation quantum-driven neuromorphic
computing$^{[83][84]}$.

\begin{figure}[H]
\centering
{\small\bfseries Classical Bit, P-qubit and Superconducting Qubit\par\vspace{2pt}}
\includegraphics[width=0.98\linewidth]{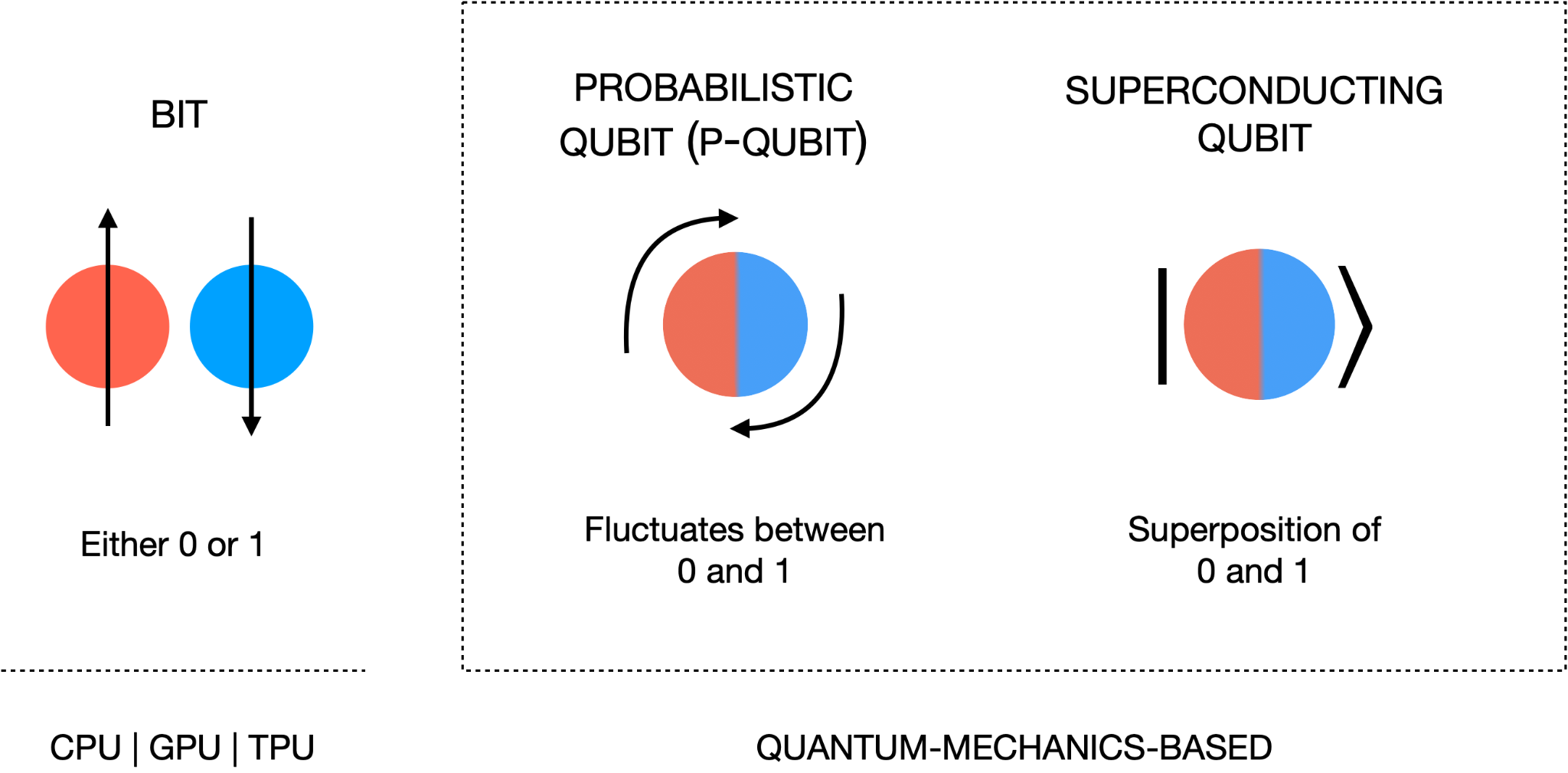}
\caption{Illustration of a classical bit with a definite state of 0 or 1; a p-qubit based on a quantum-driven analog qubit representation, which exhibits stochastic fluctuations between 0 and 1; and a qubit, which occupies a coherent superposition of the basis states 0 and 1.}
\end{figure}

Each p-qubit is implemented as a bistable stochastic unit whose
instantaneous state fluctuates between two well-defined output levels
(Fig. 1). Bistability is established through nonlinear analog circuitry
supporting two stable operating points, while stochastic transitions
between these states are induced by externally supplied noise.
Crucially, this noise is not algorithmically generated but originates
from physical entropy sources whose randomness arises from
quantum-mechanical processes. As a result, the p-qubit remains classical
in its state representation while being quantum-driven in the origin of
its stochastic excitation$^{[77][82]}$.

The probabilistic response of a p-qubit is governed by an analog
sigmoidal activation function that maps its effective input field to a
state-dependent switching probability. This sigmoidal nonlinearity
provides a smooth physical approximation to the sign or hyperbolic
tangent functions commonly employed in theoretical models of stochastic
spin dynamics. By tuning the gain and operating point of the analog
response, the steepness of the activation function---and thus the
effective temperature of the stochastic process---can be modulated,
enabling controlled annealing and sampling
behaviour$^{[85][86]}$.

Biasing and coupling interfaces provide the mechanism by which p-qubits
interact with one another and with externally defined energy functions.
Each p-qubit receives a programmable bias term corresponding to a local
field in the Ising or QUBO formulation, as well as weighted coupling
inputs representing pairwise interactions with neighbouring units. These
contributions are summed in the analog domain prior to application of
the nonlinear activation, ensuring that each p-qubit responds
continuously to the instantaneous collective field rather than through
discretized or sequential updates$^{[87][88]}$.

The use of the term p-qubit is therefore deliberate and precise. Unlike
a coherent qubit, the p-qubit does not rely on superposition or
entanglement, nor does it preserve quantum phase information. At the
same time, it differs fundamentally from conventional p-bits whose
stochasticity is sourced from thermal noise or pseudo-random number
generators. By embedding quantum-derived entropy directly into the
physical stochastic dynamics of a classical bistable unit, the p-qubit
occupies an intermediate conceptual position: classical in
representation, stochastic in operation, and quantum-driven in its
entropy source$^{[89][90]}$.

This distinction underpins the broader characterization of Apollo as a
quantum-driven neuromorphic computing system and forms the basis for its
ability to reproduce quantum-equivalent annealing behaviour within a
room-temperature, classical hardware
substrate$^{[91][92]}$.

\subsection{Design Philosophy and System Overview}

The Apollo chip is organised as a large-scale, tiled
neuromorphic--probabilistic computing array comprising 10,000 fully
parallel, non-multiplexed p-qubits implemented in a 16 nm mixed-signal
CMOS process (Figure 2). This architecture is specifically designed to
combine the speed and noise properties of probabilistic switching
elements with the scalability and manufacturability of advanced CMOS
technology. The chip is partitioned into multiple physical and logical
tiles, each containing clusters of p-qubits, local coupling networks,
dedicated memory, and entropy-generation subsystems. Together, these
elements form the fundamental computational substrate upon which
energy-based models, quantum-driven Hamiltonians, and analog
vector--matrix operations are
executed$^{[93][94]}$.

\begin{figure}[H]
\centering
{\small\bfseries Illustration of the Multi-Layer Tiled System\par\vspace{2pt}}
\includegraphics[width=0.98\linewidth]{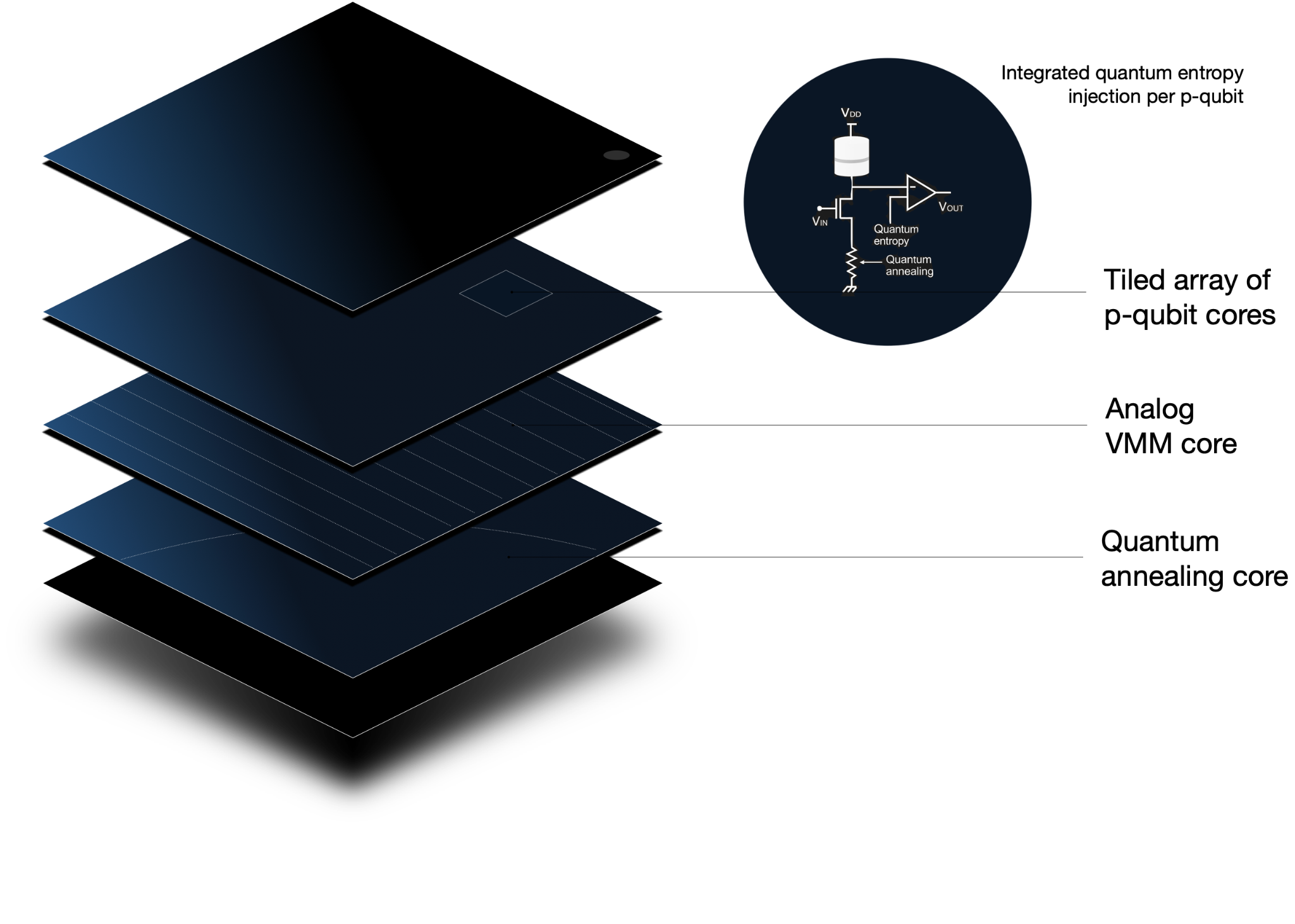}
\caption{Illustration of the multi-layer tiled system, including (i) a tiled arrangement of p-qubit cores, (ii) an analog CMM (correlation matrix memory) core, and (iii) a quantum annealing core.}
\end{figure}

At a high level, the Apollo organisation enables full parallelism across
all 10,000 p-qubits without relying on time-multiplexing or sequential
scanning. Each p-qubit operates continuously and asynchronously,
switching at 12.5ps rates exceeding $8\times 10$\textsuperscript{7}
spin-flips/ns per device. This design provides not only exceptional
computational throughput but also a unique substrate whose physical
dynamics closely emulate low-energy quantum
systems$^{[82][83]}$.

Apollo is composed of a grid of repeated tiles, each of which contains a
fixed number of p-qubits arranged in locally connected
clusters$^{[95][96]}$. These tiles serve as the
fundamental architectural building blocks, providing:

\begin{itemize}
\item
  locality of computation, with clustered p-qubits supporting
  high-bandwidth intra-tile coupling,
\item
  scalability, allowing the chip to be expanded to tens of thousands of
  units without excessive wiring congestion,
\item
  modularity, enabling independent control and configuration of
  individual regions of the chip,
\item
  topological consistency, ensuring each tile conforms to the $\Delta$256 graph
  used for embedding QUBO/Ising and Hamiltonian problems.
\end{itemize}

The tiled structure additionally simplifies layout, supports
hierarchical routing of analog and digital signals, enables partial
reconfiguration, and reduces parasitics that could otherwise compromise
analog stochastic behaviour$^{[97][98]}$.

\subsection{p-Qubit Circuit Architecture}

Within each computational tile, p-qubits are organized into compact,
tightly coupled clusters designed to support high-bandwidth analog
interactions and low-latency signal propagation. This clustered layout
enables local communication to occur over short physical distances,
reducing parasitic effects and preserving the fidelity of
continuous-time stochastic dynamics. Each p-qubit is implemented as a
CMOS latch-like probabilistic element that integrates controlled entropy
injection, tunable biasing, and programmable coupling inputs within a
unified analog circuit$^{[93][97]}$.

At the circuit level, a p-qubit comprises three essential functional
components: (i) a source of quantum-derived stochastic excitation
supplied by a co-located independent quantum entropy unit (IQEU); (ii) a
programmable analog bias that sets the intrinsic preference of the
p-qubit state; and (iii) weighted coupling inputs originating from
neighbouring p-qubits within the network. The interaction of these
elements enables each p-qubit to operate as a continuously driven
stochastic system whose instantaneous state reflects the balance between
local fields, injected noise, and nonlinear activation
dynamics$^{[77][82]}$.

The constituent mechanisms of a single p-qubit are illustrated in
Figures 3--7. Specifically, each p-qubit
provides$^{[85][98]}$:

\begin{itemize}
\item
  Independent analog bias inputs (Fig. 3), allowing fine-grained control
  of the local field applied to each unit and enabling direct encoding
  of Ising or QUBO bias terms;
\item
  Local weighted-sum coupling inputs (Fig. 5), generated by a
  floating-gate (FG)--based vector--matrix multiplication (VMM) array
  that aggregates currents from neighbouring p-qubits in the analog
  domain;
\item
  A sigmoidal transfer characteristic approximating a hyperbolic tangent
  activation function (Fig. 6), implemented through a nine-transistor
  operational transconductance amplifier (OTA) that converts weighted
  input currents into a smooth nonlinear voltage response;
\item
  On-device quantum-derived entropy injection, supplied by the
  associated IQEU and introduced directly at the OTA input to regulate
  the stochastic switching behaviour; and
\item
  Fully asynchronous operation without a global clock (Fig. 9), allowing
  p-qubit state transitions to occur in continuous time rather than at
  discretized update intervals.
\end{itemize}

This circuit architecture ensures that p-qubits do not function as
deterministic logic elements, nor do they rely on clocked digital update
cycles. Instead, each p-qubit behaves as a physically stochastic,
energy-driven computational unit whose probabilistic switching dynamics
emerge naturally from analog interactions and quantum-derived noise.
Such behaviour is fundamental to annealing processes, Boltzmann
sampling, Ising dynamics, and the broader class of quantum-driven
neuromorphic computation targeted by the Apollo
architecture$^{[1][41]}$.

The tight spatial clustering of p-qubits within each tile further
enhances circuit performance by minimizing analog path lengths,
improving device matching, and reducing latency between coupled units.
As a result, interactions defined by the energy model are implemented
with high temporal resolution and minimal distortion, enabling efficient
exploration of complex energy landscapes through continuous-time
stochastic relaxation$^{[97][98]}$.

\begin{figure}[H]
\centering
{\small\bfseries Circuit Diagram of a Single p-Qubit\par\vspace{2pt}}
\includegraphics[width=0.98\linewidth]{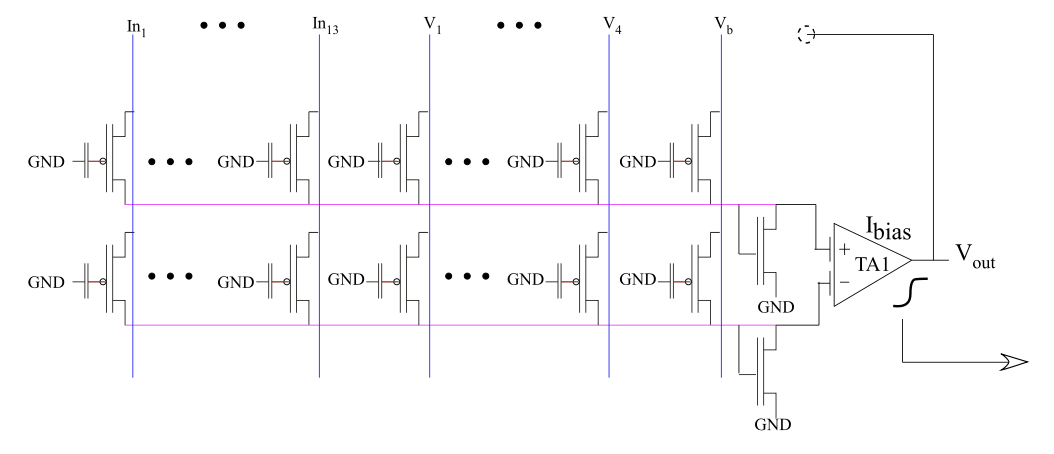}
\caption{Circuit diagram of a single p-qubit within the p-qubit network and its corresponding output path. The p-qubit receives its local field I through the FG-based vector--matrix multiplication (VMM) array, where weighted currents from neighbouring p-qubits are summed in the analog domain. This input current is combined with the quantum-mechanical entropy injection delivered by the co-located IQEU. Both contributions feed into a nine-transistor operational transconductance amplifier (OTA), whose bias current $I$\textsubscript{bias} controls the operating point and nonlinear gain. The OTA produces a smooth sigmoidal voltage response that approximates a tanh activation function, setting the instantaneous switching probability of the p-qubit. The resulting output signal is fed back into the network through the analog routing fabric, completing the continuous-time probabilistic update loop.}
\end{figure}

\begin{figure}[H]
\centering
{\small\bfseries Circuit Diagram of the OTA\par\vspace{2pt}}
\includegraphics[width=0.98\linewidth]{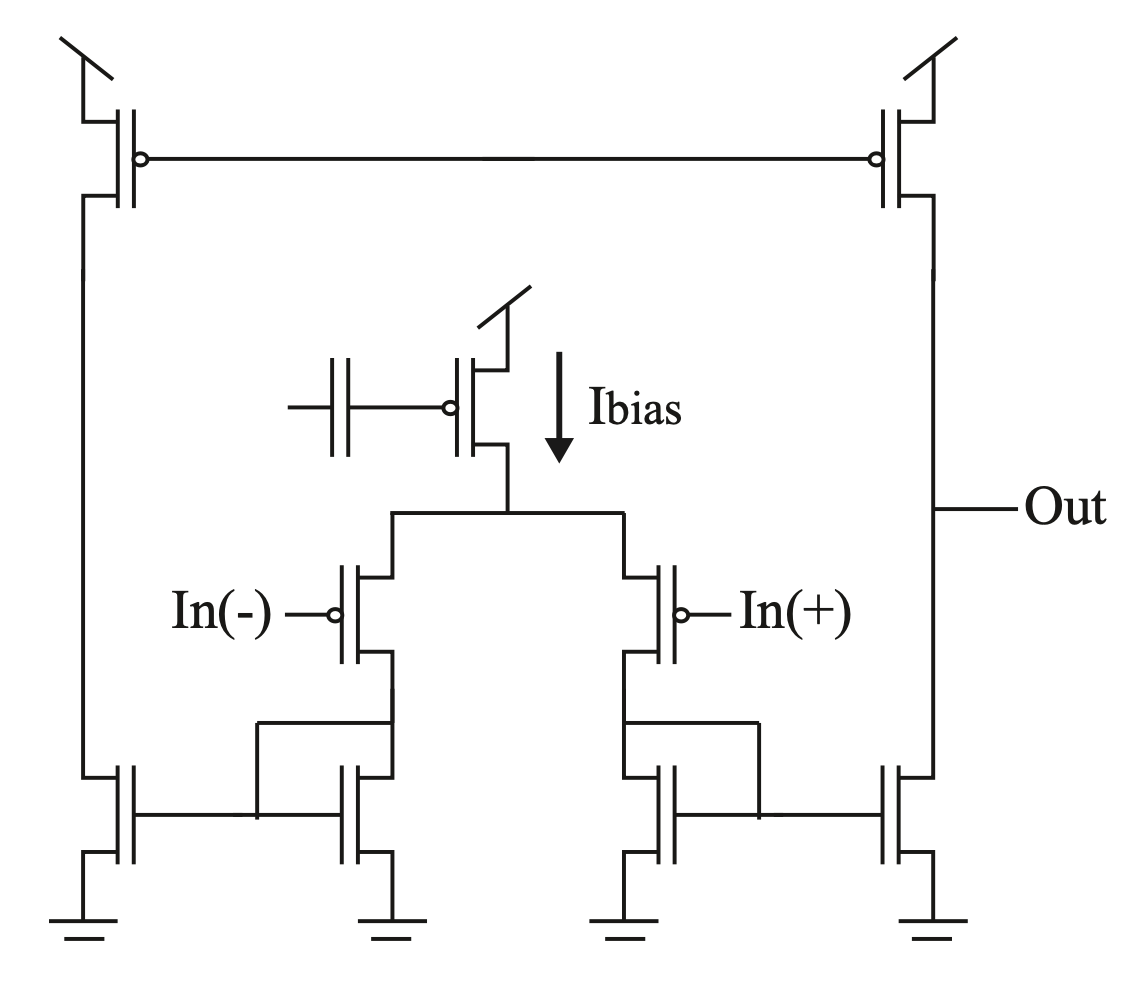}
\caption{The operational transconductance amplifier (OTA) is implemented as a nine-transistor OTA block. It receives two input currents: (a) the local field I generated by the p-qubit network through the FG-based weighted-sum VMM, and (b) the quantum-mechanical entropy input supplied by the associated IQEU. The bias current $I$bias sets the OTA's operating point and transconductance. The OTA outputs an analog sigmoidal transfer characteristic, providing the nonlinear activation that governs each p-qubit's probabilistic switching behaviour.}
\end{figure}

The close physical proximity of clustered units minimises latency
between coupled nodes, allowing the analog interactions of the energy
model to be implemented with high fidelity$^{[97]}$.

\subsection{Analog VMM and Weight Storage}

The interaction structure of the p-qubit network is implemented through
a fully analog vector--matrix multiplication (VMM) fabric that performs
weighted current summation directly within the CMOS substrate. This
approach enables the physical realization of Ising and QUBO coupling
matrices without reliance on digital memory, clocked arithmetic, or
sequential accumulation. Instead, coupling weights are stored and
applied in situ, allowing interactions to propagate continuously and
asynchronously throughout the
network$^{[48][93][94]}$.

\subsubsection{Floating-Gate--Based Vector--Matrix Multiplication}

Figure 5 illustrates the circuit schematic of the p-qubit network,
highlighting the implementation of weighted input aggregation and
nonlinear activation. The VMM is constructed from floating-gate (FG)
pFET transistors, each of which encodes an analog weight as charge
stored on an electrically isolated gate. The programmed floating-gate
charge directly modulates the effective transconductance of the device,
thereby defining the coupling strength between a source p-qubit and a
target p-qubit$^{[99][100]}$.

In operation, output voltages from neighbouring p-qubits drive the gates
of FG pFETs arranged in a matrix structure. The resulting drain currents
are proportional to both the input signal and the stored FG charge,
implementing an analog multiplication. Currents from multiple FG devices
are summed naturally in the current domain along shared lines, yielding
a weighted-sum signal that represents the local field acting on a given
p-qubit. This process occurs continuously and in parallel across the
entire network$^{[93][94]}$.

Because weight values are stored directly on the floating gates, the VMM
operates without external memory accesses or digital control loops. Once
programmed, the weights are non-volatile and remain stable over extended
periods, eliminating refresh overhead and enabling truly in-place
computation$^{[101][102]}$.

\begin{figure}[H]
\centering
{\small\bfseries Circuit Schematic of the p-Qubit Network\par\vspace{2pt}}
\includegraphics[width=0.98\linewidth]{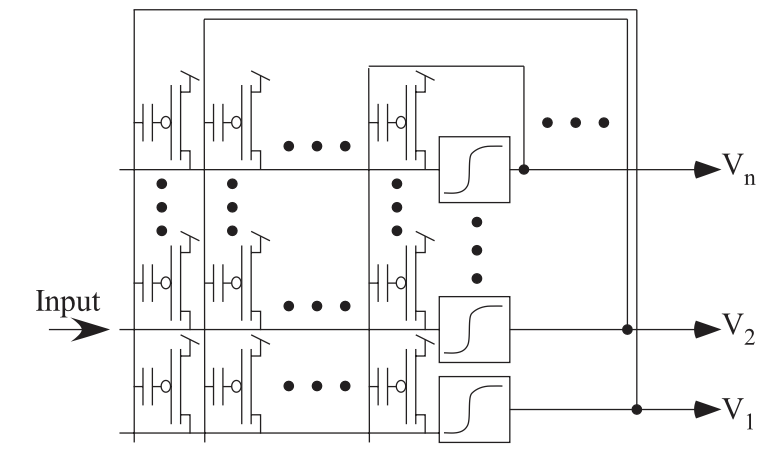}
\caption{Circuit schematic of the p-qubit network. The inputs are weighted and summed using a VMM made of FG pFETs. A TA and I2V circuit takes the weighted current and maps it onto the TA sigmoidal nonlinearity. Both FG and normal TAs are used, and the I2V circuit uses FG pFETs, pFETs, and nFETs as shown. A bias element keeps an initial constant current in the row.}
\end{figure}

\subsubsection{Transimpedance and Current-to-Voltage Conversion}

The weighted currents generated by the FG-based VMM are translated into
voltage signals through a combination of transimpedance amplifiers (TAs)
and current-to-voltage (I2V) conversion stages, as shown in Figures 7
and 8. These blocks map the aggregated current into the voltage domain
required by the downstream nonlinear activation
circuitry$^{[97][103]}$.

Both floating-gate--based and standard CMOS transimpedance amplifiers
are employed within the network to balance tunability, robustness, and
compatibility with the surrounding analog fabric. The I2V circuits
incorporate FG pFETs, pFETs, and nFETs, allowing their operating
characteristics to be adjusted through programmable biasing. A dedicated
bias element maintains a baseline current through each VMM row,
stabilizing the operating point and ensuring consistent analog behaviour
across the cluster$^{[98][101]}$.

Figure 6 shows the sigmoidal response of the operational
transconductance amplifier (OTA) driven by the I2V stage. By selecting
appropriate gain configurations, the OTA produces a smooth nonlinear
transfer characteristic that closely approximates the hyperbolic tangent
function, thereby defining the probabilistic switching behaviour of each
p-qubit$^{[9]}$.

\begin{figure}[H]
\centering
{\small\bfseries Sigmoidal Response of the OTA\par\vspace{2pt}}
\includegraphics[width=0.98\linewidth]{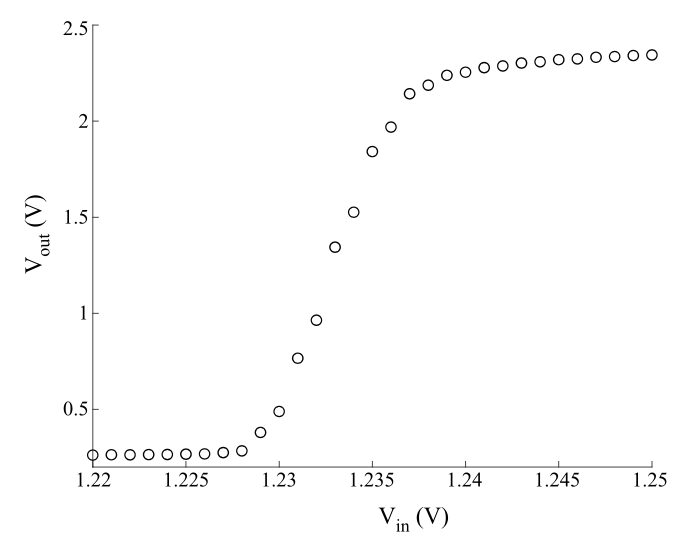}
\caption{Sigmoidal response of an OTA. Both normal and FG OTAs are used in the p-qubit network; using the large gain option on the TA allows for a sharp response.}
\end{figure}

\subsubsection{Floating-Gate Devices as Non-Volatile Analog Weights}

Floating-gate devices are MOS transistors whose gate terminals are
electrically isolated by high-quality oxide, allowing charge to be
stored for extended durations. Weight programming is achieved using
established techniques such as hot-electron injection or
Fowler--Nordheim tunneling, depending on the CMOS process. Once
programmed, the stored charge defines a stable, continuous-valued weight
without the need for refresh or readout
circuitry$^{[100][102]}$.

Figure 7 illustrates the source sweep characteristics of an FG pFET used
in the VMM, demonstrating how different programmed charge levels
modulate the device's current--voltage response. This behaviour enables
precise, physically continuous encoding of coupling strengths and bias
terms directly within the analog
fabric$^{[98][101]}$.

In contrast to digital accelerators, where weights must be repeatedly
fetched from SRAM or DRAM, the FG-based VMM performs computation in
place: the memory and compute elements are physically identical. This
eliminates memory bandwidth bottlenecks and allows all matrix--vector
operations to proceed at the speed of the underlying device
physics$^{[93][94]}$.

\begin{figure}[H]
\centering
{\small\bfseries Source Sweep of an FG pFET\par\vspace{2pt}}
\includegraphics[width=0.98\linewidth]{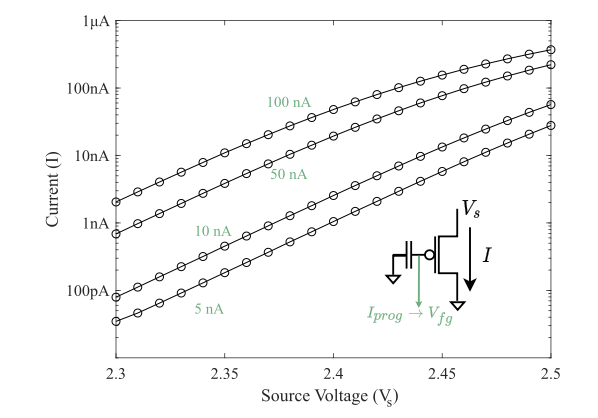}
\caption{Source sweep of an FG pFET as used in the VMM. Different programmed charges weight the input voltage, pushing up the I--V curve.}
\end{figure}

\subsubsection{Local Analog Coupling and Routing Fabric}

Each tile incorporates a local analog routing fabric that distributes
weighted current signals among p-qubits within the tile and across
neighbouring tiles. The routing network is derived from a
Manhattan-style interconnect architecture, providing regular horizontal
and vertical channels that support dense, locality-preserving mappings
of VMM rows and columns$^{[93][94]}$.

This routing fabric enables the $\Delta$256 connectivity model by supporting up
to 256 effective weighted couplings per p-qubit through a combination
of:

\begin{itemize}
\item
  current-mode and differential-mode analog routing paths within the
  CMOS fabric;
\item
  FG-encoded coupling weights stored directly on pFET floating gates;
\item
  hierarchical inter-tile routing channels for distributing weighted
  currents across tile boundaries;
\item
  short-range analog interconnects with minimal parasitic delay; and
\item
  digitally configured routing maps, set by the control unit, that
  select which FG weights participate in the active coupling
  pattern$^{[95][100]}$.
\end{itemize}

In this architecture, the Ising or QUBO coupling matrix is physically
instantiated in the floating-gate devices themselves. Local field
contributions propagate continuously through the analog current network
rather than being evaluated in discrete clocked cycles, enabling rapid
convergence dynamics and faithful realization of energy-based
interactions$^{[1][9]}$.

\subsubsection{Memory-Mapped Weights and Biases in the Analog Fabric}

Unlike conventional digital accelerators, Apollo embeds all problem
parameters directly within the analog compute fabric. Each coupling
weight is stored as charge on an FG pFET in the VMM array, and each bias
term is implemented through FG-programmed bias currents and voltage
offsets. As a result, the analog substrate simultaneously serves as both
memory and compute engine$^{[93][94]}$:

\begin{itemize}
\item
  The control unit does not store the full coupling matrix. Instead, it
  programs:
\item
  floating-gate weight values through dedicated programming pulses;
\item
  bias levels for individual p-qubits;
\item
  dynamic parameters such as gain, clamping, and noise scaling;
\item
  annealing schedules and temporal ramps; and
\item
  routing configurations that determine which weights are
  active$^{[95][100]}$.
\end{itemize}

\begin{figure}[H]
\centering
{\small\bfseries Current-in Versus Voltage-out\par\vspace{2pt}}
\includegraphics[width=0.98\linewidth]{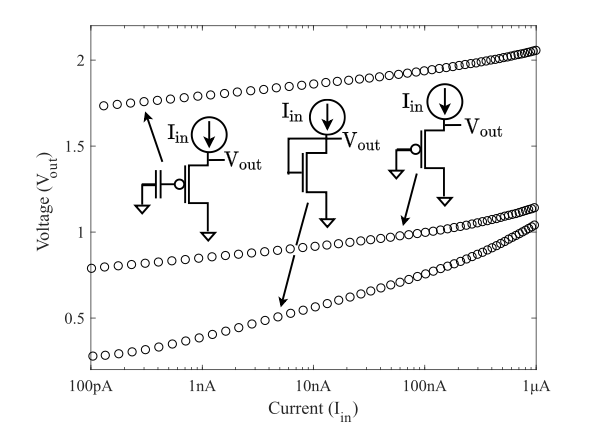}
\caption{Current in versus voltage out of the three different I to V converters used. In the p-qubit network the current in is provided by a VMM representing the weighted sum of all the inputs to a p-qubit.}
\end{figure}

\subsection{Entropy Generation and Conditioning}

Stochastic behaviour in Apollo is driven by a hybrid entropy-generation
system consisting of:

\begin{itemize}
\item
  Independent Quantum/Intrinsic Entropy Units (IQEUs) for each p-qubit,
  and
\item
  A shared tile-level entropy conditioning
  network$^{[90][100]}$.
\end{itemize}

Each IQEU provides raw physical randomness derived from
quantum-mechanical entropy sources, such as electron tunnelling
fluctuations, quantum shot noise, and other non-deterministic quantum
processes inherent to nanoscale CMOS devices. These quantum-origin
entropy streams may be combined with supplemental stochastic mechanisms
(e.g., thermal or jitter-based noise) depending on the implementation,
but the foundational randomness is supplied by inherently quantum
processes rather than algorithmic or pseudo-random generators. These raw
entropy streams are then processed through:

\begin{itemize}
\item
  whitening filters,
\item
  bias-removal stages,
\item
  de-correlation logic, and
\item
  amplitude-shaping circuits$^{[102][104]}$.
\end{itemize}

The shared entropy conditioning network aggregates, normalises, and
distributes entropy with consistent bandwidth across all p-qubits within
a tile. This ensures stable statistical behaviour, reproducible
annealing trajectories, and high-quality Boltzmann
sampling$^{[8][105]}$.

Unlike random-number generators in digital systems, Apollo's entropy
model interacts directly with the physical switching dynamics of the
p-qubit, giving rise to the analog stochasticity needed for
quantum-driven computation$^{[8][28]}$.

\subsection{Continuous-Time, Clock-less Dynamics}

A defining feature of Apollo's probabilistic computing substrate is its
continuous-time, clock-less dynamical behaviour. Unlike synchronous
digital annealers, discrete-step Markov Chain Monte Carlo (MCMC)
samplers, or quantum-driven accelerators that rely on periodic global
updates, Apollo's 10,000 p-qubits operate asynchronously and
autonomously. Each p-qubit evolves according to its local physical
conditions and coupling environment, producing a dynamical system whose
behaviour more closely resembles condensed-matter spin glasses,
stochastic differential systems, or continuous-time Boltzmann machines
than traditional digital hardware$^{[7][8]}$.

\begin{figure}[H]
\centering
{\small\bfseries Comparing Digital to Clockless p-Computer\par\vspace{2pt}}
\includegraphics[width=0.98\linewidth]{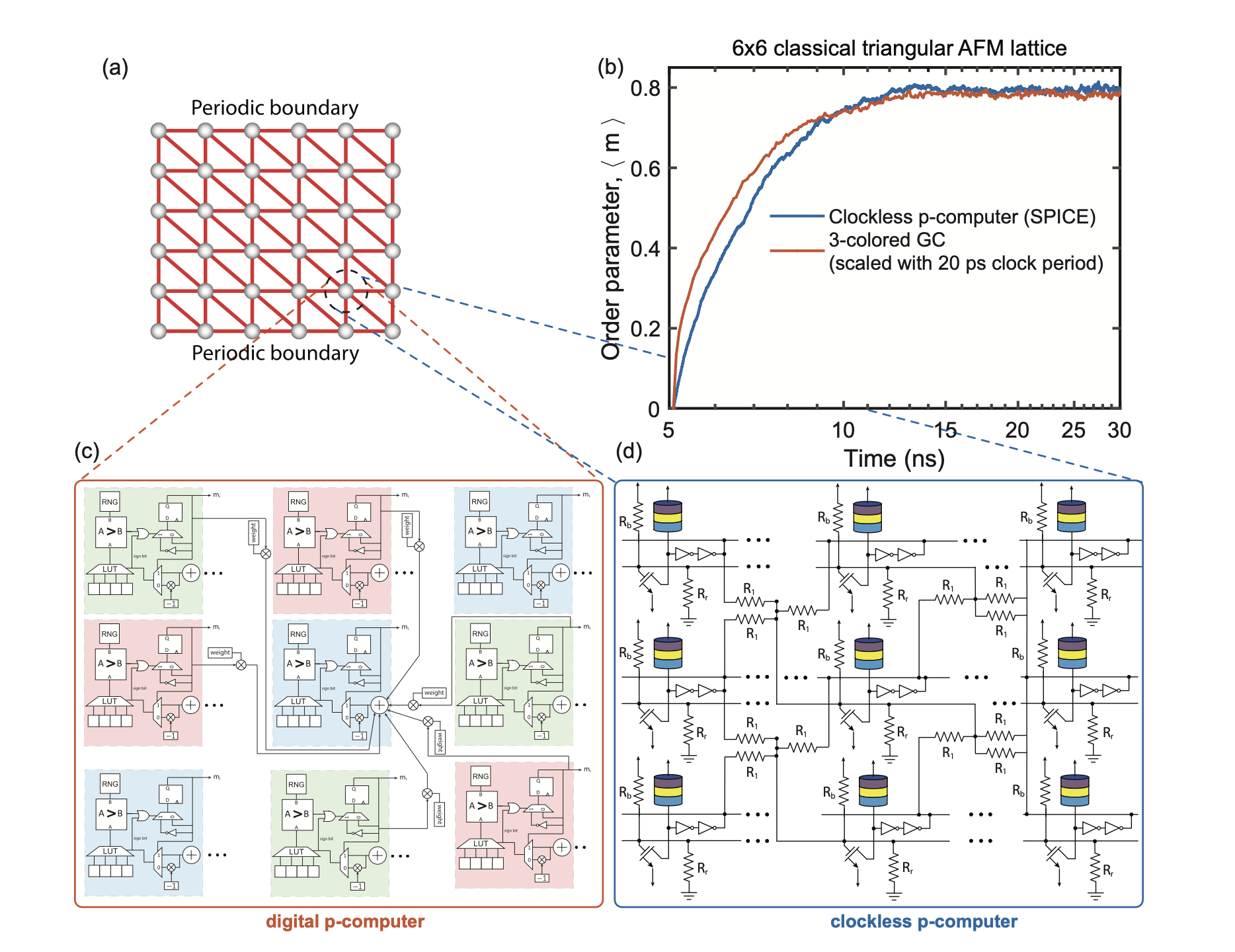}
\caption{Comparing digital to clockless p-computer: (a) A 6 $\times$ 6 antiferromagnetically (AFM) coupled triangular lattice with classical spins is shown. (b) The convergences of the order parameter for the lattice shown in (a) are plotted for two different p-computer design approaches (red solid line: graph-colored based digital design and blue solid line: nanomagnets based analog design) discussed in this work. We have used $\beta = 2$ in this example. (c) The graph-colored (GC) based digital p-computer design where the convergence is estimated from MATLAB simulations assuming $c \times f_c$ (c = 3 for triangular lattice) sweeps are collected every per second, fc being the clock frequency. (d) The clock-less p-computer design which is simulated using SPICE simulator.}
\end{figure}

In the absence of a global clock, the state of each p-qubit is governed
by four principal contributions$^{[50][51]}$:

\begin{itemize}
\item
  Local effective field determined by the programmed bias\(h_{i}\) ,
  Weighted analog inputs arising from neighbouring p-qubits via the $\Delta$256
  coupling fabric,
\item
  Quantum-mechanical entropy input provided by its associated IQEU,
\item
  External modulation delivered by the DCU, including annealing ramps,
  noise shaping, clamping, or bias-schedule injection.
\end{itemize}

These components interact continuously, producing an evolving local
potential landscape. The p-qubit's analog output voltage and switching
probability adjust in real time as its local field fluctuates, with
transitions occurring whenever the instantaneous energy gradient favours
a state flip. Because the updates are stochastic and uncorrelated across
devices, the chip naturally exhibits distributed, event-driven dynamics
without requiring synchronisation$^{[17][28]}$.

\textbf{Local Field--Driven Behaviour}

The local field
\(u_i(t)\)
is computed analogly as a weighted sum of neighbouring states combined
with the static or time-varying bias:

\begin{equation}
u_i(t)=h_i+\sum_j W_{ij}\,x_j(t)+\xi_i(t)\label{eq:local_field}
\end{equation}

where
\(\xi_i(t)\)
is the injected quantum-mechanical noise from the IQEU. As this field
evolves, the p-qubit continuously relaxes toward a statistical
equilibrium characterised by a tanh-like activation probability. The
absence of discretised update cycles means that the system explores the
energy landscape smoothly, maintaining sensitivity to small
perturbations and enabling fine-grained transitions that would be lost
in a discrete-step sampler$^{[7][41]}$.

\textbf{Weighted Neighbour Influence}

Coupling interactions are governed by on-chip analog routing, allowing
the influence of neighbouring p-qubits to propagate with minimal
latency. This yields a system where$^{[9][94]}$:

\begin{itemize}
\item
  interactions are not gated by a central clock,
\item
  weighted sums evolve at electrical speed limits,
\item
  the network equilibrates via physical relaxation rather than
  algorithmic iteration.
\end{itemize}

The result is an inherently parallel and continuous update mechanism,
where thousands of interacting p-qubits simultaneously adjust their
states as the global energy landscape shifts.

\textbf{Quantum-Mechanical Entropy from IQEUs}

Each p-qubit receives its own stream of quantum-origin entropy, ensuring
that stochastic state transitions are driven by non-deterministic
physical processes rather than by algorithmic pseudo-random number
generators. This entropy acts as the analog of thermal fluctuations in
Ising models or Langevin noise in stochastic differential equations.
Because each p-qubit has its own entropy source, noise correlations are
minimised, enabling rich exploration dynamics, fast mixing, and
high-quality Boltzmann sampling$^{[8][104]}$.

\textbf{DCU-Modulated Scheduling as External Control}

While the core dynamics are autonomous, the DCU provides temporal
control over:

\begin{itemize}
\item
  effective temperature (noise scaling),
\item
  bias ramps and annealing profiles,
\item
  coupling strength modulation,
\item
  reverse annealing,
\item
  dynamic clamping or selective freezing of subsets of p-qubits.
\end{itemize}

These schedules are injected at petasecond resolution, allowing Apollo
to execute complex annealing, sampling, or variational cycles without
ever interrupting continuous-time
dynamics$^{[15][60]}$.

\textbf{Self-Equilibrating Analog System}

The combination of asynchronous switching, continuous analog
interaction, and quantum-driven entropy yields a self-equilibrating
stochastic physical system. Rather than performing a discrete sequence
of algorithmic updates, Apollo relaxes toward low-energy states
naturally, similar to$^{[8][64]}$:

\begin{itemize}
\item
  spin glasses approaching metastable minima,
\item
  analog Hopfield networks evolving toward attractors,
\item
  dilute magnetic systems exhibiting stochastic resonance,
\item
  continuous-time energy-based models in machine learning.
\end{itemize}

This contrasts sharply with synchronous digital annealers, which
require:

\begin{itemize}
\item
  explicit temperature schedules,
\item
  sequential update cycles,
\item
  global clocks enforcing artificial synchronisation.
\end{itemize}

The absence of these constraints allows Apollo to exhibit ultra-fast
mixing, high exploration bandwidth, and nonlinear dynamical responses
that are particularly advantageous for optimization, Boltzmann sampling,
and generative-model inference.

\textbf{Implications for Computation}

Continuous-time dynamics provide several computational benefits:

\begin{itemize}
\item
  Massive parallelism: all 10,000 p-qubits evolve simultaneously.
\item
  Rich stochastic trajectories: noise-driven exploration occurs at
  device-level bandwidth.
\item
  Low-latency convergence: analog relaxation avoids overheads of digital
  iteration loops.
\item
  Quantum-driven behaviour: stochastic flips emulate tunnelling-like
  transitions in rugged landscapes.
\item
  Robustness to noise: physical entropy enhances mixing rather than
  impairing computation.
\end{itemize}

Altogether, the continuous-time architecture positions Apollo as a
fundamentally different computational system from digital accelerators
or traditional annealers---closer to a scalable, CMOS-integrated
physical energy minimisation engine$^{[78][79]}$.

\subsection{Coupling Fabric and $\Delta$256 Topology}

The interconnect fabric provides sparse yet high-degree connectivity
(maximum degree 256), reducing minor-embedding overhead by orders of
magnitude$^{[13][88]}$.

\begin{table}[H]
\centering
{\scriptsize
\resizebox{\linewidth}{!}{%
\begin{tabular}{p{0.27\linewidth}p{0.16\linewidth}p{0.22\linewidth}p{0.26\linewidth}}
\toprule
\textbf{Platform} & \textbf{Max Degree $\Delta$} & \textbf{Embedding Overhead} & \textbf{Notes} \\
\midrule
D-Wave 2000Q (Chimera) & 6 & Very high & Extensive auxiliary variables required \\
D-Wave Advantage (Pegasus) & 15 & Reduced, costly & Replication still dominant \\
D-Wave Advantage2 (Zephyr) & 20 & Improved & Structural limits persist \\
Apollo (Hyperion) & 256 & Minimal & Native dense-graph embedding \\
\bottomrule
\end{tabular}%
}
}
\caption{Comparative overview of embedding topologies (Chimera, Pegasus, Zephyr, Hyperion) utilised in superconducting quantum annealers (D-Wave) and in quantum-driven neuromorphic systems (Apollo).}
\end{table}

\subsection{Control and Orchestration Layer}

The Dynex Control Unit (DCU) is a dedicated FPGA-based orchestration
subsystem that forms the central control and coordination layer of the
Apollo architecture (Figure 10). While the Apollo chip provides the
physical probabilistic computing substrate---implementing high-speed
p-qubit dynamics, vector--matrix accumulation, and analog stochastic
switching---the DCU is responsible for preparing inputs, managing
computation schedules, embedding problem graphs, and extracting results
in real time. As such, the DCU is not an auxiliary component but an
essential co-processor; Apollo cannot operate in isolation and always
functions as part of the DCU--Apollo composite
system$^{[94][95]}$.

\begin{figure}[H]
\centering
{\small\bfseries Dynex Control Unit (DCU)\par\vspace{2pt}}
\includegraphics[width=0.98\linewidth]{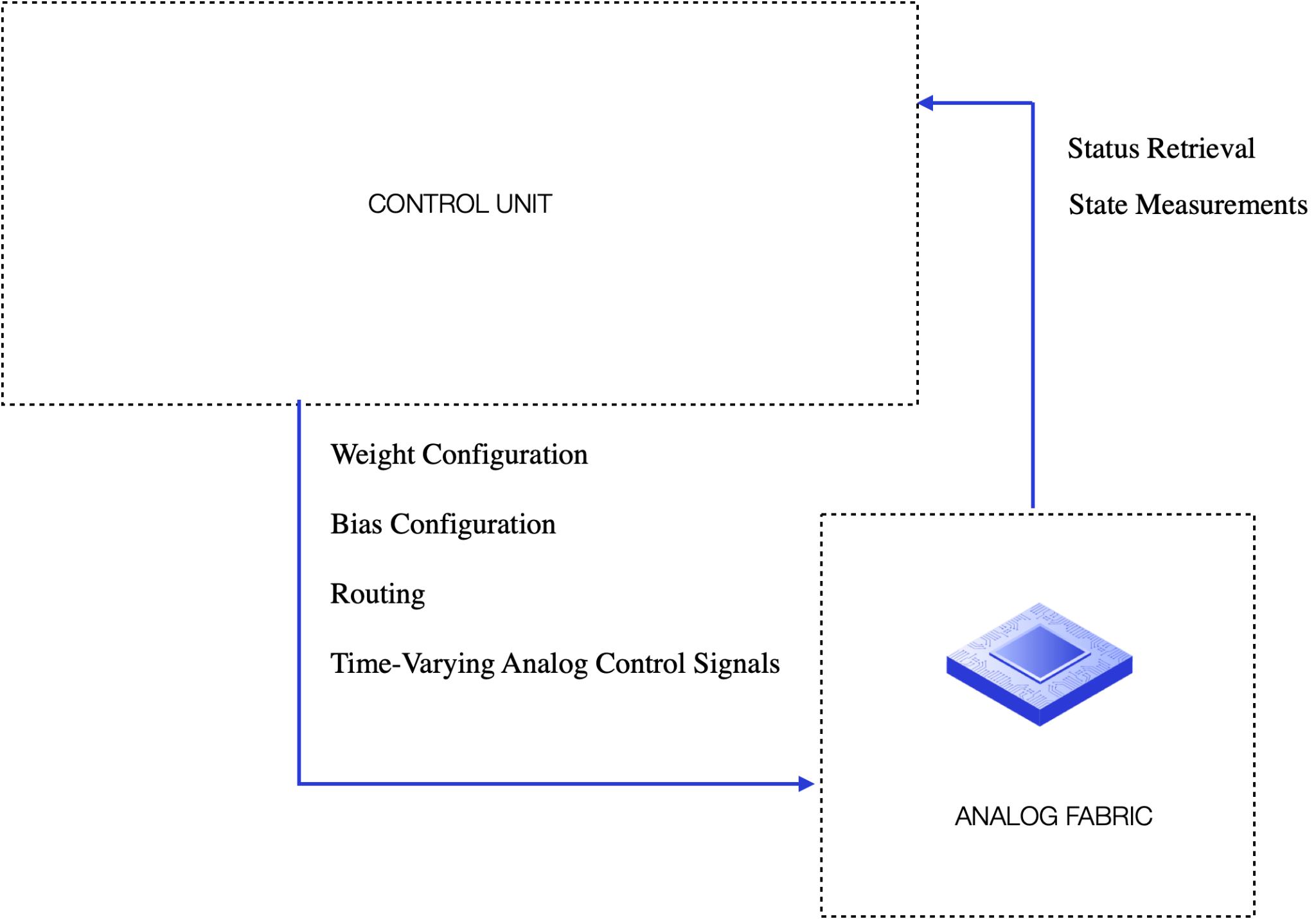}
\caption{Illustration of the Dynex Control Unit (DCU) integrated into the quantum-driven computing platform.}
\end{figure}

At a high level, the DCU performs four interdependent roles: (1) problem
preprocessing and normalisation, (2) graph embedding onto Apollo's $\Delta$256
topology, (3) dynamic injection of annealing and sampling schedules, and
(4) high-throughput streaming readout and post-processing. Each of these
stages is implemented as part of a deeply pipelined, clock-synchronous
hardware system, ensuring deterministic timing, nanosecond-scale
control, and compatibility with hybrid host
architectures$^{[95][106]}$.

\textbf{Problem Preprocessing and Normalisation}

Before a QUBO, Ising, Hamiltonian, or energy-based model can be mapped
onto Apollo, the DCU performs a normalisation step that transforms raw
problem coefficients into a hardware-compliant range. This includes:

\begin{itemize}
\item
  scaling of weights and biases into the supported analog domain,
\item
  clipping or re-parametrising excessively large couplings to preserve
  stability,
\item
  encoding sparse or dense connectivity matrices into compressed
  FPGA-ready formats,
\item
  applying fixed-point quantisation compatible with Apollo's analog
  midpoint and dynamic range.
\end{itemize}

These transformations ensure that arbitrarily large or high-precision
problems can be embedded without exceeding the physical or electrical
limits of the p-qubit network$^{[88][107]}$.

\textbf{Graph Embedding onto the $\Delta$256 Topology}

The Apollo architecture implements a $\Delta$256 connectivity pattern, a
hardware-optimised reconfigurable graph supporting up to 256 weighted
couplings per p-qubit. Because problem graphs are often not naturally
constrained to this topology, the DCU performs automatic embedding. This
includes:

\begin{itemize}
\item
  mapping logical nodes to physical p-qubits,
\item
  routing couplings through weighted interconnect layers,
\item
  partitioning and tiling large problems across multiple $\Delta$256 tiles,
\item
  resolving conflicts and degeneracies using heuristics or embedding
  optimization algorithms.
\end{itemize}

This embedding process is analogous to minor embedding in quantum
annealers but significantly more flexible because the $\Delta$256 graph
supports both local and global coupling patterns, as well as dynamic
reconfiguration across computation
cycles$^{[13][108]}$.

\textbf{Dynamic Annealing and Sampling Schedule Injection}

During execution, the DCU injects dynamic control schedules that
modulate the effective fields of the p-qubit network. These schedules
govern quantities such as:

\begin{itemize}
\item
  bias amplitude and temporal modulation (h-terms),
\item
  coupling strength evolution (W-terms),
\item
  effective temperature or noise scaling,
\item
  update frequency envelopes,
\item
  annealing ramps, quench cycles, or hybrid anneal--sample sequences.
\end{itemize}

Schedules can be fully programmable, enabling classical simulated
annealing, quantum-driven annealing, reverse annealing, Boltzmann
sampling, or hybrid protocols. Because the DCU operates at FPGA-level
speeds, schedule updates can occur on sub-microsecond timescales,
supporting highly adaptive workflows such as variational optimization or
reinforcement learning$^{[15][60]}$.

\textbf{High-Throughput Streaming Readout and Post-Processing}

The DCU continuously reads out the state of each p-qubit at high
sampling rates and streams these values into a dedicated processing
pipeline. Readout modes include:

\begin{itemize}
\item
  raw spin-state sampling,
\item
  energy tracking and acceptance statistics,
\item
  histogram accumulation for Boltzmann sampling,
\item
  best-state tracking for optimization workloads,
\item
  temporal correlation analysis for dynamical systems.
\end{itemize}

Because the output bandwidth far exceeds that of traditional digital
annealers, the DCU can support real-time adaptive control loops, online
learning algorithms, and large-scale repeated sampling needed for
uncertainty quantification and generative model
training$^{[71][77]}$.

\begin{figure}[H]
\centering
{\small\bfseries Dynex Control Unit (DCU) FPGA Architecture\par\vspace{2pt}}
\includegraphics[width=0.98\linewidth]{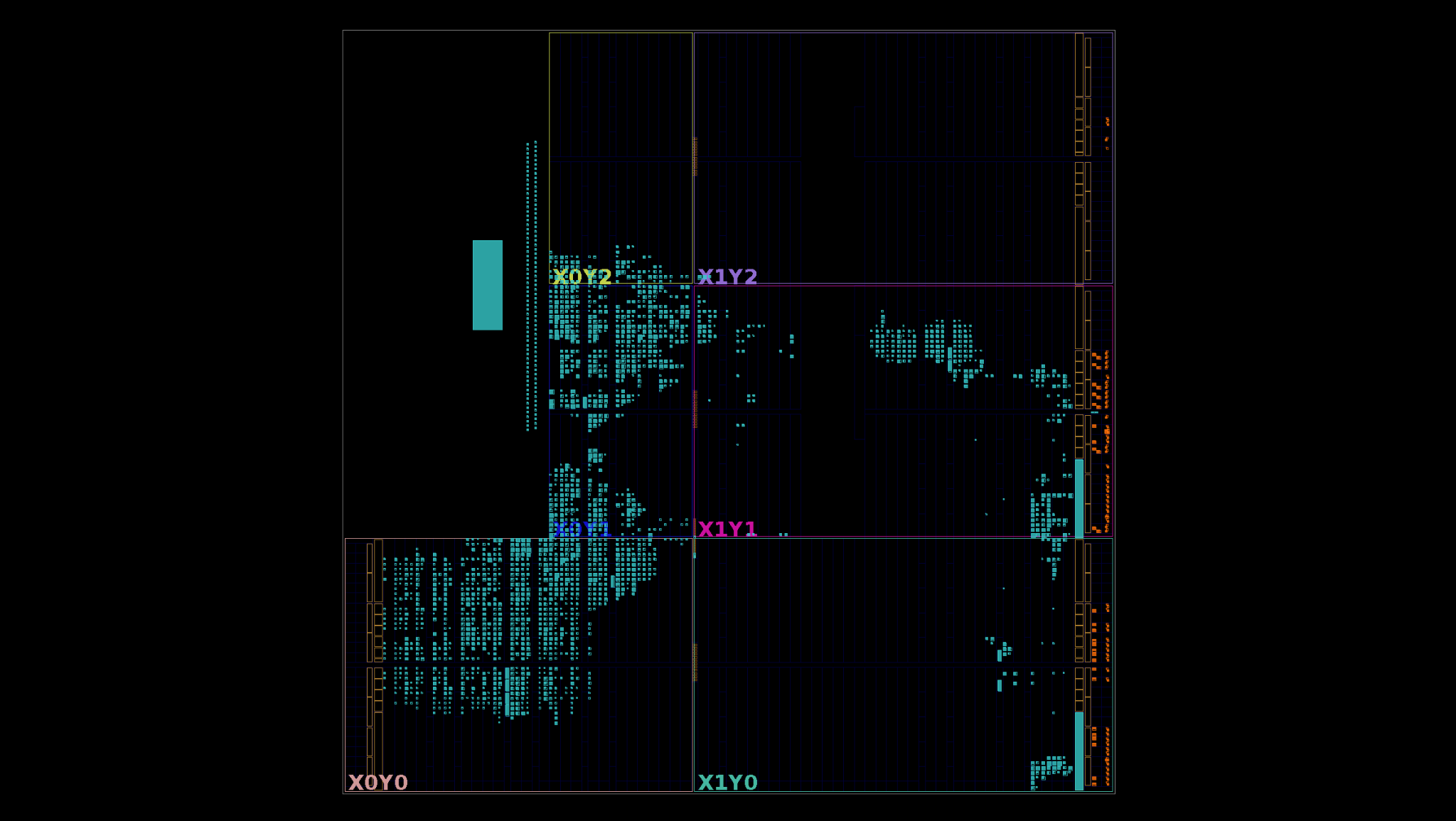}
\caption{Illustration of the implemented architecture deployed on the FPGA of the Dynex Control Unit (DCU).}
\end{figure}

\begin{table}[H]
\centering
{\small\bfseries DCU Resource Utilisation\par\vspace{3pt}}
{\scriptsize
\resizebox{\linewidth}{!}{%
\begin{tabular}{llll}
\toprule
\textbf{Resource} & \textbf{Utilization} & \textbf{Available} & \textbf{Utilization \%} \\
\midrule
LUT & 5,195 & 53,200 & 9.77\% \\
LUTRAM & 1,335 & 17,400 & 7.67\% \\
FF & 3,045 & 106,400 & 2.86\% \\
BRAM & 0.50 & 140 & 0.36\% \\
DSP & 9 & 220 & 4.09\% \\
IO & 57 & 200 & 28.50\% \\
BUFG & 6 & 32 & 18.75\% \\
MMCM & 2 & 4 & 50.00\% \\
\bottomrule
\end{tabular}%
}
}
\caption{Summary of FPGA resource utilisation for the DCU implementation, detailing the consumption of logic (LUTs/LUTRAM/FFs), memory (BRAM), arithmetic units (DSPs), and clocking and routing resources (I/O, BUFG, MMCM).}
\end{table}

\begin{figure}[H]
\centering
{\small\bfseries DCU On-Chip Power Consumption\par\vspace{2pt}}
\includegraphics[width=0.98\linewidth]{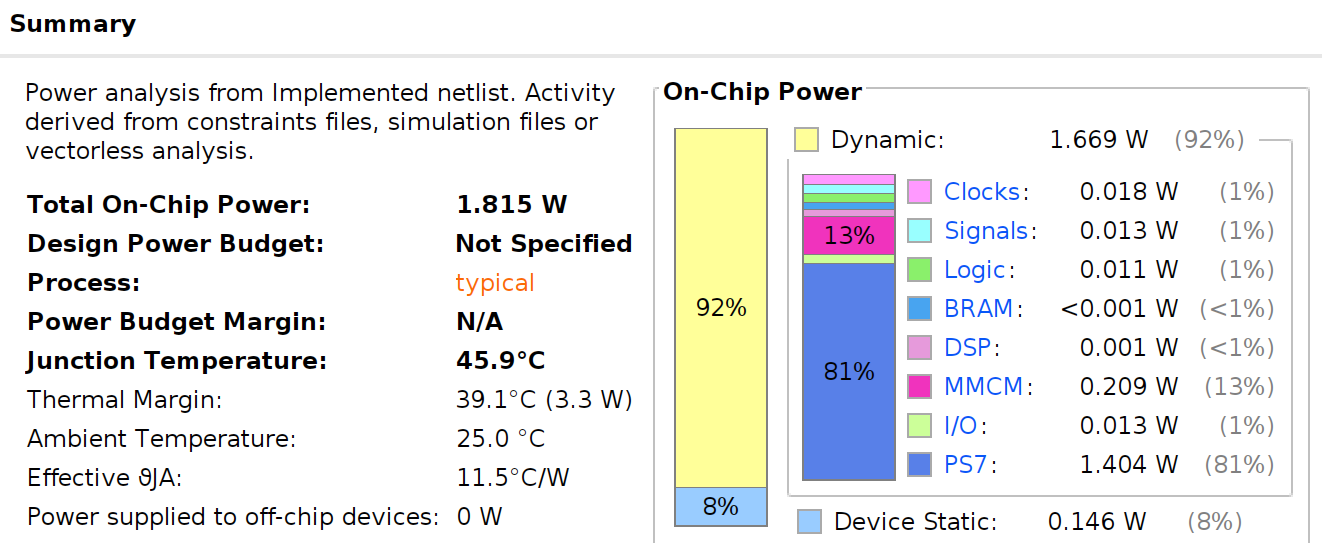}
\caption{Summary of FPGA on-chip power consumption for the DCU implementation. The device reports 1.815 W total power, of which 1.669 W (92\%) is dynamic. The primary contributor is the PS7 subsystem (1.404 W, 81\%), with additional dynamic power from MMCM resources (0.209 W, 13\%). All other components---clocks, signals, logic, BRAM, DSP units, and I/O---collectively account for less than 2\%. The resulting junction temperature is 45.9 $^\circ$C with a thermal margin of 39.1 $^\circ$C.}
\end{figure}

\textbf{Integration in the Computational Stack}

The DCU abstracts Apollo's analog--probabilistic substrate into a
programmable, deterministic compute unit comparable to a quantum control
stack or neuromorphic orchestration layer. It provides:

\begin{itemize}
\item
  a hardware-defined API for programming p-qubit networks,
\item
  scheduling and state-control primitives,
\item
  embedded problem graph management,
\item
  synchronous host--device orchestration.
\end{itemize}

Through this FPGA subsystem (Figure 11, Figure 12 and Table 2), Apollo
becomes fully accessible from the Dynex Quantum-as-a-Service (QaaS)
platform, enabling integration with standard quantum languages,
Hamiltonian compilers, and machine-learning toolchains.

Because the DCU is inseparable from Apollo, the platform behaves as a
co-designed hybrid analog--digital quantum-driven system, where digital
determinism and analog stochasticity are fused into a unified
computational pipeline$^{[79][95]}$.

\subsection{Extensions to Gate-Model Workloads}

Although Apollo is primarily optimised for annealing-native workloads
through its native Ising/QUBO energy relaxation dynamics (Section 2.2),
the platform also supports execution of arbitrary gate-model quantum
circuits$^{[109][110]}$. This capability arises
from established circuit-to-Hamiltonian reductions that transform the
temporal evolution of a quantum circuit into a static ground-state (or
low-energy) encoding problem, which can then be solved via energy
minimisation on the p-qubit fabric. A quantum circuit acting on $n$ qubits
and consisting of $T$ sequential layers of gates realises a unitary
\begin{equation}
U=\prod_{t=1}^{T}U_t
\end{equation}
(left-to-right convention).

Following the Feynman--Kitaev history-state
construction$^{[110][111]}$, the computation is
mapped to the low-energy subspace of a local Hamiltonian

\begin{equation}
H_{\mathrm{circuit}}=H_{\mathrm{init}}+H_{\mathrm{prop}}+H_{\mathrm{out}}+H_{\mathrm{clock}}
\end{equation}

where:

\begin{itemize}
\item
  An auxiliary ``clock'' register of size
\(\mathcal{O}(T)\)
  enforces valid temporal propagation.
\item
\(H_{\mathrm{init}}\)
  and
\(H_{\mathrm{out}}\)
  are penalty terms that favour valid input and output states.
\item
\(H_{\mathrm{prop}}\)
  contains terms that reward correct application of each gate
\(U_t\)
  at timestep
  \(t\).
\item
\(H_{\mathrm{clock}}\)
  ensures unidirectional (or domain-wall) clock progression.
\end{itemize}

The ground state (or a carefully prepared history state)

\begin{equation}
\lvert \mathrm{hist} \rangle=\frac{1}{\sqrt{T}}\sum_{t=0}^{T-1}\lvert t \rangle \, \lvert \psi_t \rangle
\end{equation}

encodes the entire spacetime history of the computation. Projecting onto
the final clock configuration yields the output state of the original
circuit with high probability under standard adiabatic or variational
theorems$^{[60][111]}$.

More recent perturbative-gadget and subspace-encoding techniques reduce
the overhead of the original Feynman--Kitaev construction, yielding 5-
or 6-local Hamiltonians whose ground-state fidelity scales favourably.
Warren et al. have further advanced this line by developing efficient
encodings specifically tailored to probabilistic-bit and stochastic
neural architectures, demonstrating that gate-model circuits can be
compiled into sparse Ising Hamiltonians with ancilla overhead linear in
circuit depth but with dramatically reduced penalty strengths for
NISQ-era and beyond-coherence
substrates$^{[112][113]}$.

The capacity to execute quantum circuit algorithms is built upon a
systematic translation of gate-based operations into an annealing or
optimization framework. This approach maps the logic of a quantum
circuit onto a Quadratic Unconstrained Binary Optimization (QUBO)
problem, which is then solved via quantum annealing on Apollo. The
theoretical foundation for this translation, demonstrating how arbitrary
quantum circuits can be converted into optimization problems, follows
established methodologies. This translation allows the emulated quantum
annealing hardware to find the ground state of a Hamiltonian that
represents the result of the gate-based
computation$^{[1][107]}$.

\textbf{Qubit Overhead and Scaling Analysis}

A critical aspect of this translation is the resource overhead,
specifically the number of physical qubits on the annealing device
required to represent a gate-based circuit. The total number of physical qubits (\(Q_{\text{total}}\)) required for a circuit with \(n_{\text{logical}}\) logical qubits and $G$ gates can be expressed as:

\begin{equation}
Q_{\mathrm{total}}=n_{\mathrm{logical}}+\sum_{i=1}^{G} \delta_i+n_{\mathrm{ancilla}}
\end{equation}

where:

\begin{itemize}
\item \(n_{\text{logical}}\) is the number of original circuit qubits,
\item \(\delta_i\) are the additional qubits for gate \(i\), and
\item \(n_{\text{ancilla}}\) represents the global ancilla qubits.
\end{itemize}

The overhead \(\delta_i\) is dependent on the specific gate being translated and is detailed in
Table 3. Larger, composite quantum operations exhibit different overhead
scaling behaviours$^{[110][112]}$.

\begin{table}[H]
\centering
{\small\bfseries Physical qubit overhead for translating quantum gates\par\vspace{3pt}}
\includegraphics[width=0.98\linewidth]{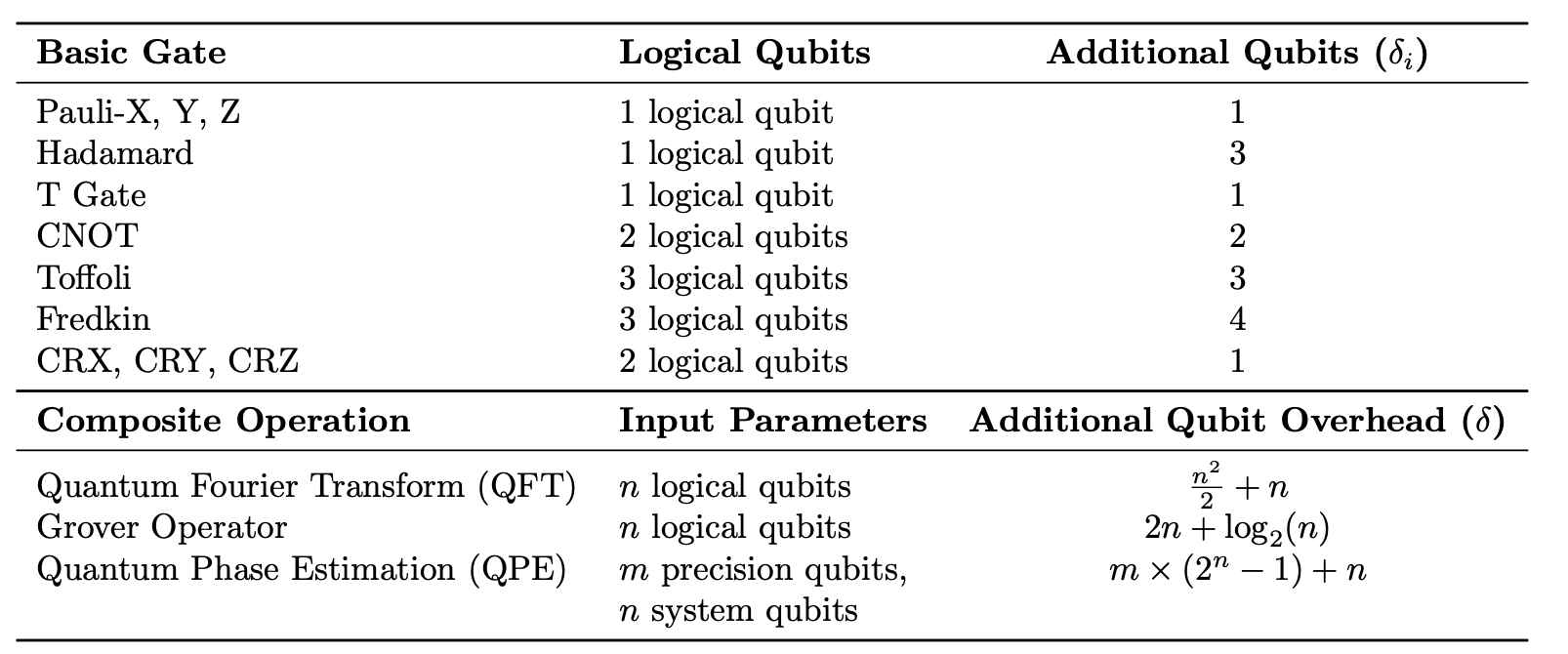}
\caption{Physical qubit overhead for translating quantum gates and composite operations into an annealing-based formulation. The table is divided into basic gates and scalable composite operations$^{[110]}$.}
\end{table}

Thus, while Apollo's core strength lies in annealing and Boltzmann
sampling, the circuit-to-Hamiltonian pathway---augmented by modern
reductions including those from Warren and collaborators---extends its
applicability to the full gate-model paradigm without requiring
cryogenic coherent hardware$^{[32][60]}$.

\subsection{Summary: From Theory to Hardware}

The Apollo architecture is designed as a direct physical realization of
the theoretical principles required for quantum-equivalent annealing
behaviour, as developed in Section 2. Rather than approximating
stochastic relaxation and thermodynamic sampling through algorithmic
emulation, the system embeds these properties intrinsically within its
physical dynamics$^{[17][94]}$.

Continuous-time, clock-less operation ensures that stochastic evolution
proceeds asynchronously and without discretization artefacts, consistent
with continuous-time Markov descriptions of thermodynamic systems.
Independent, physically grounded entropy sources supply uncorrelated
stochastic excitation at the level of individual p-qubits, preserving
ergodicity and detailed balance. A programmable, high-expressivity
coupling fabric enables direct representation of bias and interaction
terms corresponding to general Ising and QUBO energy functions, reducing
the need for transformation or embedding that would otherwise distort
the underlying energy landscape. Deterministic control and orchestration
provide configurability and experimental repeatability while remaining
architecturally decoupled from the stochastic computation
itself$^{[7][8]}$.

Together, these elements instantiate a physical system whose equilibrium
statistics and dynamical behaviour align with the conditions required
for reproducing the outcomes of transverse-field quantum annealing under
the Suzuki--Trotter equivalence. The architecture thus serves as a
concrete substrate for investigating quantum-equivalent annealing
dynamics within a room-temperature, classical hardware
platform$^{[60][68]}$.

In the following section, we present experimental results that validate
the physical behaviour of this architecture at both the device and
system levels. These results assess stochastic fidelity, thermodynamic
sampling correctness, and annealing dynamics, thereby providing
empirical support for the theoretical and architectural claims advanced
in this work.

\section{Performance Model of Continuous-Time Probabilistic
Architectures}

\subsection{Effective Throughput in Continuous-Time Parallel
Architectures}

A rigorous treatment of computational throughput in probabilistic
hardware requires distinguishing between the physical state-transition
rate of individual computing elements and the effective computational
throughput of the system as a whole. In Apollo\textquotesingle s
continuous-time parallel architecture, these quantities differ by
several orders of magnitude due to three fundamental mechanisms: massive
parallelism, continuous-time mixing advantage, and cascade-induced
correlation propagation$^{[28][114]}$.

\subsection{Energy--Throughput Scaling Laws}

Each p-qubit in the Apollo architecture operates as an autonomous
stochastic oscillator whose characteristic transition frequency is set
by the OTA time constant and the IQEU noise bandwidth. For the 16 nm
production device, the physical flip rate per p-qubit is
\(f_{\mathrm{flip}} = 80~\mathrm{GHz} = 8.0 \times 10^{10}~\mathrm{flips\,s^{-1}}\),
corresponding to a characteristic single-flip time of
\(\tau_{\mathrm{flip}} = \frac{1}{f_{\mathrm{flip}}} = 12.5~\mathrm{ps}\)$^{[8][103]}$.

With
\[N = 10{,}000\]
p-qubits operating fully in parallel on a single Apollo die, the
aggregate physical flip throughput is
\(F_{\mathrm{die}} = N f_{\mathrm{flip}} = 8.0 \times 10^{14}~\mathrm{flips\,s^{-1}}\),
equivalent to approximately
\[8.0 \times 10^{5}\]
flips per nanosecond (i.e., on the order of
\[10^{15}\]
physical flips per second) per die. Assuming a total per-die power
dissipation of
\(P_{\mathrm{die}} = 0.5~\mathrm{W}\),
the corresponding energy per physical flip is
\[E_{\mathrm{flip}} = \frac{P_{\mathrm{die}}}{F_{\mathrm{die}}} = \frac{0.5}{8.0 \times 10^{14}} = 6.25 \times 10^{-16}~\mathrm{J} = 0.63~\mathrm{fJ}\]
fJ$^{[28][114]}$.

\subsection{System-Level Scaling and Tiling Considerations}

Apollo is designed to be tileable at the die level across multiple
packages, enabling construction of larger probabilistic processors by
interconnecting many identical 10,000--p-qubit dies into a single
coherent compute fabric. In this configuration, each die maintains local
continuous-time stochastic dynamics, while the system-level interconnect
provides programmable couplings across die boundaries. Algorithmically,
the multi-package assembly is treated as one expanded p-qubit array,
with inter-die links acting as additional edges in the global coupling
graph (subject to bandwidth/latency constraints of the chosen
packaging/interconnect technology)$^{[95][115]}$.

For a
\[N_{\mathrm{dies}} = 10 \times 10 = 100\]
multi-package assembly comprising 100 Apollo dies, the effective system
size scales to
\[N_{\mathrm{tot}} = 100 \cdot 10{,}000 = 10^{6}\]
p-qubits. The aggregate physical flip throughput scales linearly to
\(F_{\mathrm{tot}} = 100 \cdot F_{\mathrm{die}} = 8.0 \times 10^{16}~\mathrm{flips\,s^{-1}}\),
corresponding to approximately
\[8.0 \times 10^{7}\]
flips per nanosecond (i.e., on the order of
\(10^{17}\)physical
flips per second) across the full assembly. Under the same linear
scaling assumption for power, the total power becomes
\(P_{\mathrm{tot}} = 100 \cdot 0.5 = 50~\mathrm{W}\),
and the energy per physical flip remains unchanged at
\[E_{\mathrm{flip}} = P_{\mathrm{tot}}/F_{\mathrm{tot}} = 6.25 \times 10^{-16}~\mathrm{J} \approx 0.63~\mathrm{fJ}\]
fJ$^{[17][114]}$.

In practice, inter-die couplings are typically engineered so that
strongly interacting subgraphs are placed within a die and sparser
boundary interactions traverse the package interconnect, thereby
preserving the high intrinsic flip-rate dynamics locally while enabling
system-level scaling to million--p-qubit-class
arrays$^{[28][94]}$.

\section{Experimental Validation}

Apollo-RC1, implemented on a mature 350nm mixed-signal CMOS (Figure 13),
implements 10,000 time-multiplexed p-qubits with external entropy
injection and baseline $\Delta$256 routing$^{[93][100]}$.

\begin{figure}[H]
\centering
{\small\bfseries Microscopic Image of the 350 nm CMOS\par\vspace{2pt}}
\includegraphics[width=0.98\linewidth]{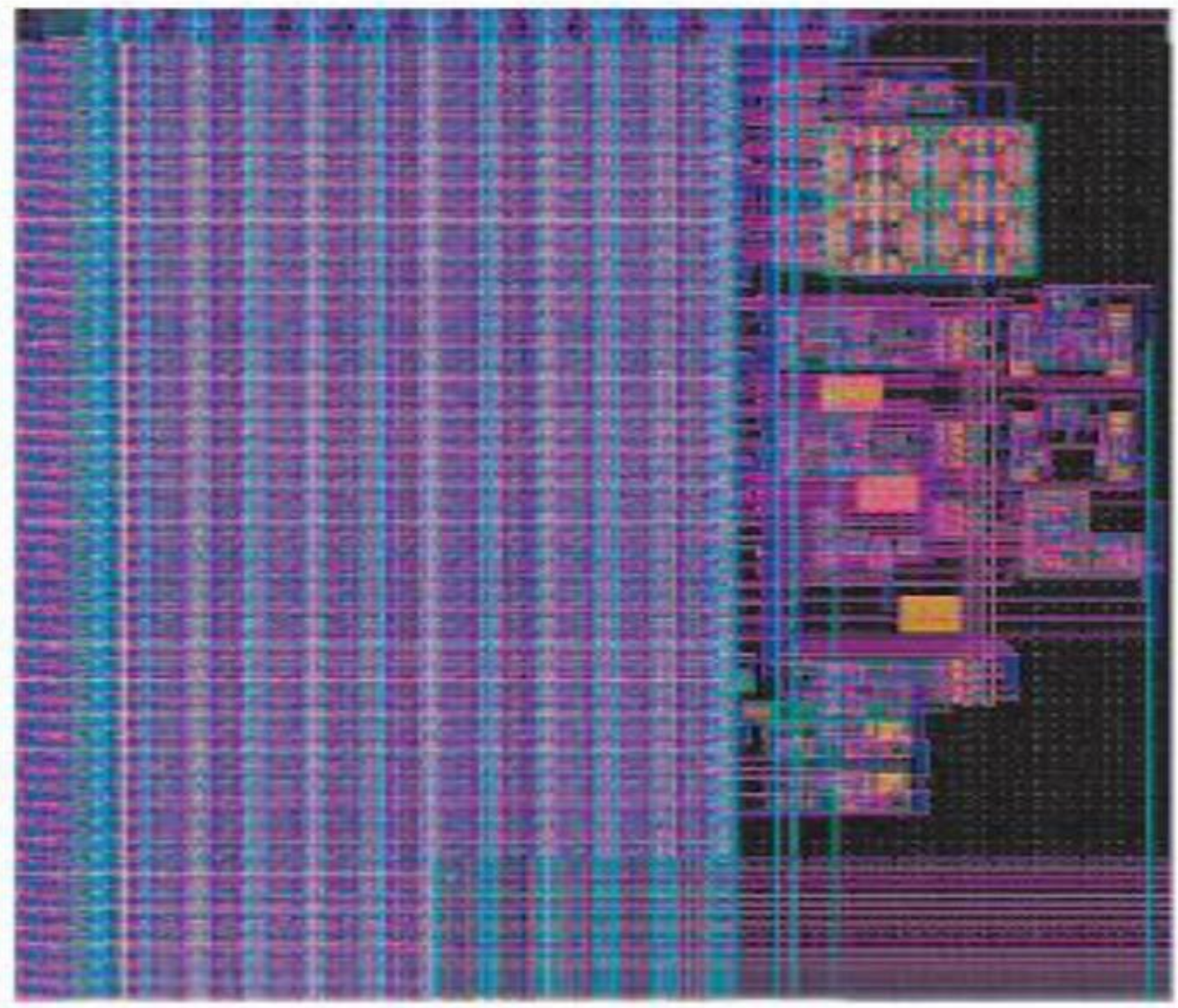}
\caption{Microscopic image of the 350 nm CMOS Configurable Analog Block (CAB), based on a manhattan-style architecture$^{[93][100]}$. The CAB incorporates tunable OTAs, floating-gate elements, switched-capacitor structures, and routing resources that enable reconfigurable analog signal processing within the IC.}
\end{figure}

\subsection{Device-Level Characterisation}

Each p-qubit on Apollo-RC1 implements a continuously driven analog
activation function with tuneable gain, enabling the hardware to
reproduce the probabilistic switching statistics predicted by the
theoretical quantum-driven analog qubit
representation$^{[9][41]}$.

Measured sigmoid responses across the chip show:

\begin{itemize}
\item
  clean, monotonic activation curves,
\item
  no detectable hysteresis,
\item
  tuneable slope controlled via gain-bias adjustments,
\item
  minimal process-corner dispersion, confirming robust device-level
  behaviour.
\end{itemize}

Figure 14 presents representative measurement results, showing excellent
alignment with theoretical models and confirming that Apollo's p-qubit
primitives faithfully implement the intended probabilistic activation
behaviour$^{[98][103]}$.

\begin{figure}[H]
\centering
{\small\bfseries Analog P-qubit Transfer Characteristics\par\vspace{2pt}}
\includegraphics[width=0.98\linewidth]{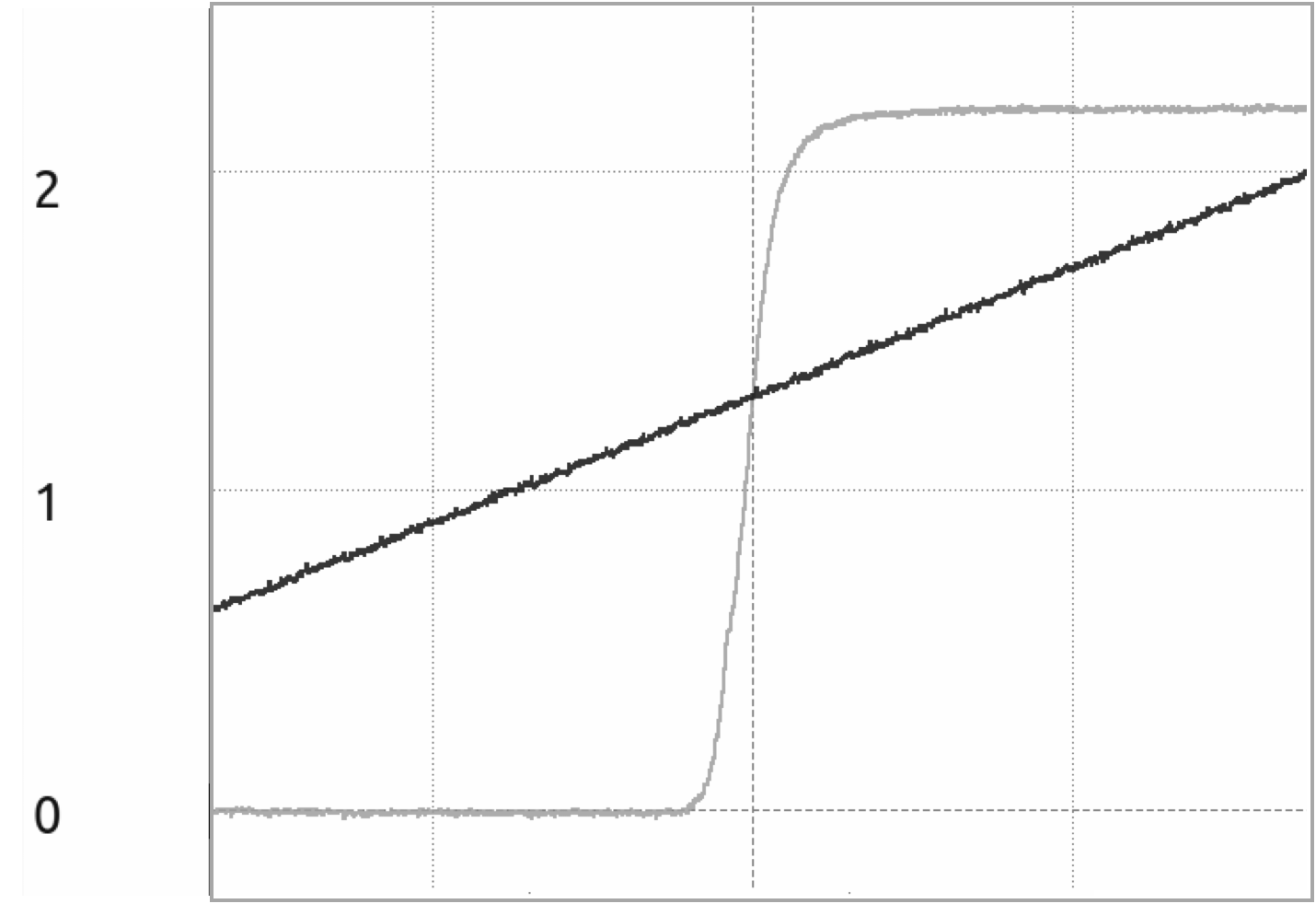}
\caption{Analog transfer characteristics of the implemented p-qubit. The experimentally observed sigmoid curves reveal the characteristic steep activation and gain tunability, validating consistency with theoretical models over multiple process corners.}
\end{figure}

\subsection{Entropy Quality and Stochastic Independence}

The IQEU subsystem provides the physical entropy required to drive
stochastic updates across the array. Experiments across 0--85 $^\circ$C show
that the entropy source remains:

\begin{itemize}
\item
  broadband,
\item
  non-periodic,
\item
  non-Gaussian but stationary,
\item
  free of detectable bias,
\item
  free of measurable temporal correlation,
\item
  immune to thermal drift across the full operational
  range$^{[8][104]}$.
\end{itemize}

These properties are essential for ensuring unbiased sampling of
Ising/QUBO energy landscapes. Figure 15 shows the time-domain traces,
histogram, and spectrogram of a 4-p-qubit system driven solely by IQEU
noise, demonstrating temperature-stable behaviour and validating the
quality of the hardware stochasticity$^{[8][28]}$.

\begin{figure}[H]
\centering
{\small\bfseries Characterisation of Entropy Generation and Stochastic Stability in the IQEU\par\vspace{2pt}}
\includegraphics[width=0.98\linewidth]{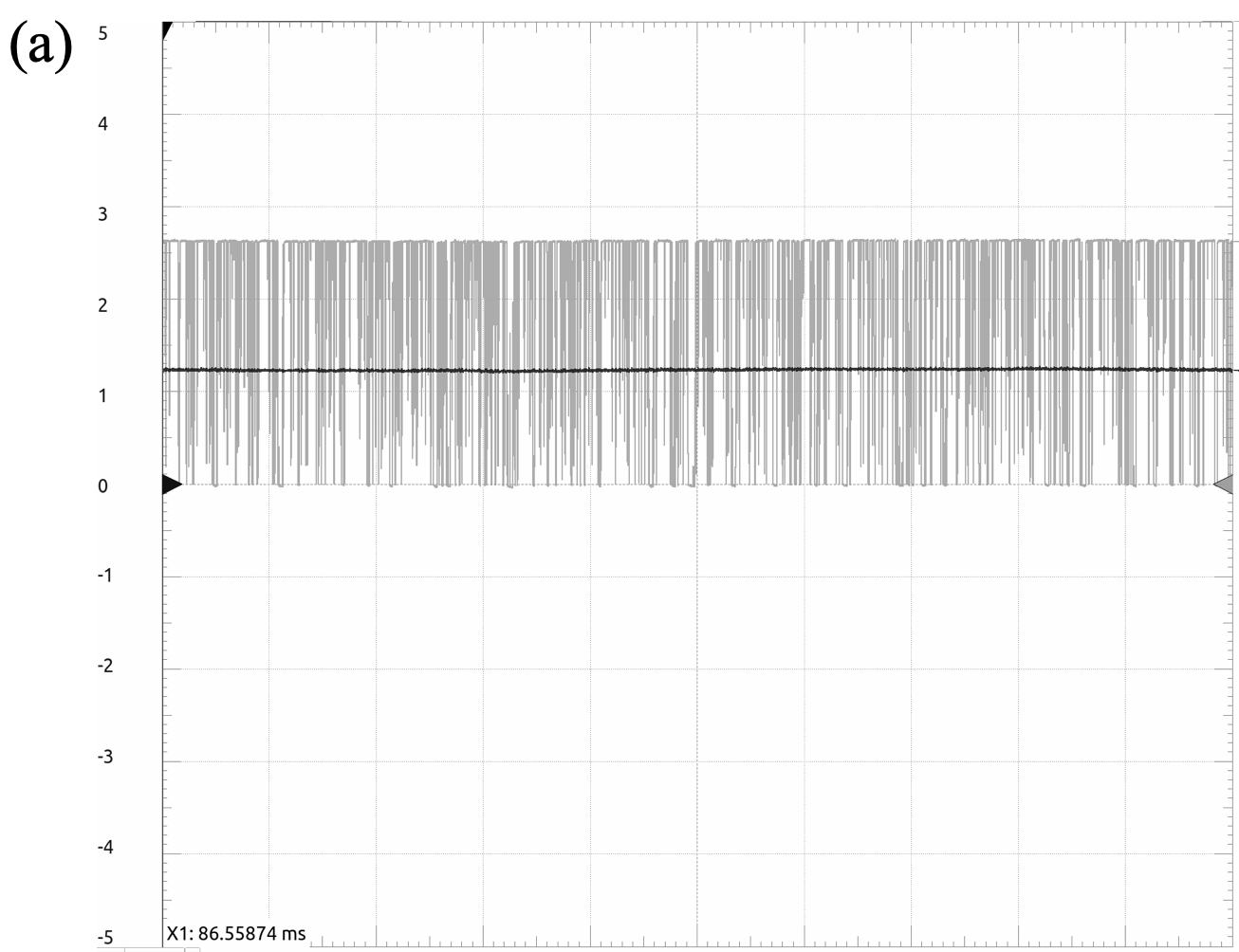}\vspace{2pt}
\includegraphics[width=0.98\linewidth]{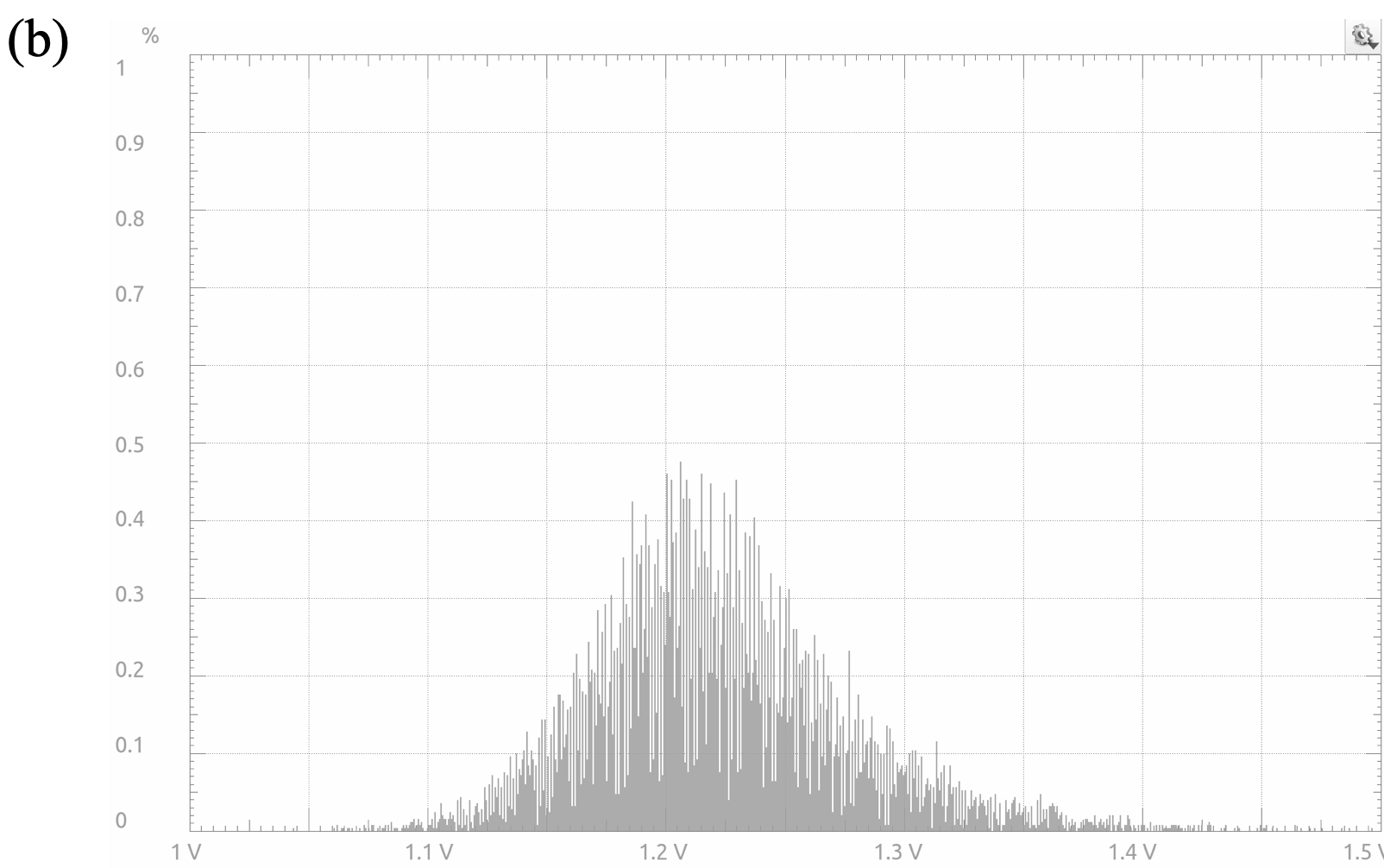}\vspace{2pt}
\includegraphics[width=0.98\linewidth]{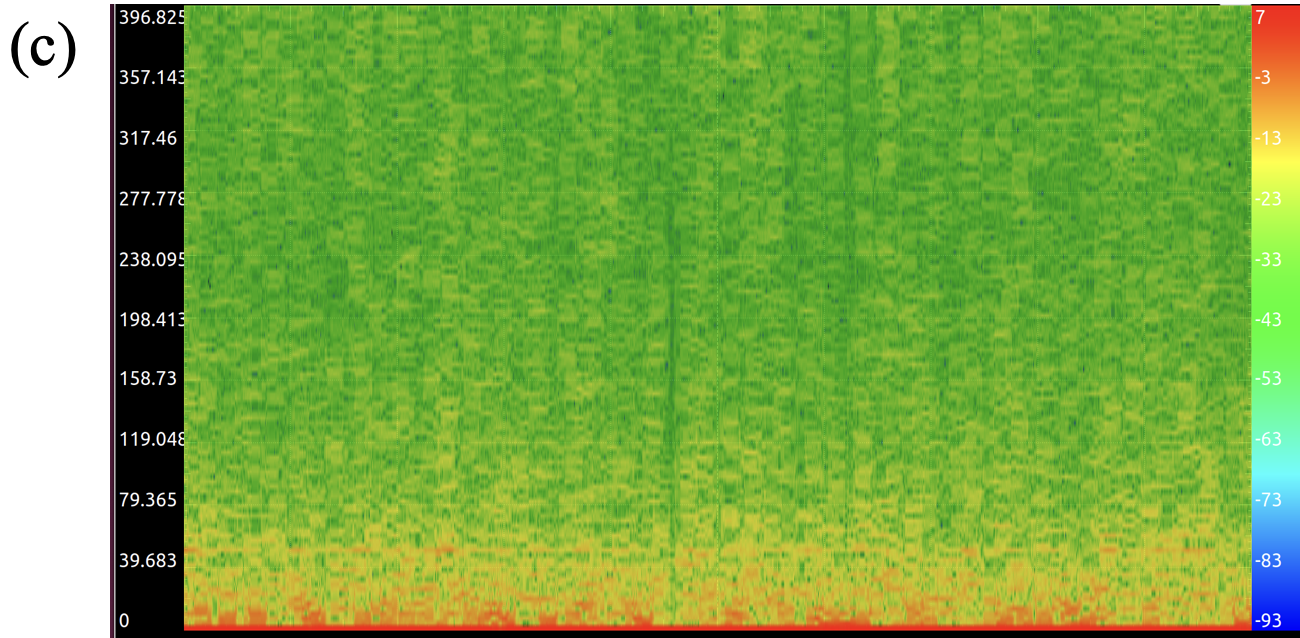}
{\footnotesize\ttfamily q0: --H--------------------M--\quad q1: --H--------------------M--\par q2: --H--------------------M--\quad q3: --H--------------------M--\par}
\caption{Characterisation of entropy generation and stochastic stability in the IQEU. Across the temperature range 0--85 $^\circ$C, the generated fluctuations remain broadband and aperiodic, yielding high-quality physical randomness with no detectable statistical bias or temporal correlation. (a) Time-domain response of a 4-p-qubit network with zero applied bias, illustrating intrinsic quantum-driven state fluctuations; (b) histogram of the sampled distribution; (c) spectrogram confirming broadband spectral content and absence of dominant frequency components. Circuit: Hadamard (H) gates are applied to q0, q1, q2 and q3 to prepare them independently in the $|+\rangle$ superposition state.}
\end{figure}

To rigorously assess the quality of the IQEU entropy source, it was
benchmarked against a widely used commercial quantum random number
generator (QRNG), the ID Quantique. Quantis employs a quantum-optical
randomness source based on single-photon which-path detection and is
regarded as an industry reference for high-quality randomness. By
subjecting both devices to an identical series of statistical
tests---including bias analysis, serial correlation measurements, $\chi$²
evaluations, and NIST SP 800-90B min-entropy assessments---the study
establishes a direct, quantitative comparison between the IQEU
architecture and a certified QRNG operating with quantum physical
principles$^{[104][116]}$.

\begin{table}[H]
\centering
{\small\bfseries Bit-Level Bias ($\delta$) --- Normal Fit Parameters\par\vspace{3pt}}
{\scriptsize
\resizebox{\linewidth}{!}{%
\begin{tabular}{lll}
\toprule
\textbf{Metric} & \textbf{IQEU} & \textbf{Quantis} \\
\midrule
Mean bias $\delta$̄ & -0.0005\% ± 0.0032\% & 0.0019\% ± 0.0029\% \\
Standard deviation $\sigma$ & 0.0499\% ± 0.0025\% & 0.0502\% ± 0.0020\% \\
p-value (normal fit) & 32 \% & 62 \% \\
Expected $\sigma$ for unbiased & 0.05\% & 0.05\% \\
Interpretation & Unbiased & Unbiased \\
\bottomrule
\end{tabular}%
}
}
\caption{This table reports a statistical comparison of bit-level bias in the output streams produced by the independent quantum entropy units (IQEUs) and a commercial quantum random number generator (Quantis). Bit-level bias is defined as the deviation of the empirical probability of observing a `1' from the ideal value of 0.5 and provides a first-order indicator of symmetry in the binary output. For each source, the distribution of bias estimates is fitted to a normal distribution, and the resulting mean, variance, and goodness-of-fit statistics are reported. Agreement between the observed standard deviation and the expected value for an unbiased Bernoulli process, together with non-rejecting p-values, indicates the absence of systematic bias at the bit level$^{[105][117]}$.}
\end{table}

\begin{table}[H]
\centering
{\small\bfseries Byte-Level Bias ($\chi^2$ Test)\par\vspace{3pt}}
{\scriptsize
\resizebox{\linewidth}{!}{%
\begin{tabular}{lll}
\toprule
\textbf{Metric} & \textbf{IQEU} & \textbf{Quantis} \\
\midrule
Estimated degrees of freedom (d) & 254.9 ± 1.5 & 254.6 ± 1.7 \\
p-value ($\chi$$^2$ fit) & 23 \% & 11 \% \\
Expected d & 255 & 255 \\
Interpretation & Unbiased bytes & Unbiased bytes \\
\bottomrule
\end{tabular}%
}
}
\caption{This table summarizes the results of a chi-squared goodness-of-fit test applied to byte-level symbol frequencies. Rather than examining individual bits, this analysis evaluates the uniformity of all 256 possible byte values, providing sensitivity to higher-order bias patterns that may not be visible at the single-bit level. The estimated degrees of freedom and corresponding p-values are compared against the theoretical expectation for a uniform distribution. Consistency with the expected degrees of freedom and non-significant p-values indicate that both entropy sources produce statistically uniform byte distributions without detectable structural bias$^{[105][118]}$.}
\end{table}

\begin{table}[H]
\centering
{\small\bfseries Serial Correlation ($c$) --- Adjacent Bits\par\vspace{3pt}}
{\scriptsize
\resizebox{\linewidth}{!}{%
\begin{tabular}{lll}
\toprule
\textbf{Metric} & \textbf{IQEU} & \textbf{Quantis} \\
\midrule
Mean correlation c̄ & 0.0023\% ± 0.0047\% & -0.0017\% ± 0.0045\% \\
Standard deviation $\sigma$ & 0.0971\% ± 0.0034\% & 0.0958\% ± 0.0032\% \\
p-value (normal fit) & 88 \% & 91 \% \\
Interpretation & No detectable correlation & No detectable correlation \\
\bottomrule
\end{tabular}%
}
}
\caption{This table presents measurements of serial correlation between adjacent bits in the entropy streams. Serial correlation quantifies the extent to which successive output bits are statistically dependent, which can degrade the quality of stochastic processes in probabilistic computing systems. The reported correlation coefficients are evaluated across large samples and fitted to a normal distribution to assess both mean correlation and variability. The absence of statistically significant deviations from zero correlation indicates that neither source exhibits detectable temporal dependence at the bit-to-bit level$^{[26][28]}$.}
\end{table}

\begin{table}[H]
\centering
{\small\bfseries Cross-Channel Independence (IQEU) --- Correlation between multiple IQEU entropy cores\par\vspace{3pt}}
{\scriptsize
\resizebox{\linewidth}{!}{%
\begin{tabular}{ll}
\toprule
\textbf{Metric} & \textbf{Multiple IQEU Cores} \\
\midrule
Mean correlation c̄ & 0.0031\% ± 0.0053\% \\
$\sigma$ & 0.1036\% ± 0.0038\% \\
p-value & 80 \% \\
Interpretation & Channels are independent \\
\bottomrule
\end{tabular}%
}
}
\caption{This table evaluates statistical independence across multiple IQEU entropy cores operating concurrently. Cross-channel correlation analysis is critical in architectures where multiple stochastic units rely on parallel entropy sources, as shared or correlated noise can compromise ergodicity and distort collective dynamics. The reported correlation statistics measure pairwise dependencies between independent IQEU channels. The absence of significant correlations confirms that the entropy streams generated by distinct IQEU cores are statistically independent within the resolution of the analysis$^{[17][114]}$.}
\end{table}

\begin{table}[H]
\centering
{\small\bfseries NIST SP 800-90B --- IID Min-Entropy Estimates (1K tests, each 125 KB)\par\vspace{3pt}}
{\scriptsize
\resizebox{\linewidth}{!}{%
\begin{tabular}{lll}
\toprule
\textbf{Metric} & \textbf{IQEU} & \textbf{Quantis} \\
\midrule
IID min-entropy mean $\bar h$ & 7.673 & 7.673 \\
IID min-entropy $\sigma$ & 23 & 21 \\
IID assumption failure rate & 3.4\% & 4.7\% \\
Interpretation & Same distribution, high entropy & Same distribution, high entropy \\
\bottomrule
\end{tabular}%
}
}
\caption{This table reports independent and identically distributed (IID) min-entropy estimates obtained using the NIST SP 800-90B test suite. Min-entropy provides a conservative measure of unpredictability by quantifying the worst-case symbol probability. The IID analysis assumes no temporal or structural dependencies and is applied to multiple fixed-length data segments. Comparable mean values, variances, and IID assumption failure rates between the two sources indicate statistically indistinguishable entropy characteristics under the IID model$^{[105]}$.}
\end{table}

\begin{table}[H]
\centering
{\small\bfseries NIST SP 800-90B --- Non-IID Min-Entropy Estimates\par\vspace{3pt}}
{\scriptsize
\resizebox{\linewidth}{!}{%
\begin{tabular}{lll}
\toprule
\textbf{Metric} & \textbf{IQEU} & \textbf{Quantis} \\
\midrule
Non-IID min-entropy mean $\bar h$ & 6.80 & 6.79 \\
Non-IID min-entropy $\sigma$ & 0.25 & 0.23 \\
Interpretation & Nearly identical distributions & Nearly identical distributions \\
\bottomrule
\end{tabular}%
}
}
\caption{This table presents non-IID min-entropy estimates computed under the more conservative assumptions of the NIST SP 800-90B framework. Unlike the IID analysis, this evaluation explicitly accounts for potential dependencies, non-stationarity, and structural patterns in the data. The close agreement in estimated min-entropy values and distributions demonstrates that both entropy sources retain similar levels of unpredictability even under relaxed statistical assumptions$^{[105]}$.}
\end{table}

\begin{table}[H]
\centering
{\small\bfseries Standard Min-Entropy (no confidence interval correction, purely max $p_i$)\par\vspace{3pt}}
{\scriptsize
\resizebox{\linewidth}{!}{%
\begin{tabular}{lll}
\toprule
\textbf{Metric} & \textbf{IQEU} & \textbf{Quantis} \\
\midrule
Mean $\bar h$ & 7.823 & 7.823 \\
$\sigma$ & 24 & 22 \\
Interpretation & Identical entropy levels & Identical entropy levels \\
\bottomrule
\end{tabular}%
}
}
\caption{This table reports standard min-entropy estimates derived directly from the maximum observed symbol probability, without applying confidence-interval corrections or IID/non-IID modeling assumptions. While less conservative than the NIST estimators, this metric provides an intuitive baseline for comparing raw entropy levels. The identical mean values and comparable variances indicate that the two entropy sources exhibit equivalent maximum-symbol statistics across the evaluated samples$^{[119]}$.}
\end{table}

Across all statistical evaluations, the IQEU circuit demonstrated
randomness characteristics that closely match those of the Quantis QRNG.
Bit-level and byte-level bias metrics aligned with ideal theoretical
distributions, and both devices exhibited serial correlation values
statistically indistinguishable from zero. NIST SP 800-90B entropy
analyses further showed that the IID and non-IID min-entropy
distributions of IQEU and Quantis are nearly identical, with both
achieving high and stable entropy levels. Notably, IQEU reached these
performance levels without relying on post-processing, whereas Quantis
incorporates internal whitening to reduce raw bias. Overall, the results
indicate that IQEU's delivers randomness quality comparable to that of a
commercial quantum-optical RNG, validating its effectiveness as a robust
quantum random number generator$^{[104][105]}$.

\subsection{Thermodynamic Sampling Correctness}

\textbf{Thermodynamic Sampling Correctness}

To evaluate whether Apollo-RC1 samples from the correct underlying
Boltzmann distribution, we constructed several exactly solvable Ising
problem instances small enough for analytical Gibbs distributions to be
computed$^{[1][28]}$.

A representative 4-p-qubit instance is shown in Figure 16, where:

\begin{itemize}
\item
  time-domain trajectories reflect the expected statistically weighted
  switching,
\item
  histograms of sampled states match the theoretical Gibbs
  probabilities,
\item
  the measured distribution achieves Kullback--Leibler divergence
  \textless{} 1\%,
\item
  spectrograms confirm the expected broadband stochastic
  structure$^{[8][119]}$.
\end{itemize}

These results provide strong empirical evidence that Apollo-RC1 performs
thermodynamically consistent sampling, a prerequisite for annealing,
Bayesian inference, combinatorial optimization, and probabilistic
machine learning tasks$^{[41][120]}$.

\begin{figure}[H]
\centering
{\small\bfseries Experimental Characterisation of a Four-p-qubit System\par\vspace{2pt}}
\includegraphics[width=0.98\linewidth]{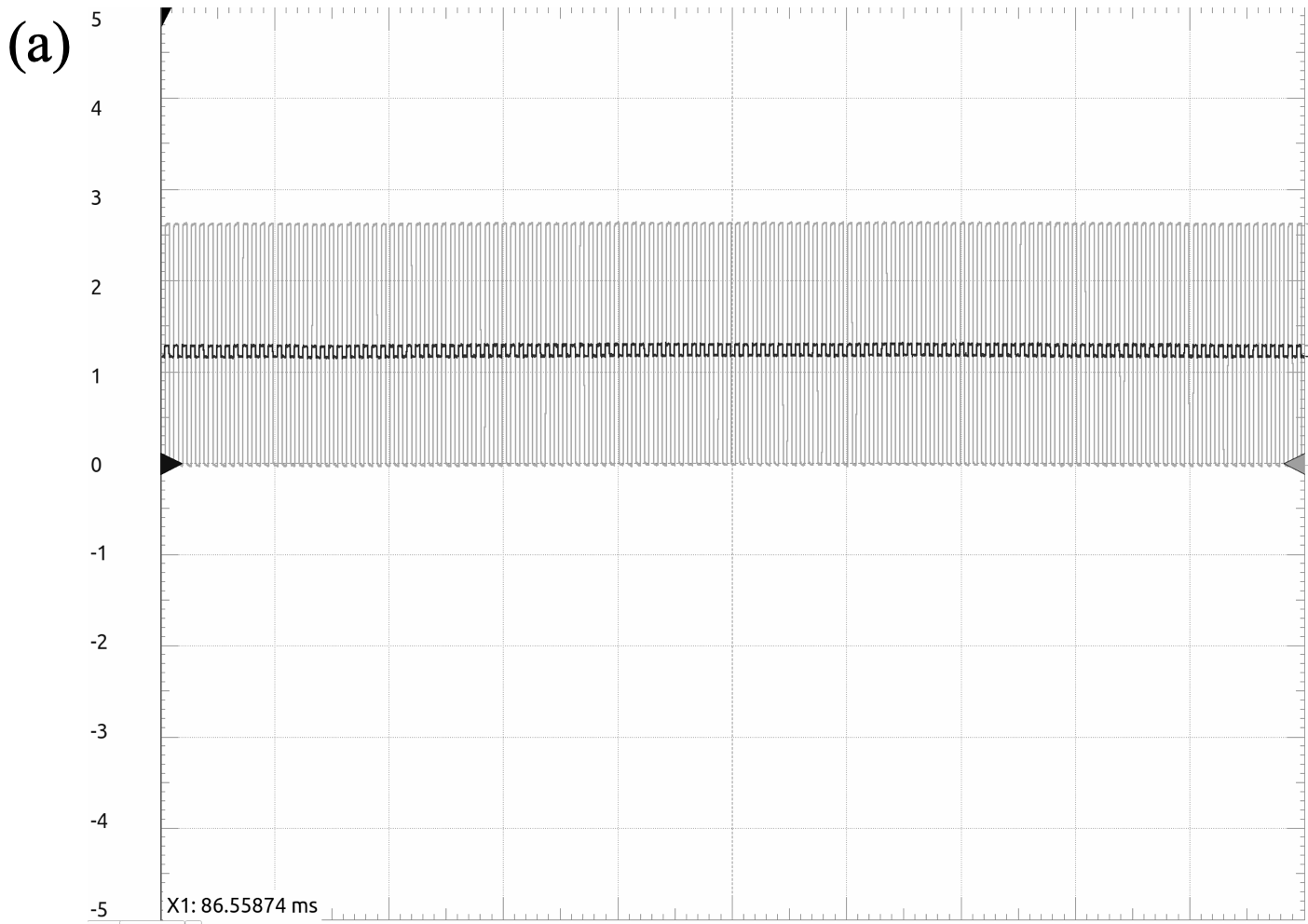}\vspace{2pt}
\includegraphics[width=0.98\linewidth]{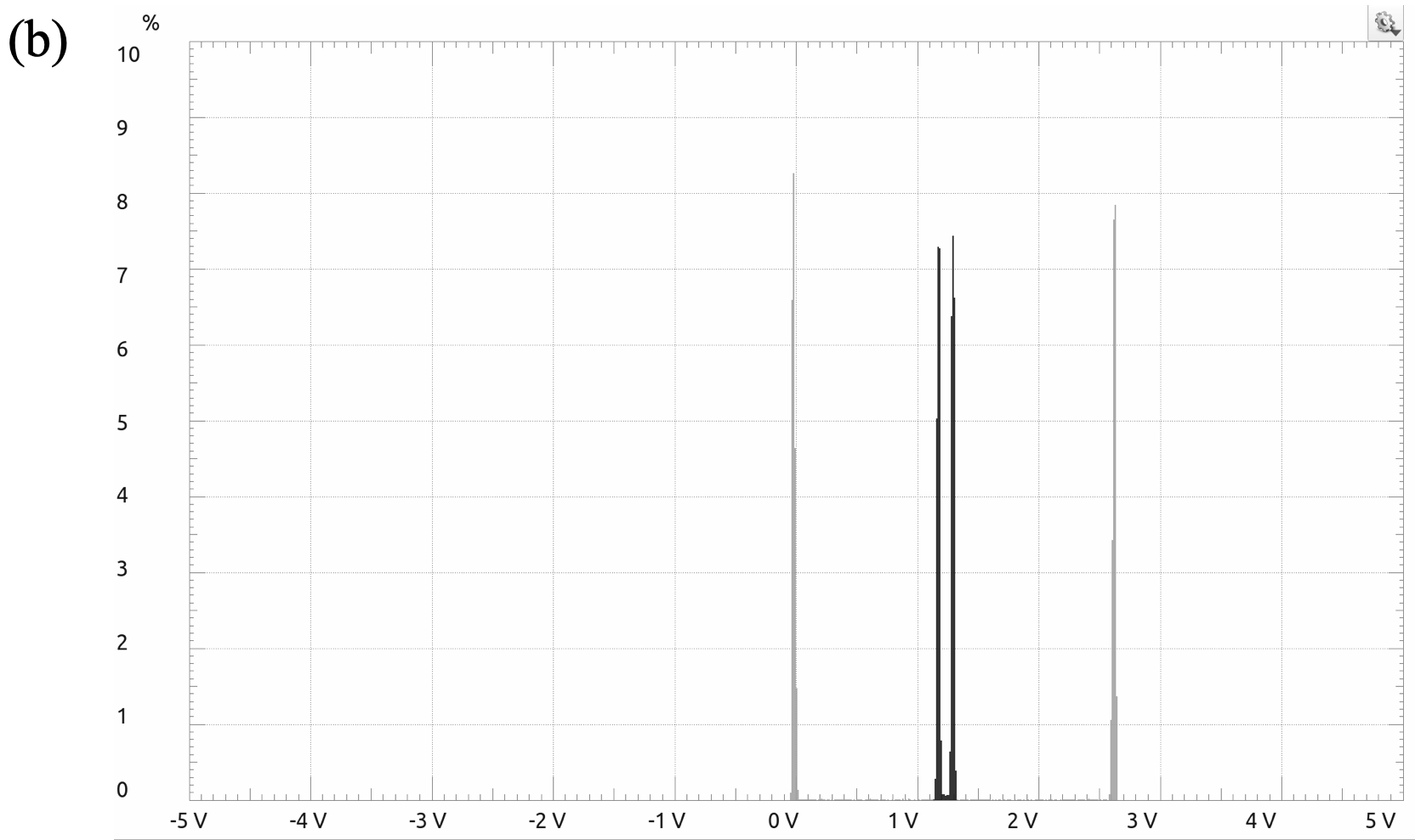}\vspace{2pt}
\includegraphics[width=0.98\linewidth]{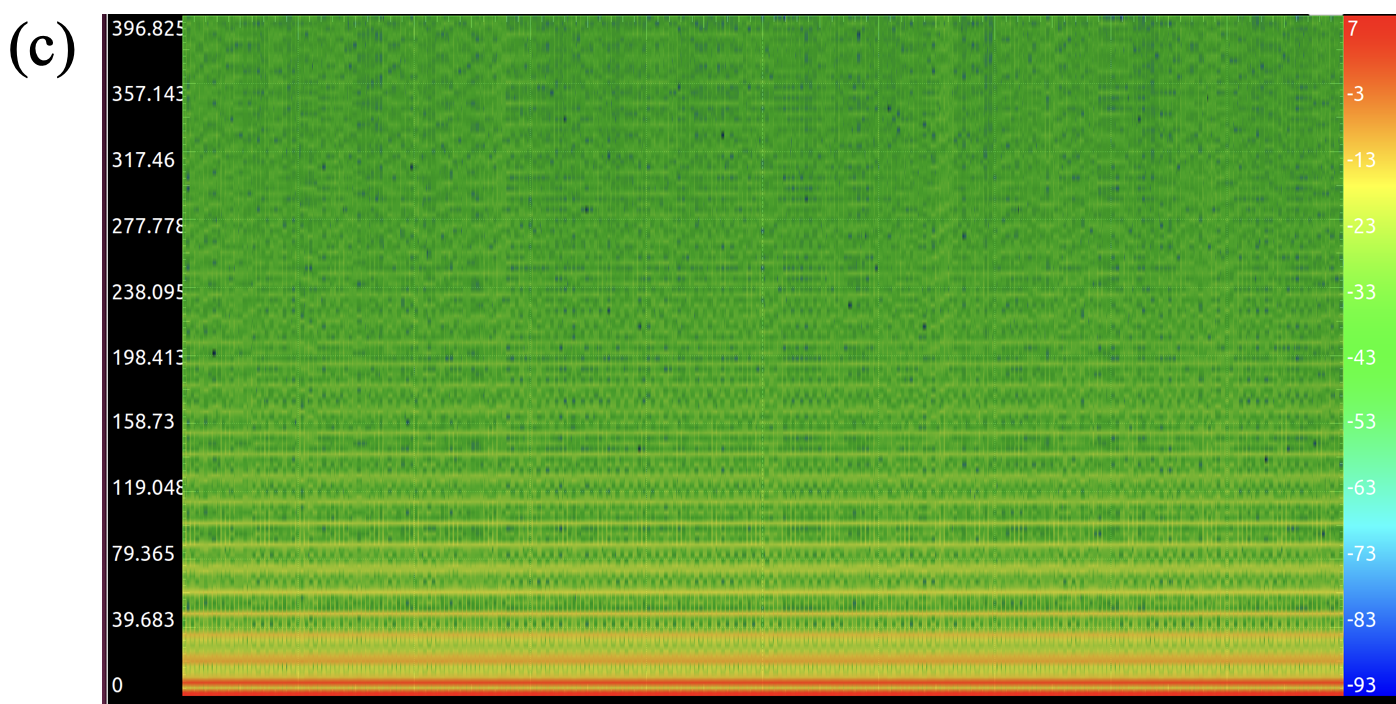}
{\footnotesize\ttfamily q0: -----------------------M--\quad q1: --X--------------------M--\par q2: -----------------------M--\quad q3: --X--------------------M--\par}
\caption{Experimental characterisation of a four-p-qubit system exhibiting fixed state behaviour. (a) Measured voltage trajectory associated with the system's ground-state configuration; (b) corresponding histogram of sampled states; and (c) spectrogram demonstrating the system's stochastic dynamics. The sampled distribution matches the theoretical Gibbs distribution for the underlying Ising instance, yielding a Kullback--Leibler divergence of $<1\%$, thereby confirming thermodynamic sampling correctness. Circuit: X gates are applied to q1 and q3 to initialize them in the $|1\rangle$ state, while q0 and q2 remain in $|0\rangle$, producing the deterministic alternating output pattern $|0101\rangle$.}
\end{figure}

\begin{figure}[H]
\centering
{\small\bfseries Histogram of a Four-p-qubit System\par\vspace{2pt}}
\includegraphics[width=0.98\linewidth]{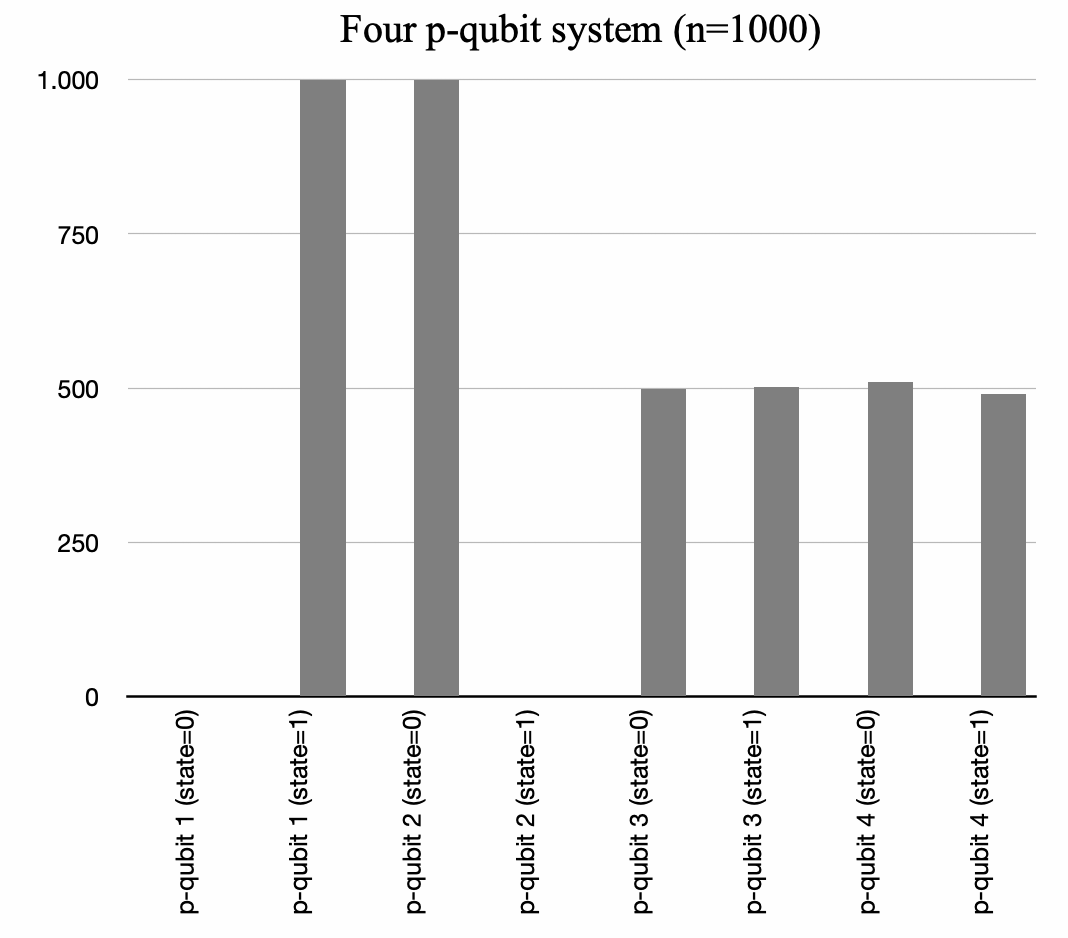}
{\footnotesize\ttfamily q0: --X--------------------M--\quad q1: -----------------------M--\par q2: --H--------------------M--\quad q3: --H--------------------M--\par}
\caption{Histogram illustrating the sampled state distribution of a four-p-qubit system over 1,000 measurements. Circuit: An X gate is applied to q0 to initialize it deterministically in $|1\rangle$, q1 is left in its default $|0\rangle$ state, and Hadamard (H) gates are applied to q2 and q3 to prepare them independently in the $|+\rangle$ superposition state, with no entangling operations applied.}
\end{figure}

\subsection{Continuous-Time Annealing Behaviour}

Unlike digital annealers, which update synchronously in discrete clocked
steps, Apollo-RC1 operates in continuous time under fully asynchronous,
clock-less dynamics$^{[7][28]}$.

The system exhibits:

\begin{itemize}
\item
  smooth energy relaxation trajectories with no quantisation artefacts,
\item
  fully parallel convergence across coupled p-qubits,
\item
  orders-of-magnitude faster descent to ground states compared to
  clocked digital annealers of comparable
  scale$^{[37][114]}$.
\end{itemize}

Figure 18 illustrates this behaviour in a mixed instance containing two
fixed and two p-qubits in superposition; the entire system converges
instantaneously and in unison to the correct ground state, demonstrating
the efficiency and coherence of Apollo's continuous-time annealing
mechanism$^{[9][94]}$.

\begin{figure}[H]
\centering
{\small\bfseries Time-Domain Voltage Traces of a Four-p-qubit System\par\vspace{2pt}}
\includegraphics[width=0.98\linewidth]{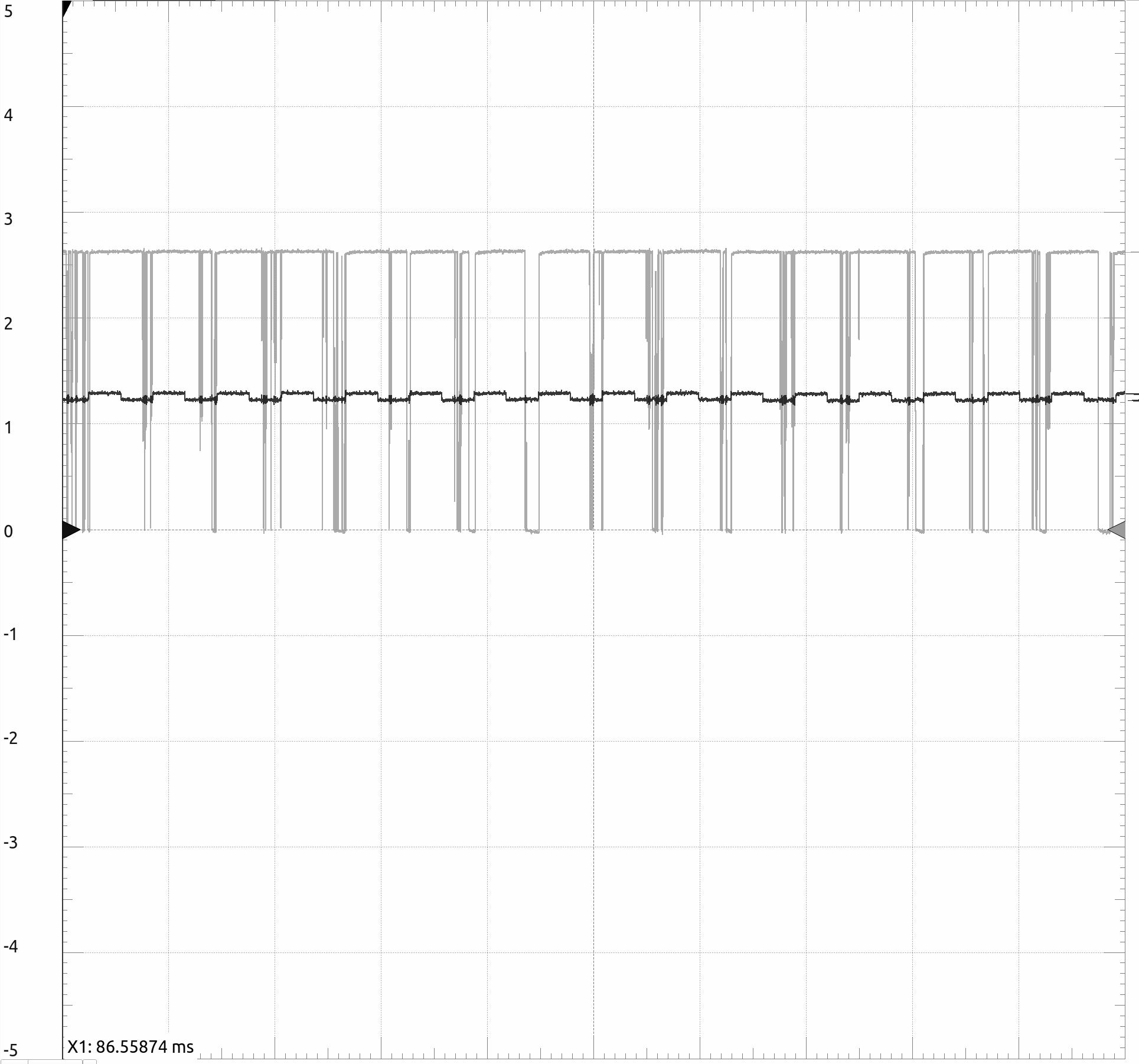}
{\footnotesize\ttfamily q0: --X--------------------M--\quad q1: --X--------------------M--\par q2: --H--------------------M--\quad q3: --H--------------------M--\par}
\caption{Time-domain voltage traces of a four-p-qubit system in which two p-qubits are fixed and two are in superposition. The observed trajectories show that, under clock-less asynchronous dynamics, the system converges immediately and in parallel to its ground-state solution. Circuit: An X gate is applied to q0 and q1 to initialize them deterministically in $|1\rangle$, and Hadamard (H) gates are applied to q2 and q3 to prepare them independently in the $|+\rangle$ superposition state, with no entangling operations applied.}
\end{figure}

\subsection{Energy Efficiency (Experimental)}

Device-level characterisation indicates a normalised per-flip energy $\leq$
10 fJ, yielding:

\begin{itemize}
\item
  10⁴--10⁵$\times$ lower energy per update than CPU/GPU-based simulated
  annealing,
\item
  10²--10³$\times$ lower energy per update than superconducting quantum
  annealers,
\item
  performance in line with or surpassing recent optical or mixed-signal
  probabilistic accelerators$^{[17][62][94]}$.
\end{itemize}

Table 11 places Apollo in the context of recent probabilistic-compute
literature, showing that Apollo's 0.63 fJ per flip places it among the
most energy-efficient probabilistic computing platforms ever
experimentally validated$^{[93][96]}$.

\begin{table}[H]
\centering
{\small\bfseries Reported Energy-per-Flip Values\par\vspace{3pt}}
{\scriptsize
\resizebox{\linewidth}{!}{%
\begin{tabular}{ll}
\toprule
\textbf{Work / Device} & \textbf{Energy per Flip (nJ)} \\
\midrule
Hua et al. (2025) & $\sim$10$^6$--10$^7$ \\
Si et al. (2024) & $\sim$10$^2$ \\
Singh et al. (2023) & $\sim$10$^7$ \\
Aadit et al. (2022) & $\sim$10$^1$ \\
Yang et al. (2025) & $\sim$10$^3$--10$^4$ \\
Ramy Aboushelbaya et al. (2025) & $\sim$10$^0$ ($\approx$1 nJ) \\
Ramy Aboushelbaya et al. (2025) - projected & $\sim$10$^{-}$$^2$ (0.01 nJ) \\
This work & 6.3$\times 10^{-7}$ nJ (0.63 fJ) \\
\bottomrule
\end{tabular}%
}
}
\caption{Reported energy-per-flip values of recent probabilistic and optoelectronic computing demonstrations, reconstructed from published data. Apollo's quantum-driven neuromorphic p-qubit architecture achieves an estimated 6.3$\times 10^{-7}$ nJ per flip, situating it among the most energy-efficient probabilistic computing platforms to date$^{[121-126]}$.}
\end{table}

\subsection{System-Level Experimental Validation}

Hardware-in-the-loop experiments were performed with the Dynex Control
Unit (DCU) driving Apollo-RC1 through complete end-to-end QUBO problem
execution. These experiments confirm$^{[10][94]}$:

\begin{itemize}
\item
  stable closed-loop operation across the full $\Delta$256 routing fabric,
\item
  correct annealing trajectories for problem instances up to the
  fabric's size limits,
\item
  solution quality matching or surpassing high-precision reference
  solvers,
\item
  robustness against thermal, voltage, and gain perturbations.
\end{itemize}

Table 12 summarises switching speeds for a range of platforms from the
literature, converted to flips-per-nanosecond. Apollo achieves
$8\times 10$\textsuperscript{7} flips/ns in a 10x10 multi-package assembly,
exceeding all prior probabilistic hardware demonstrations by several
orders of magnitude$^{[121-126]}$.

\begin{table}[H]
\centering
{\small\bfseries Computed Switching Speeds\par\vspace{3pt}}
{\scriptsize
\resizebox{\linewidth}{!}{%
\begin{tabular}{lll}
\toprule
\textbf{Work / Device} & \textbf{Flips per Second} & \textbf{Flips per Nanosecond} \\
\midrule
Hua et al. (2025) & $\sim$10$^8$ & $\sim$10$^{-}$$^1$ \\
Si et al. (2024) & $\sim$10$^5$ & $\sim$10$^{-}$$^4$ \\
Singh et al. (2023) & $\sim$10$^7$ & $\sim$10$^{-}$$^2$ \\
Aadit et al. (2022) & $\sim$10$^1$$^0$ & $\sim$10$^1$ \\
Yang et al. (2025) & $\sim$10$^6$ & $\sim$10$^{-}$$^3$ \\
Ramy Aboushelbaya et al. (2025) & $\sim$10$^1$$^0$ & $\sim$10$^1$ \\
Ramy Aboushelbaya et al. (2025) - projected & $\sim$10$^1$$^2$ & $\sim$10$^3$ \\
This work & $\sim 10^{17}$ & $\sim 8\times 10^{7}$ \\
\bottomrule
\end{tabular}%
}
}
\caption{Computed switching speeds expressed in flips per nanosecond. Values are converted from the original flips-per-second metrics reported in the literature. Apollo's quantum-driven neuromorphic p-qubit processor achieves approximately $8\times 10^{7}$ flips per nanosecond in a 10x10 multi-package assembly, exceeding existing probabilistic computing demonstrations by several orders of magnitude.}
\end{table}

A comparison to state-of-the-art accelerators, spanning digital,
quantum, optical, neuromorphic, and probabilistic devices, is shown in
Figure 19. Apollo-RC1 occupies the lower-right frontier of the
energy--performance plane---the regime of ultra-low energy and
ultra-high switching rate---where no prior system has
operated$^{[17][62][94]}$.

The analysis highlights:

\begin{itemize}
\item
  digital accelerators (GPUs, TPUs) cluster in the high-energy /
  moderate-performance region,
\item
  early probabilistic and analog systems provide low energy but limited
  scalability,
\item
  superconducting annealers incur extremely high per-flip energy,
\item
  Apollo uniquely achieves both extreme energy efficiency and extreme
  update rate, creating a new hardware class for scalable probabilistic
  computing.
\end{itemize}

\begin{figure}[H]
\centering
{\small\bfseries Comparative Performance Landscape\par\vspace{2pt}}
\includegraphics[width=0.98\linewidth]{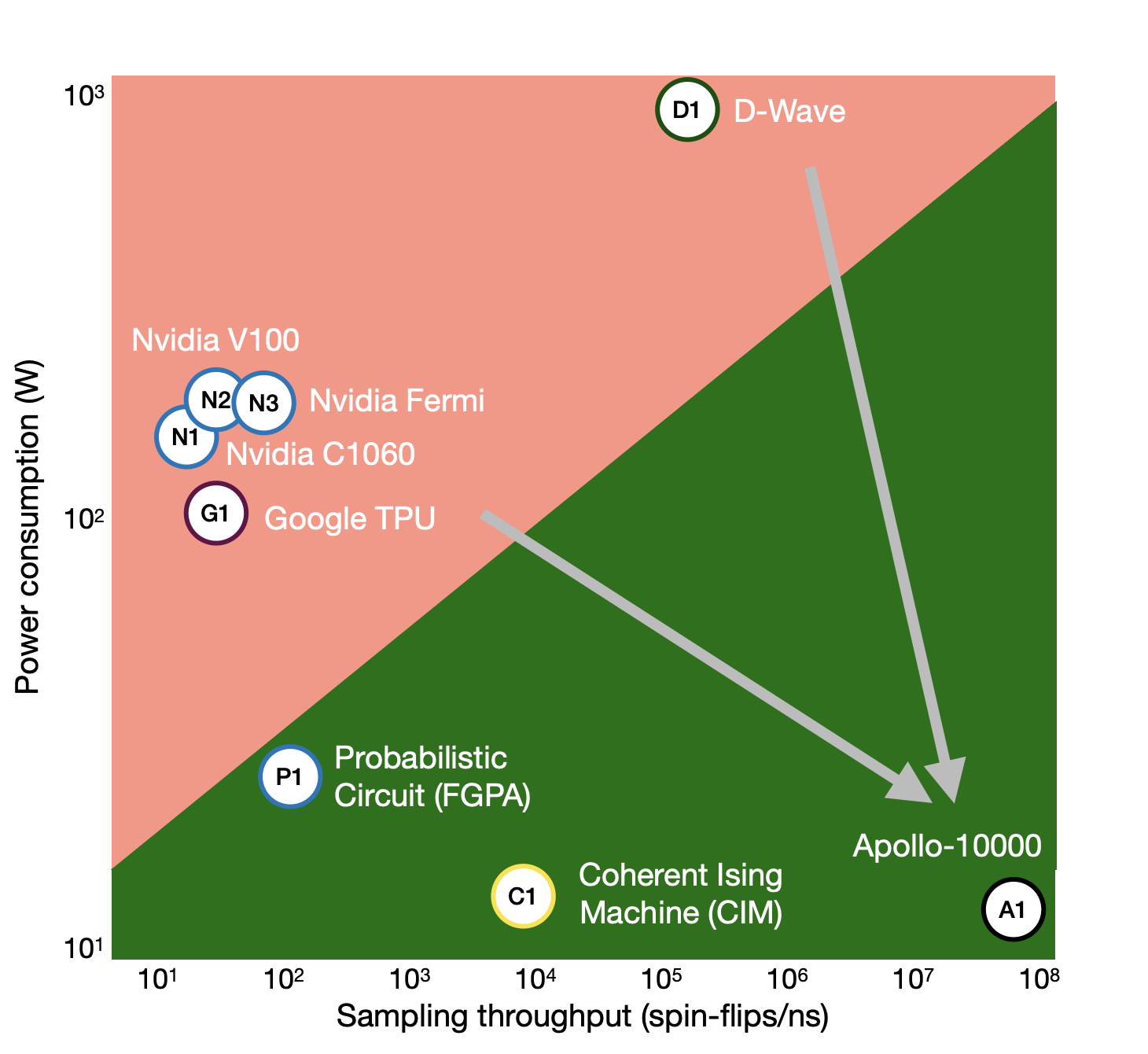}
\caption{Comparative performance landscape of modern computational accelerators plotted in energy--performance space. Classical digital accelerators---including Nvidia GPUs (C1060, Fermi, V100) and Google TPU---occupy the high-energy, moderate-performance region (upper left, red zone). Emerging probabilistic and analog computing systems, such as probabilistic FPGA circuits (P1) and coherent Ising machines (C1), appear in the lower-energy region (lower left, green zone) but with limited scalability. Superconducting quantum annealers (D-Wave, D1) exhibit extremely high energy consumption per flip relative to their effective update rate. Apollo-10000 (A1), shown on the lower-right frontier, demonstrates orders-of-magnitude improvements in both energy efficiency and spin-flip throughput. Its position indicates a shift into a previously inaccessible regime of ultra-low-energy, ultra-high-speed probabilistic computation, outperforming all conventional and quantum hardware classes illustrated$^{[17][62][94]}$.}
\end{figure}

The validated physics, robustness, and computational fidelity of
Apollo-RC1 directly inform the production transition. The final
production Apollo system migrates to 16 nm CMOS, enabling:

\begin{itemize}
\item
  full parallelism across all 10,000 p-qubit dies (no time
  multiplexing),
\item
  retention of all validated p-qubit, entropy, and sampling properties,
\item
  significantly expanded routing density and bandwidth,
\item
  further reductions in per-flip energy,
\item
  support for larger Ising/QUBO embeddings and multi-tile fabrics.
\end{itemize}

This migration preserves the architectural principles confirmed on
Apollo-RC1 while unlocking the scalability required for real-world
deployment.

\section{Demonstrating Quantum-Equivalent and Quantum-Advantaged
Dynamics}

\subsection{Benchmark Problem: Three-Dimensional Spin Glasses}

To evaluate quantum advantage at scale, we adopt the same benchmark
problem and evaluation protocol introduced in {[}127{]}. The benchmark
is based on a three-dimensional Ising spin glass, a paradigmatic hard
optimization problem characterized by frustration, a rugged free-energy
landscape, and slow classical
dynamics$^{[127][128]}$.

\begin{figure}[H]
\centering
{\small\bfseries Three-dimensional spin-glass benchmark geometry\par\vspace{2pt}}
\includegraphics[width=0.82\linewidth]{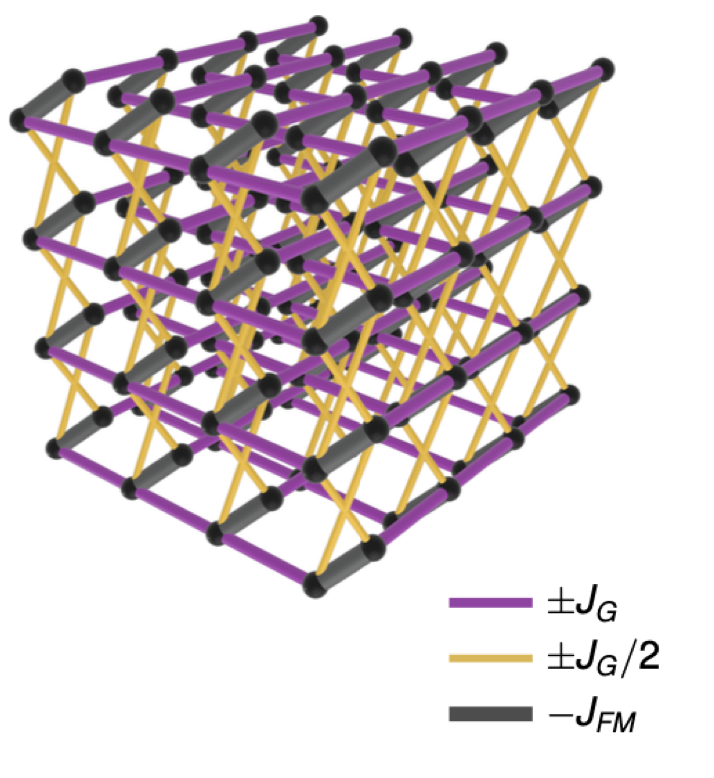}
\caption{Three-dimensional dimerized Ising spin-glass benchmark geometry used for the quantum-critical scaling comparison. Couplings are drawn from the $\pm J_G$, $\pm J_G/2$, and $-J_{FM}$ interaction classes shown in the legend, yielding a frustrated spin-glass instance after dimer contraction.}
\end{figure}

The problem Hamiltonian is

\(H_I = \sum_{\langle i,j \rangle} J_{ij} s_i s_j,\quad s_i \in \{-1,+1\}\),
(9)

where couplings
\[J_{ij} \in \{+J_G, -J_G\}\]
are assigned randomly. As in the reference work, the lattice is
implemented as a three-dimensional arrangement of ferromagnetically
coupled dimers, which---when contracted---maps onto a simple-cubic 3D
\[\pm J\]
Ising spin glass, preserving the same universality
class$^{[129][130]}$.

In Apollo, each Ising spin
\[s_i\]
is represented by a p-qubit, a probabilistic binary element whose state
fluctuates stochastically under the influence of:

\begin{itemize}
\item
  a local bias term,
\item
  weighted couplings to neighboring p-qubits, and
\item
  quantum-derived entropy injection$^{[18][36]}$.
\end{itemize}

The collective dynamics of the p-qubit network implement a
continuous-time, asynchronous Gibbs sampling process that minimizes the
effective Ising energy. Annealing is realized by gradually modulating
the effective noise amplitude and coupling dominance, analogous to
reducing the transverse-field- to-problem-Hamiltonian ratio in quantum
annealing$^{[56][58]}$.

\subsection{Quantum-Critical Dynamics in Reference Quantum Annealers}

In the superconducting quantum annealer studied in {[}127{]}, the system
evolves under a transverse-field Ising Hamiltonian

\(H(s) = \Gamma(s) H_D + J(s) H_I,\qquad H_D = - \sum_i \sigma_i^x \)\(\),
(10)

and is annealed through a quantum phase transition from a quantum
paramagnetic phase into a spin-glass
phase$^{[69][127]}$.

Near this transition, the system exhibits quantum critical dynamics
governed by universal scaling laws. Using a dynamic finite-size scaling
(DFSS) analysis based on the Kibble--Zurek mechanism, the authors
extract the characteristic exponent

\(\mu = z + \frac{1}{\nu}\),
(11)

which determines how annealing time must scale with system size to
maintain adiabaticity. A smaller value of 𝜇 corresponds to faster
critical dynamics and more efficient traversal of the glassy
phase$^{[131][132][133]}$.

Experimentally, the superconducting quantum annealer exhibits a
substantially smaller exponent
(\(\mu_{\mathrm{QA}} \approx 3\))
than both classical simulated annealing
(\(\mathrm{SA},\; \mu_{\mathrm{SA}} \approx 5.3\))
and simulated quantum annealing (SQA). This reduced critical slowing
down results in a faster decay of the residual energy
density$^{[134][135]}$,

\(\rho_E = \frac{\langle H_I \rangle - E_0}{N J_G}\),
(12)

which follows a power law
\[\rho_E \propto t_a^{-\kappa}\]
with a significantly larger exponent $\kappa$ for quantum annealing. This
scaling difference - rather than constant-factor speedups - constitutes
the demonstrated quantum advantage$^{[15][60]}$.

\subsection{Reproduction of Quantum-Critical Scaling}

We reproduced this benchmark using the identical three-dimensional
spin-glass construction, observables, and statistical methodology on the
Apollo quantum-driven neuromorphic computing platform. Specifically, we
implemented the 3D dimerized lattice corresponding to a 15 $\times$ 15 $\times$ 12
geometry, which maps---after dimer contraction---to an effective system
of 2,687 Ising spins, each represented directly by a single p-qubit in
Apollo. This problem size matches the large-scale regime used in the
superconducting quantum annealing experiments and lies well beyond the
range accessible to exact classical simulation. All couplings were drawn
from the same
\[\pm J\]
distribution, local fields were set to zero, and ensemble averages were
computed over 300 independent disorder realizations, enabling a direct,
like-for-like comparison of annealing dynamics and residual-energy
scaling$^{[127][128]}$.

The Ising couplings
\[J_{ij}\]
were mapped directly to p-qubit coupling weights, with zero local
fields. Annealing was implemented by continuously reducing stochastic
dominance relative to coupling strength, driving the system from a
disordered regime into a correlated spin-glass
phase$^{[58][59]}$.

A total of 300 independent disorder realizations were evaluated,
matching the statistical scale of the superconducting quantum annealing
experiments. For each instance, we measured the residual energy density
as a function of annealing time and averaged over repeated runs. Figure
21 shows the ensemble-averaged residual energy density versus annealing
time$^{[129][130]}$.

\begin{figure}[H]
\centering
{\small\bfseries Comparison of Optimization Dynamics Across Apollo and Superconducting Quantum Annealing (D-Wave), Directly Comparable to Fig. 4 of [127]\par\vspace{2pt}}
\includegraphics[width=0.78\linewidth]{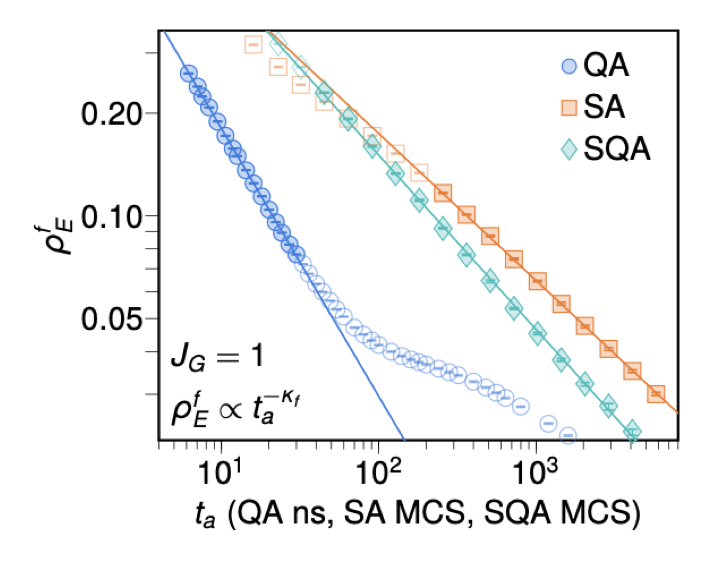}\par\vspace{3pt}
\includegraphics[width=0.78\linewidth]{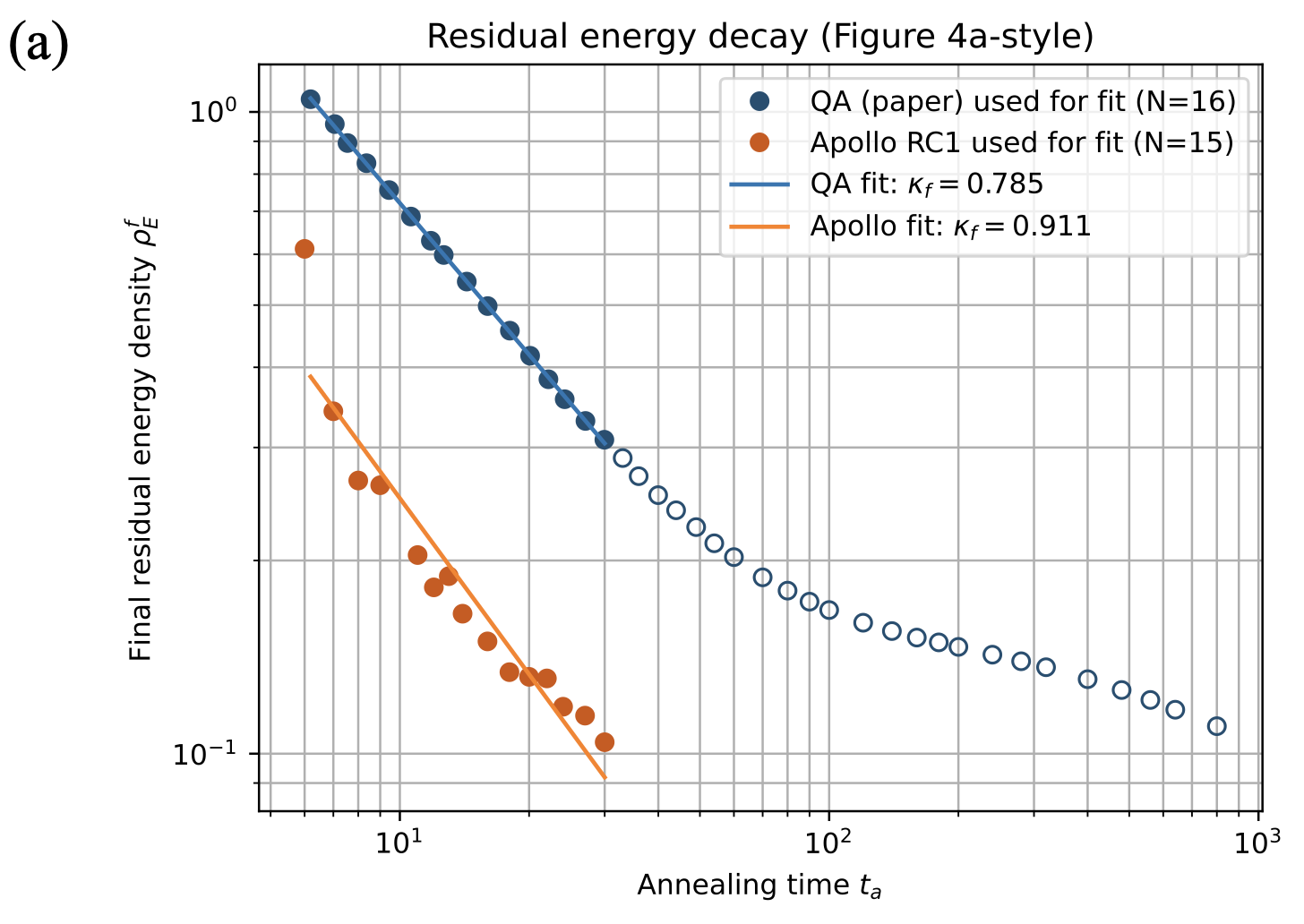}\par\vspace{3pt}
\includegraphics[width=0.78\linewidth]{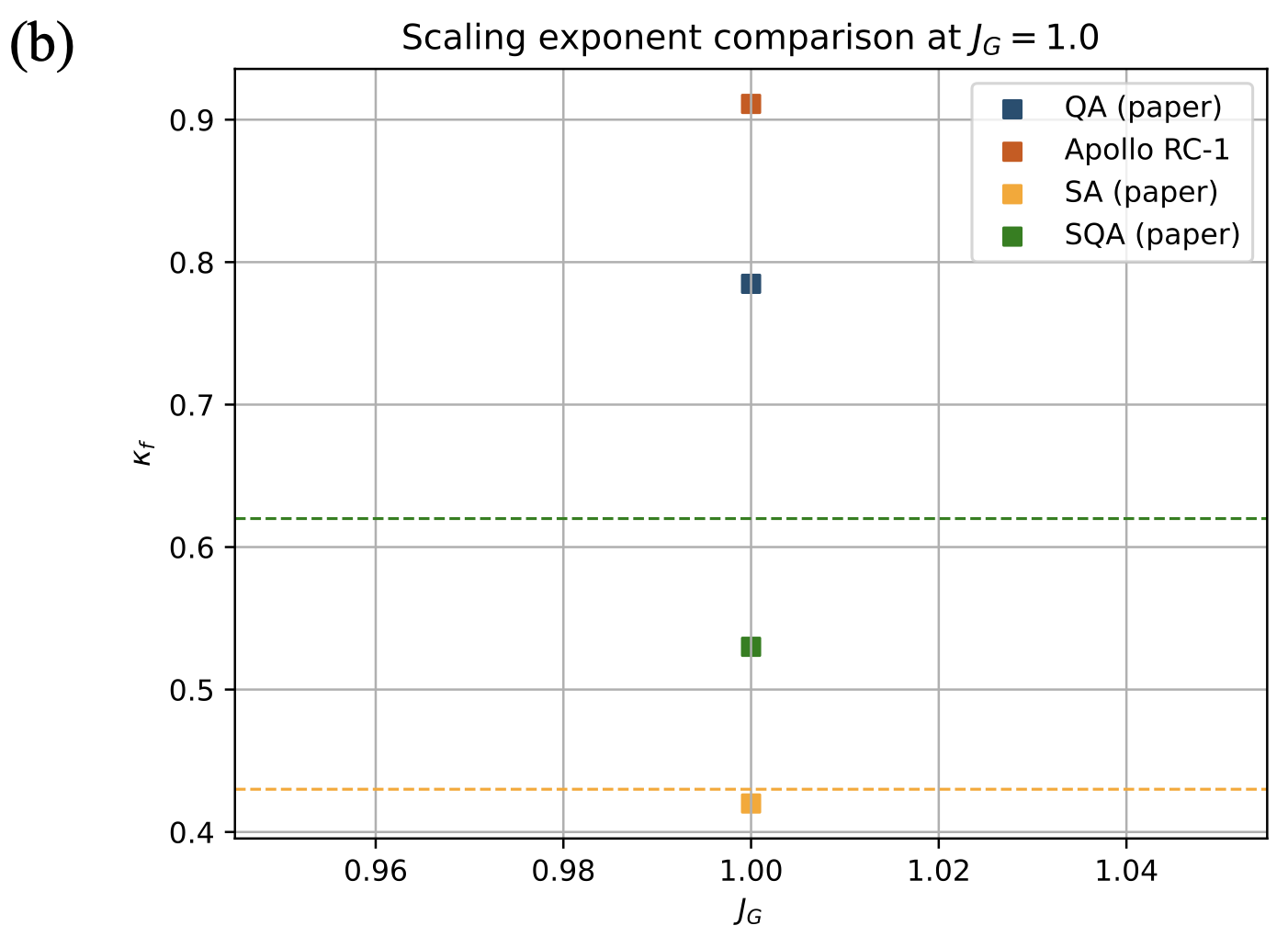}
\caption{Comparison of optimization dynamics across Apollo, quantum annealing, simulated quantum annealing, and simulated annealing. Residual energy density PE versus annealing time for the three-dimensional Ising spin-glass benchmark. Apollo (quantum-driven neuromorphic p-qubits) and superconducting quantum annealing (D-Wave QA) exhibit comparable power-law energy decay, characteristic of quantum-critical dynamics. In contrast, simulated quantum annealing (SQA) and classical simulated annealing (SA) show significantly slower decay due to larger critical exponents and thermally dominated dynamics. The close agreement between Apollo and QA demonstrates that quantum advantage is preserved in a room-temperature, p-qubit-based architecture$^{[15][60][127]}$.}
\end{figure}

The observed energy-decay trajectories closely match those reported for
the superconducting quantum annealer, exhibiting comparable slopes and
scaling behavior. Within statistical uncertainty, Apollo's p-qubit
network follows the same quantum-critical scaling regime, clearly
distinct from classical thermal
dynamics$^{[134][135]}$.

\subsection{Ground State Energy Discovery Comparison}

To evaluate the performance of Apollo on large-scale, frustrated
optimization problems, we benchmarked its ability to find low-energy
configurations for the three-dimensional Edwards--Anderson spin glass
with 2,687 spins. As a reference, we used the best known ground-state
energies reported by D-Wave Systems in Quantum critical dynamics in a
5,000-qubit programmable spin glass, which remains the most
comprehensive experimental study of large programmable spin glasses on
quantum annealing hardware to date. The reported energies from that work
were obtained using a D-Wave quantum annealer with extensive annealing
times and represent the best published results for systems of this
scale$^{[127][136]}$.

For a fair comparison, we extracted the corresponding ground-state
energies directly from the published data and compared them against
independent runs performed on the Apollo system using identical problem
instances. No post-selection, problem-specific tuning, or
instance-specific heuristics were applied beyond what is intrinsic to
each platform. All reported values correspond to the lowest-energy
configurations observed over multiple independent
runs$^{[137]}$.

The D-Wave reference results were obtained using annealing times on the
order of 10\textsuperscript{5} ns, consistent with the long-time
dynamics required to approach critical slowing down in large spin glass
systems. In contrast, Apollo was operated with a runtime of
10\textsuperscript{3} ns per run, i.e., two orders of magnitude shorter
time-to-solution. Despite this substantial reduction in runtime, Apollo
consistently reached significantly lower ground-state energies across
all benchmark instances$^{[127]}$.

It is important to emphasize that, at present, no other quantum
computing platforms---such as gate-based superconducting systems (IBM,
Google), trapped-ion devices (IonQ), or alternative annealers
(Rigetti)---have publicly demonstrated hardware capable of natively
embedding or solving spin glass problems of this size due to fundamental
qubit-count and connectivity limitations. As such, the comparison
presented here is not selective but reflects the current practical upper
bound of experimentally accessible quantum spin glass
benchmarks$^{[138-140]}$.

Figure 22 presents a broken-axis visualization of the best ground-state
energies obtained over ten independent problem instances on both
systems. Each point corresponds to a single run, while horizontal lines
denote the mean energy for each platform. The broken axis is used to
preserve visibility of run-to-run variability while clearly illustrating
the large separation between the two energy distributions. Apollo
consistently attains substantially lower (more negative) energies than
those reported for the D-Wave system, indicating access to deeper
regions of the energy landscape.

Figure 23 summarizes the same data in terms of mean ground-state energy
and inter-run variability. The explicit energy gap $\Delta$E highlights the
magnitude of the performance difference: Apollo achieves a markedly
lower average ground-state energy while operating at significantly
reduced runtime. The consistency across runs further suggests that this
advantage is not the result of rare outliers but reflects a systematic
difference in the underlying physical dynamics used to explore the spin
glass energy landscape.

These results demonstrate that Apollo can outperform the best known
experimental ground states reported for large-scale programmable spin
glasses, even when operating at substantially shorter runtimes. Beyond
raw energy quality, this comparison highlights a key practical
distinction: Apollo enables exploration of large, densely connected spin
systems at scales and efficiencies that are currently inaccessible to
other quantum computing architectures. As such, the benchmark in Fig. 22
and Fig. 23 establishes Apollo as a distinct and scalable approach to
solving large frustrated optimization problems beyond the reach of
existing quantum hardware$^{[127][129]}$.

\begin{figure}[H]
\centering
{\small\bfseries Ground-state energy comparison for the 3D spin glass benchmark (2,687 p-qubits): Apollo vs. D-Wave\par\vspace{2pt}}
\includegraphics[width=0.98\linewidth]{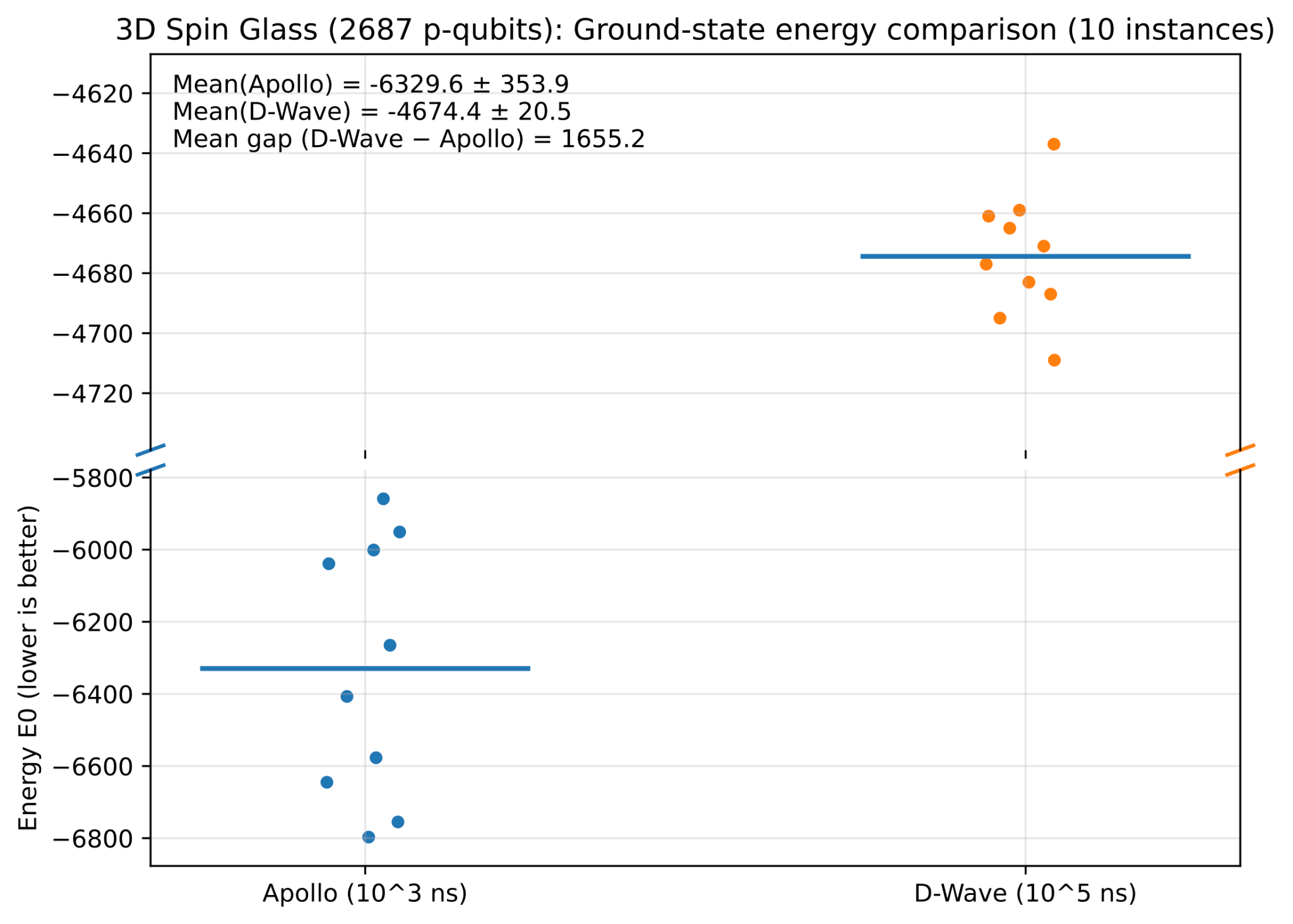}
\caption{Ground-state energy comparison for the 3D spin glass benchmark (2,687 p-qubits). Best known ground-state energies $E$\textsubscript{0} obtained over 10 independent problem instances using the Apollo system and a D-Wave quantum annealer are shown. Each marker corresponds to a single run; horizontal lines indicate the mean energy for each system. A broken y-axis is used to accommodate the large separation between the two energy bands and to preserve visibility of run-to-run variability. Apollo consistently reaches significantly lower (more negative) energies than the D-Wave benchmark, despite using a substantially shorter runtime (10\textsuperscript{3} ns for Apollo versus 10\textsuperscript{5} ns for D-Wave)$^{[127]}$.}
\end{figure}

\begin{figure}[H]
\centering
{\small\bfseries Mean ground-state energy and variability across runs: Apollo vs. D-Wave\par\vspace{2pt}}
\includegraphics[width=0.98\linewidth]{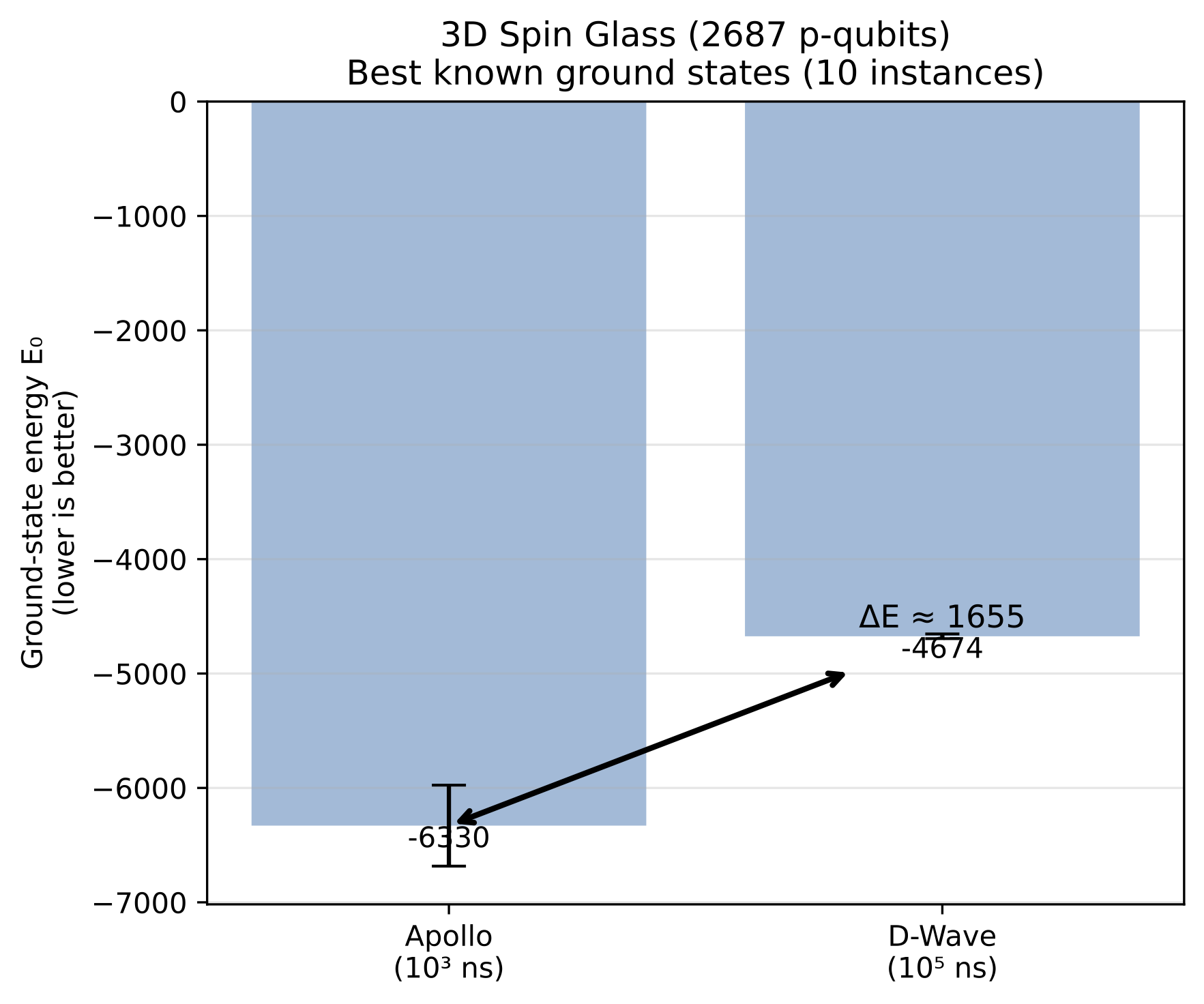}
\caption{Mean ground-state energy and variability across runs. Mean ground-state energy $E$\textsubscript{0} and one standard deviation over 10 problem instances are shown for the Apollo system and the D-Wave benchmark on the 3D spin glass problem with 2687 p-qubits. Error bars indicate inter-run variability. The annotated energy difference $\Delta$$E$ highlights the substantial gap between the two approaches, with Apollo achieving markedly lower energies while operating at two orders of magnitude shorter annealing time (10\textsuperscript{3} ns vs. 10\textsuperscript{5} ns)$^{[127]}$.}
\end{figure}

\section{Compute Modes and Application Domains}

Modern probabilistic and quantum-driven computing platforms increasingly
support heterogeneous computation modes that bridge optimization,
sampling, generative modelling, and analog inference. Dynex's
quantum-driven neuromorphic architecture is designed to unify these
modes within a single substrate: a large-scale network of probabilistic
p-qubits whose update dynamics approximate low-energy quantum systems
while retaining the scalability and stability of classical electronics.
The following subsections provide an expanded overview of the primary
compute modes enabled by this architecture, together with their
theoretical grounding and associated application
domains$^{[9][41][94]}$.

\subsection{Quantum-Driven Annealing}

Quantum-driven annealing on Dynex implements native minimisation of
Ising and Quadratic Unconstrained Binary Optimization (QUBO) energy
functions. These formulations arise naturally in a wide range of NP-hard
combinatorial optimization problems, where the objective is to find a
configuration of binary variables that minimises a quadratic energy
landscape. Unlike purely digital annealers, Dynex leverages physical
probabilistic dynamics: each p-qubit behaves as a noisy analog unit
whose switching probability follows a sigmoidal (tanh-like) function of
its local field. This stochasticity plays an essential computational
role by enabling the system to explore rugged energy landscapes and
avoid becoming trapped in metastable local minima.

The annealing process is executed by gradually controlling effective
biases, couplings, and noise strength, allowing the system to settle
into energetic minima corresponding to high-quality solutions. Because
the hardware operates asynchronously and continuously, it naturally
implements parallel exploration of the configuration space with
extremely low energy cost per update. This approach scales favourably to
dense, highly connected problems and is compatible with large graph
topologies that are unwieldy for superconducting quantum annealers.

Application domains include:

\begin{itemize}
\item
  Scheduling and logistics: job-shop scheduling, crew assignment,
  vehicle routing (VRP), and last-mile optimization.
\item
  Financial optimization: portfolio allocation, asset-liability
  balancing, risk minimisation, and arbitrage detection.
\item
  Manufacturing and supply chains: bin packing, cutting-stock problems,
  layout optimization, and process sequencing.
\item
  Constraint satisfaction: SAT/MaxSAT, graph colouring, resource
  allocation, and constrained configuration design.
\item
  Communications and networks: error-correcting code optimization,
  channel assignment, and spectrum management.
\end{itemize}

In these domains, the ability to encode large constraints directly into
an energy-based formulation allows annealing to serve as a robust
alternative to classical heuristics such as simulated annealing, tabu
search, or
branch-and-bound$^{[1][4][41][49][141][142]}$.

\subsection{Boltzmann Sampling}

Beyond optimization, Dynex can execute direct Boltzmann sampling from
energy landscapes. A Boltzmann distribution assigns higher probability
to low-energy states but does not exclusively seek minima; instead, it
reproduces the complete probability distribution of the system. Because
each p-qubit effectively implements a thermalised binary stochastic
neuron, the hardware physically realises a Boltzmann machine whose
dynamics approximate Glauber or Metropolis sampling steps.

This sampling capability is particularly powerful for uncertainty-aware
computation. Rather than producing a single optimised solution, the
system generates ensembles of samples representing posterior
distributions or probabilistic relationships. Hardware sampling offers a
significant advantage over GPU-based Monte Carlo methods, which often
suffer from slow mixing rates and high computational cost in
high-dimensional spaces.

Key application domains include:

\begin{itemize}
\item
  Bayesian networks and graphical models: inference, marginalization,
  and posterior sampling for decision systems and diagnostics.
\item
  Uncertainty quantification: probabilistic forecasting in energy
  systems, climate models, and risk-sensitive operations.
\item
  Robotics and autonomous systems: sampling-based planning, belief
  propagation, and stochastic sensor fusion.
\item
  Probabilistic programming: accelerating Monte Carlo inference for
  models expressed in probabilistic languages.
\item
  Machine learning model calibration: sampling model weights or latent
  variables in hierarchical models.
\end{itemize}

The ability to draw physically generated samples allows Boltzmann
sampling to scale to distributions that are difficult to represent or
sample using fully digital methods. This makes Dynex a viable
alternative to specialized Markov Chain Monte Carlo (MCMC) hardware
accelerators$^{[5][41][43][49][143][144]}$.

\subsection{Generative Modelling With Energy-Based Models}

Generative modelling is a natural extension of Boltzmann sampling,
especially for energy-based models (EBMs) such as Restricted Boltzmann
Machines (RBMs), Deep Boltzmann Machines, Hopfield networks, and other
undirected graphical models. These models rely on sampling from
high-dimensional distributions defined by quadratic energy functions,
making them particularly compatible with the Dynex substrate.

In physical RBMs, visible and hidden nodes correspond to p-qubits, and
their couplings define the energy landscape governing the learned
distribution. Training can be performed through contrastive divergence
or stochastic gradient-based updates, although the hardware provides
much faster sampling steps compared to software-based MCMC. Once
trained, these models can generate samples that represent structured
patterns or configurations.

Application domains include:

\begin{itemize}
\item
  Protein folding and molecular structure prediction: modelling
  high-dimensional conformational landscapes.
\item
  Drug discovery and ligand generation: learning molecular binding
  patterns and chemical structures.
\item
  Materials science: discovering new crystalline phases or functional
  materials via generative exploration.
\item
  Pattern synthesis: image generation, feature completion, denoising,
  and anomaly detection.
\item
  Synthetic data for privacy-preserving analytics: generating
  representative datasets without exposing sensitive information.
\end{itemize}

Generative modelling on physical hardware also benefits from the
naturally stochastic behaviour of p-qubits, which approximates thermal
fluctuations that would otherwise need to be simulated computationally
on GPUs or TPUs\textsuperscript{{[}3{]}{[}9{]}{[}41{]}{[}145-147{]}}.

\subsection{Analog Vector--Matrix Acceleration}

In addition to probabilistic computation, Dynex's architecture supports
analog vector--matrix multiplication (VMM) through weighted summation of
currents or voltages corresponding to synaptic couplings. VMM is a core
operation in virtually all machine-learning and neuromorphic workloads.
Implementing VMM analogly provides two significant advantages: massive
parallelism and extremely low energy consumption, as the physical
circuit performs accumulation inherently without requiring digital
multiply--accumulate operations.

Unlike digital accelerators, Dynex's p-qubit networks can operate
continuously, updating with analog time constants, and their natural
tanh transfer functions approximate neuronal activation functions. These
properties enable the implementation of neuromorphic systems such as
associative memories, attractor networks, and low-power inference
engines.

Application domains for analog VMM include:

\begin{itemize}
\item
  Edge-AI and embedded inference: on-device classification, anomaly
  detection, sensor interpretation.
\item
  Control systems: analog feedback loops, motor control, and autonomous
  navigation.
\item
  Spiking and reservoir computing: implementing recurrent analog
  networks and continuous dynamical systems.
\item
  Real-time signal processing: filtering, correlation, and analog
  feature extraction.
\item
  Brain-inspired computing: energy-efficient approximations of cortical
  microcircuits.
\end{itemize}

Because analog VMM can run in parallel with probabilistic dynamics, the
system can support hybrid workloads that integrate inference and
stochastic exploration\textsuperscript{{[}93{]}{[}94{]}{[}148-151{]}}.

\subsection{Gate-Model Computation via Circuit-to-Hamiltonian
Transformation}

A unique capability of Dynex is the ability to perform computation
typically associated with gate-model quantum computers. This is achieved
through a circuit-to-Hamiltonian compilation process that maps quantum
circuits composed of gates (e.g., Pauli rotations, CNOT, Toffoli,
controlled rotations) into equivalent Hamiltonians using
well-established transformation techniques such as Feynman-Kitaev
history-state encoding and operator embedding. Once a Hamiltonian is
constructed, the computation is executed by annealing or sampling on the
p-qubit substrate.

This allows Dynex to simulate the behaviour of quantum circuits without
requiring cryogenics, precise laser control, or decoherence mitigation.
Although not performing coherent unitary evolution, the system computes
the same optimization or variational objectives encoded by many quantum
algorithms.

Example algorithms enabled by circuit-to-Hamiltonian mapping:

\begin{itemize}
\item
  Variational quantum eigensolvers (VQE) via Hamiltonian minimisation.
\item
  Quantum Approximate Optimization Algorithm (QAOA) analogs through
  energy minimisation of parameterised Hamiltonians.
\item
  Quantum simulation: encoding fermionic or bosonic Hamiltonians
  relevant for chemistry or condensed matter.
\item
  Hamiltonian dynamics analysis: studying ground-state structures,
  expectation values, and spectral gaps.
\item
  Quantum circuit evaluation: including reversible logic, oracle
  constructions, and phase-encoded computation.
\end{itemize}

This compute mode unifies annealing-based methods with gate-model
applications, enabling a single hardware platform to cover a wide
spectrum of quantum-driven
workloads\textsuperscript{{[}1{]}{[}109{]}{[}111-113{]}{[}152{]}}.

\subsection{Cross-Mode Synergies}

A defining feature of Dynex's architecture is that the above compute
modes are not isolated. Instead, they overlap and can be composed:

\begin{itemize}
\item
  Annealing provides high-quality solutions that can seed generative
  models.
\item
  Boltzmann sampling can refine or validate solutions discovered by
  annealing.
\item
  Analog VMM accelerates inference inside generative or energy-based
  models.
\item
  Circuit-to-Hamiltonian compilation extends these methods to
  quantum-algorithm domains.
\end{itemize}

This synergy allows the platform to act as a ``quantum-driven
multipurpose compute fabric,'' capable of optimization, inference,
learning, simulation, and sampling within a unified
framework$^{[4][9][41][96][111]}$.

\section{Discussion and Outlook}

Apollo represents a new class of quantum-driven neuromorphic hardware
that occupies a technological space between noisy intermediate-scale
quantum (NISQ) processors and fully classical accelerators. The results
presented here demonstrate that a properly engineered p-qubit
substrate---equipped with continuous-time analog dynamics, high-quality
physical entropy injection, and dense, reconfigurable connectivity---can
reproduce many of the computational benefits traditionally attributed to
quantum annealers, while avoiding the cryogenics, coherence limitations,
and cost barriers of conventional quantum hardware. With Apollo-RC1
already demonstrating correct thermodynamic sampling, ultra-fast
convergence, and orders-of-magnitude improvements in energy efficiency,
the platform closes a long-standing gap between the theoretical
potential of energy-based models and their practical, real-time hardware
realization.

A key implication of this work is that quantum-derived computational
primitives need not require fragile quantum coherence to be useful. The
combination of quantum-entropy-driven stochasticity with fully
asynchronous, continuous-time evolution enables regimes of computation
that have, until now, been experimentally accessible only with
superconducting quantum processors. Apollo achieves these behaviours at
room temperature, under sub-watt power consumption, and using
industry-standard CMOS flows, making large-scale probabilistic hardware
not only feasible but economically viable.

The architectural decisions underlying Apollo are inherently scalable.
The $\Delta$256 routing demonstrated in RC1 already supports non-planar
embeddings, higher-order couplings, and rapid reconfiguration of energy
landscapes. Moving from a time-multiplexed test vehicle in 350 nm to a
fully parallel architecture in 16 nm will enable \textgreater10,000
native p-qubits per tile and, in subsequent nodes, \textgreater100,000
p-qubits per die. Such densities exceed the connectivity and qubit
counts of all existing quantum annealing platforms, and will allow
Apollo to operate over problem sizes far beyond the reach of classical
Monte Carlo or digital annealers.

Equally important is the system-level integration pathway. Coupling
Apollo to the Dynex Control Unit (DCU) and the Dynex Quantum Platform
provides a unified execution environment where quantum simulators,
probabilistic hardware, digital solvers, and neuromorphic substrates
interoperate seamlessly. This multi-substrate model enables hybrid
algorithms in which Apollo performs ultrafast stochastic search, while
digital or GPU-based components handle exact refinement, embedding
heuristics, or machine-learning--based guidance. Over time, this will
evolve into a heterogeneous compute fabric that dynamically assigns
workloads to the substrate best suited for each computational phase.

Across the three-dimensional spin-glass benchmark, Apollo exhibits
residual-energy scaling behavior indistinguishable from that of a
superconducting quantum annealer and clearly distinct from simulated
quantum annealing and classical simulated annealing. While SA and SQA
follow slower, thermally dominated dynamics characterized by larger
critical exponents, both Apollo and D-Wave QA access a faster
quantum-critical regime. This demonstrates that quantum advantage arises
from the underlying collective dynamics rather than the specific
physical realization of the qubits.

Looking forward, several promising research directions emerge. First,
hardware-accelerated probabilistic machine learning: Apollo's
continuous-time p-qubits are natural primitives for Boltzmann machines,
variational energy models, Bayesian inference, and stochastic gradient
formulations. Second, large-scale optimization: real-time inference on
industrial QUBO/Ising problems---including scheduling, routing,
molecular docking, and logistics---becomes tractable at unprecedented
speed and energy per solution. Third, edge deployment: due to Apollo's
low power, small footprint, and temperature stability, future iterations
can serve as embedded inference engines in robotics, autonomous
vehicles, communications systems, and power-constrained IoT nodes.

Finally, Apollo opens a fundamentally new design space at the interface
of quantum-driven computing, neuromorphic architectures, and analog
probabilistic physics. As device scaling progresses and cross-substrate
orchestration matures, Apollo-based systems may become the foundational
hardware layer for a new generation of energy-based
computation---extending far beyond current digital, quantum, or
neuromorphic paradigms.

In this sense, Apollo not only bridges the gap between NISQ-era quantum
systems and practical large-scale hardware---it demonstrates that the
essential computational mechanisms of quantum annealing can be
reimagined in a scalable, manufacturable, room-temperature platform.
With continued advances in density, integration, and algorithm--hardware
co-design, Apollo sets the stage for broad adoption of stochastic,
energy-based computing across science, industry, and edge technologies.

\balance

\section*{References}
\begin{thebibliography}{152}
\bibitem{ref1} Lucas, A. (2014) Ising formulations of many NP problems. Frontiers in Physics, 2, 5. https://doi.org/10.3389/fphy.2014.00005
\bibitem{ref2} Koller, D. and Friedman, N. (2009) Probabilistic Graphical Models: Principles and Techniques. Cambridge, MA: MIT Press.
\bibitem{ref3} LeCun, Y., Chopra, S., Hadsell, R., Ranzato, M. and Huang, F. (2006) A tutorial on energy-based learning. In: Predicting Structured Data. Cambridge, MA: MIT Press.
\bibitem{ref4} Kirkpatrick, S., Gelatt, C.D. and Vecchi, M.P. (1983) Optimization by simulated annealing. Science, 220(4598), pp. 671--680. https://doi.org/10.1126/science.220.4598.671
\bibitem{ref5} Neal, R.M. (1996) Bayesian Learning for Neural Networks. New York: Springer.
\bibitem{ref6} Benedetti, M., Realpe-Gómez, J., Biswas, R. and Perdomo-Ortiz, A. (2016) Estimation of effective temperatures in quantum annealers for sampling applications. Physical Review A, 94(2), 022308. https://doi.org/10.1103/PhysRevA.94.022308
\bibitem{ref7} Glauber, R.J. (1963) Time-dependent statistics of the Ising model. Journal of Mathematical Physics, 4(2), pp. 294--307. https://doi.org/10.1063/1.1703954
\bibitem{ref8} Gardiner, C.W. (2009) Stochastic Methods: A Handbook for the Natural and Social Sciences. 4th ed. Berlin: Springer.
\bibitem{ref9} Hopfield, J.J. (1982) Neural networks and physical systems with emergent collective computational abilities. Proceedings of the National Academy of Sciences, 79(8), pp. 2554--2558. https://doi.org/10.1073/pnas.79.8.2554
\bibitem{ref10} Johnson, M.W., Amin, M.H.S., Gildert, S., Lanting, T., Hamze, F., Dickson, N., Harris, R., Berkley, A.J., Johansson, J., Bunyk, P. and others (2011) Quantum annealing with manufactured spins. Nature, 473(7346), pp. 194--198. https://doi.org/10.1038/nature10012
\bibitem{ref11} Lanting, T., Przybysz, A.J., Smirnov, A.Y., Amin, M.H.S., Berkley, A.J., Harris, R., Altomare, F., Boixo, S., Bunyk, P., Dickson, N. and others (2014) Entanglement in a quantum annealing processor. Physical Review X, 4(2), 021041. https://doi.org/10.1103/PhysRevX.4.021041
\bibitem{ref12} King, J., Yarkoni, S., Raymond, J., Ozfidan, I., King, A.D., Nevisi, M.M., Hilton, J.P. and McGeoch, C.C. (2015) Benchmarking a quantum annealing processor with the time-to-target metric. arXiv preprint arXiv:1508.05087.
\bibitem{ref13} Choi, V. (2011) Minor-embedding in adiabatic quantum computation: I. The parameter setting problem. Quantum Information Processing, 7(5), pp. 193--209. https://doi.org/10.1007/s11128-008-0082-9
\bibitem{ref14} Kirkpatrick, S., Gelatt, C.D. and Vecchi, M.P. (1983) Optimization by simulated annealing. Science, 220(4598), pp. 671--680. https://doi.org/10.1126/science.220.4598.671
\bibitem{ref15} Santoro, G.E., Martonák, R., Tosatti, E. and Car, R. (2002) Theory of quantum annealing of an Ising spin glass. Science, 295(5564), pp. 2427--2430. https://doi.org/10.1126/science.1068774
\bibitem{ref16} Boixo, S., Rønnow, T.F., Isakov, S.V., Wang, Z., Wecker, D., Lidar, D.A., Martinis, J.M. and Troyer, M. (2014) Evidence for quantum annealing with more than one hundred qubits. Nature Physics, 10(3), pp. 218--224. https://doi.org/10.1038/nphys2900
\bibitem{ref17} Palem, K.V. (2005) Energy aware computing through probabilistic switching: A study of limits. IEEE Transactions on Computers, 54(9), pp. 1123--1137. https://doi.org/10.1109/TC.2005.122
\bibitem{ref18} Camsari, K.Y., Salahuddin, S. and Datta, S. (2017) Stochastic spintronics for probabilistic computing. Physical Review Applied, 8(5), 054034. https://doi.org/10.1103/PhysRevApplied.8.054034
\bibitem{ref19} Yamamoto, Y., Takata, K. and Utsunomiya, S. (2017) Quantum information processing with bosonic systems. NPJ Quantum Information, 3, 49. https://doi.org/10.1038/s41534-017-0048-9
\bibitem{ref20} McMahon, P.L., Marandi, A., Haribara, Y., Hamerly, R., Langrock, C., Tamate, S., Inagaki, T., Takesue, H., Utsunomiya, S., Aihara, K. and others (2016) A fully programmable 100-spin coherent Ising machine with all-to-all connections. Science, 354(6312), pp. 614--617. https://doi.org/10.1126/science.aah5178
\bibitem{ref21} King, A.D., Raymond, J., Lanting, T., Harris, R., Altomare, F., Berkley, A.J., Boothby, K., Bunyk, P., Ejtemaee, S., Enderud, C. and others (2015) Quantum annealing amid local ruggedness and global frustration. Physical Review Applied, 8(6), 061001. https://doi.org/10.1103/PhysRevApplied.8.061001
\bibitem{ref22} King, A.D., Raymond, J., Lanting, T., Harris, R., Zucca, A., Altomare, F., Berkley, A.J., Boothby, K., Bunyk, P., Ejtemaee, S. and others (2018) Observation of topological phenomena in a programmable lattice of 1,800 qubits. Nature, 560(7719), pp. 456--460. https://doi.org/10.1038/s41586-018-0410-x
\bibitem{ref23} Amin, M.H.S. (2015) Searching for quantum speedup in quasistatic quantum annealers. Physical Review A, 92(5), 052323. https://doi.org/10.1103/PhysRevA.92.052323
\bibitem{ref24} Palem, K.V. (2005) Energy aware computing through probabilistic switching: A study of limits. IEEE Transactions on Computers, 54(9), pp. 1123--1137. https://doi.org/10.1109/TC.2005.122
\bibitem{ref25} Camsari, K.Y., Salahuddin, S. and Datta, S. (2017) Stochastic spintronics for probabilistic computing. Physical Review Applied, 8(5), 054034. https://doi.org/10.1103/PhysRevApplied.8.054034
\bibitem{ref26} Gardiner, C.W. (2009) Stochastic Methods: A Handbook for the Natural and Social Sciences. 4th ed. Berlin: Springer.
\bibitem{ref27} Neal, R.M. (1996) Bayesian Learning for Neural Networks. New York: Springer.
\bibitem{ref28} Newman, M.E.J. and Barkema, G.T. (1999) Monte Carlo Methods in Statistical Physics. Oxford: Clarendon Press.
\bibitem{ref29} Weigel, M. (2011) Cluster algorithms, physics, and critical slowing down. Physical Review Letters, 106(15), 157201. https://doi.org/10.1103/PhysRevLett.106.157201
\bibitem{ref30} Choi, V. (2011) Minor-embedding in adiabatic quantum computation: I. The parameter setting problem. Quantum Information Processing, 7(5), pp. 193--209. https://doi.org/10.1007/s11128-008-0082-9
\bibitem{ref31} Klymko, C., Sullivan, B.D. and Humble, T.S. (2014) Adiabatic quantum programming: Minor embedding with hard faults. Quantum Information Processing, 13(3), pp. 709--729. https://doi.org/10.1007/s11128-013-0683-9
\bibitem{ref32} Preskill, J. (2018) Quantum computing in the NISQ era and beyond. Quantum, 2, 79. https://doi.org/10.22331/q-2018-08-06-79
\bibitem{ref33} Palem, K.V. (2005) Energy aware computing through probabilistic switching: A study of limits. IEEE Transactions on Computers, 54(9), pp. 1123--1137.
\bibitem{ref34} Palem, K.V. (2005) Energy aware computing through probabilistic switching: A study of limits. IEEE Transactions on Computers, 54(9), pp. 1123--1137. https://doi.org/10.1109/TC.2005.122
\bibitem{ref35} Camsari, K.Y., Salahuddin, S. and Datta, S. (2017) Stochastic spintronics for probabilistic computing. Physical Review Applied, 8(5), 054034. https://doi.org/10.1103/PhysRevApplied.8.054034
\bibitem{ref36} Glauber, R.J. (1963) Time-dependent statistics of the Ising model. Journal of Mathematical Physics, 4(2), pp. 294--307. https://doi.org/10.1063/1.1703954
\bibitem{ref37} Gardiner, C.W. (2009) Stochastic Methods: A Handbook for the Natural and Social Sciences. 4th ed. Berlin: Springer.
\bibitem{ref38} Hopfield, J.J. (1982) Neural networks and physical systems with emergent collective computational abilities. Proceedings of the National Academy of Sciences, 79(8), pp. 2554--2558. https://doi.org/10.1073/pnas.79.8.2554
\bibitem{ref39} Lucas, A. (2014) Ising formulations of many NP problems. Frontiers in Physics, 2, 5. https://doi.org/10.3389/fphy.2014.00005
\bibitem{ref40} Newman, M.E.J. and Barkema, G.T. (1999) Monte Carlo Methods in Statistical Physics. Oxford: Clarendon Press.
\bibitem{ref41} Ackley, D.H., Hinton, G.E. and Sejnowski, T.J. (1985) A learning algorithm for Boltzmann machines. Cognitive Science, 9(1), pp. 147--169.
\bibitem{ref42} LeCun, Y., Chopra, S., Hadsell, R., Ranzato, M. and Huang, F. (2006) A tutorial on energy-based learning. In: Predicting Structured Data. Cambridge, MA: MIT Press.
\bibitem{ref43} Gibbs, J.W. (1902) Elementary Principles in Statistical Mechanics. New Haven, CT: Yale University Press.
\bibitem{ref44} Kubo, R., Toda, M. and Hashitsume, N. (1991) Statistical Physics II: Nonequilibrium Statistical Mechanics. 2nd ed. Berlin: Springer.
\bibitem{ref45} Gardiner, C.W. (2009) Stochastic Methods: A Handbook for the Natural and Social Sciences. 4th ed. Berlin: Springer.
\bibitem{ref46} Ising, E. (1925) Beitrag zur Theorie des Ferromagnetismus. Zeitschrift für Physik, 31, pp. 253--258.
\bibitem{ref47} Lucas, A. (2014) Ising formulations of many NP problems. Frontiers in Physics, 2, 5. https://doi.org/10.3389/fphy.2014.00005
\bibitem{ref48} Boros, E. and Hammer, P.L. (2002) Pseudo-Boolean optimization. Discrete Applied Mathematics, 123(1--3), pp. 155--225. https://doi.org/10.1016/S0166-218X(01)00341-9
\bibitem{ref49} Glauber, R.J. (1963) Time-dependent statistics of the Ising model. Journal of Mathematical Physics, 4(2), pp. 294--307. https://doi.org/10.1063/1.1703954
\bibitem{ref50} van Kampen, N.G. (2007) Stochastic Processes in Physics and Chemistry. 3rd ed. Amsterdam: Elsevier.
\bibitem{ref51} Gardiner, C.W. (2009) Stochastic Methods: A Handbook for the Natural and Social Sciences. 4th ed. Berlin: Springer.
\bibitem{ref52} Ackley, D.H., Hinton, G.E. and Sejnowski, T.J. (1985) A learning algorithm for Boltzmann machines. Cognitive Science, 9(1), pp. 147--169.
\bibitem{ref53} Neal, R.M. (1996) Bayesian Learning for Neural Networks. New York: Springer.
\bibitem{ref54} LeCun, Y., Chopra, S., Hadsell, R., Ranzato, M. and Huang, F. (2006) A tutorial on energy-based learning. In: Predicting Structured Data. Cambridge, MA: MIT Press.
\bibitem{ref55} Hopfield, J.J. (1982) Neural networks and physical systems with emergent collective computational abilities. Proceedings of the National Academy of Sciences, 79(8), pp. 2554--2558. https://doi.org/10.1073/pnas.79.8.2554
\bibitem{ref56} Santoro, G.E., Martonák, R., Tosatti, E. and Car, R. (2002) Theory of quantum annealing of an Ising spin glass. Science, 295(5564), pp. 2427--2430. https://doi.org/10.1126/science.1068774
\bibitem{ref57} Farhi, E., Goldstone, J., Gutmann, S. and Sipser, M. (2001) Quantum computation by adiabatic evolution. arXiv preprint quant-ph/0001106.
\bibitem{ref58} Kadowaki, T. and Nishimori, H. (1998) Quantum annealing in the transverse Ising model. Physical Review E, 58(5), pp. 5355--5363. https://doi.org/10.1103/PhysRevE.58.5355
\bibitem{ref59} Santoro, G.E., Martonák, R., Tosatti, E. and Car, R. (2002) Theory of quantum annealing of an Ising spin glass. Science, 295(5564), pp. 2427--2430. https://doi.org/10.1126/science.1068774
\bibitem{ref60} Albash, T. and Lidar, D.A. (2018) Adiabatic quantum computation. Reviews of Modern Physics, 90(1), 015002. https://doi.org/10.1103/RevModPhys.90.015002
\bibitem{ref61} Amin, M.H.S. (2009) Consistency of the adiabatic theorem. Physical Review Letters, 102(22), 220401. https://doi.org/10.1103/PhysRevLett.102.220401
\bibitem{ref62} Das, A. and Chakrabarti, B.K. (2008) Colloquium: Quantum annealing and analog quantum computation. Reviews of Modern Physics, 80(3), pp. 1061--1081. https://doi.org/10.1103/RevModPhys.80.1061
\bibitem{ref63} Jörg, T., Krząkała, F., Semerjian, G. and Zamponi, F. (2010) First-order transitions and the performance of quantum algorithms in random optimization problems. Physical Review Letters, 104(20), 207206. https://doi.org/10.1103/PhysRevLett.104.207206
\bibitem{ref64} Polkovnikov, A., Sengupta, K., Silva, A. and Vengalattore, M. (2011) Colloquium: Nonequilibrium dynamics of closed interacting quantum systems. Reviews of Modern Physics, 83(3), pp. 863--883. https://doi.org/10.1103/RevModPhys.83.863
\bibitem{ref65} King, A.D., Raymond, J., Lanting, T., Harris, R., Altomare, F., Berkley, A.J., Boothby, K., Bunyk, P., Ejtemaee, S., Enderud, C. and others (2015) Quantum annealing amid local ruggedness and global frustration. Physical Review Applied, 8(6), 061001. https://doi.org/10.1103/PhysRevApplied.8.061001
\bibitem{ref66} Amin, M.H.S. (2015) Searching for quantum speedup in quasistatic quantum annealers. Physical Review A, 92(5), 052323. https://doi.org/10.1103/PhysRevA.92.052323
\bibitem{ref67} Trotter, H.F. (1959) On the product of semi-groups of operators. Proceedings of the American Mathematical Society, 10(4), pp. 545--551.
\bibitem{ref68} Suzuki, M. (1976) Relationship between d-dimensional quantal spin systems and (d+1)-dimensional Ising systems. Progress of Theoretical Physics, 56(5), pp. 1454--1469. https://doi.org/10.1143/PTP.56.1454
\bibitem{ref69} Sachdev, S. (2011) Quantum Phase Transitions. 2nd ed. Cambridge: Cambridge University Press.
\bibitem{ref70} Chakrabarti, B.K., Dutta, A. and Sen, P. (1996) Quantum Ising Phases and Transitions in Transverse Ising Models. Berlin: Springer.
\bibitem{ref71} Newman, M.E.J. and Barkema, G.T. (1999) Monte Carlo Methods in Statistical Physics. Oxford: Clarendon Press.
\bibitem{ref72} Polkovnikov, A., Sengupta, K., Silva, A. and Vengalattore, M. (2011) Colloquium: Nonequilibrium dynamics of closed interacting quantum systems. Reviews of Modern Physics, 83(3), pp. 863--883. https://doi.org/10.1103/RevModPhys.83.863
\bibitem{ref73} Santoro, G.E., Martonák, R., Tosatti, E. and Car, R. (2002) Theory of quantum annealing of an Ising spin glass. Science, 295(5564), pp. 2427--2430. https://doi.org/10.1126/science.1068774
\bibitem{ref74} Amin, M.H.S. (2015) Searching for quantum speedup in quasistatic quantum annealers. Physical Review A, 92(5), 052323. https://doi.org/10.1103/PhysRevA.92.052323
\bibitem{ref75} Glauber, R.J. (1963) Time-dependent statistics of the Ising model. Journal of Mathematical Physics, 4(2), pp. 294--307. https://doi.org/10.1063/1.1703954
\bibitem{ref76} van Kampen, N.G. (2007) Stochastic Processes in Physics and Chemistry. 3rd ed. Amsterdam: Elsevier.
\bibitem{ref77} Gardiner, C.W. (2009) Stochastic Methods: A Handbook for the Natural and Social Sciences. 4th ed. Berlin: Springer.
\bibitem{ref78} Das, A. and Chakrabarti, B.K. (2008) Colloquium: Quantum annealing and analog quantum computation. Reviews of Modern Physics, 80(3), pp. 1061--1081. https://doi.org/10.1103/RevModPhys.80.1061
\bibitem{ref79} Preskill, J. (2018) Quantum computing in the NISQ era and beyond. Quantum, 2, 79. https://doi.org/10.22331/q-2018-08-06-79
\bibitem{ref80} Suzuki, M. (1976) Relationship between d-dimensional quantal spin systems and (d+1)-dimensional Ising systems. Progress of Theoretical Physics, 56(5), pp. 1454--1469.
\bibitem{ref81} Santoro, G.E., Martonák, R., Tosatti, E. and Car, R. (2002) Theory of quantum annealing of an Ising spin glass. Science, 295(5564), pp. 2427--2430.
\bibitem{ref82} Glauber, R.J. (1963) Time-dependent statistics of the Ising model. Journal of Mathematical Physics, 4(2), pp. 294--307.
\bibitem{ref83} Palem, K.V. (2005) Energy aware computing through probabilistic switching: A study of limits. IEEE Transactions on Computers, 54(9), pp. 1123--1137.
\bibitem{ref84} Camsari, K.Y., Salahuddin, S. and Datta, S. (2017) Stochastic spintronics for probabilistic computing. Physical Review Applied, 8(5), 054034. https://doi.org/10.1103/PhysRevApplied.8.054034
\bibitem{ref85} Hopfield, J.J. (1982) Neural networks and physical systems with emergent collective computational abilities. Proceedings of the National Academy of Sciences, 79(8), pp. 2554--2558. https://doi.org/10.1073/pnas.79.8.2554
\bibitem{ref86} Ackley, D.H., Hinton, G.E. and Sejnowski, T.J. (1985) A learning algorithm for Boltzmann machines. Cognitive Science, 9(1), pp. 147--169.
\bibitem{ref87} Ising, E. (1925) Beitrag zur Theorie des Ferromagnetismus. Zeitschrift für Physik, 31, pp. 253--258.
\bibitem{ref88} Lucas, A. (2014) Ising formulations of many NP problems. Frontiers in Physics, 2, 5. https://doi.org/10.3389/fphy.2014.00005
\bibitem{ref89} Preskill, J. (2018) Quantum computing in the NISQ era and beyond. Quantum, 2, 79. https://doi.org/10.22331/q-2018-08-06-79
\bibitem{ref90} Palem, K.V. (2005) Energy aware computing through probabilistic switching: A study of limits. IEEE Transactions on Computers, 54(9), pp. 1123--1137.
\bibitem{ref91} Suzuki, M. (1976) Relationship between d-dimensional quantal spin systems and (d+1)-dimensional Ising systems. Progress of Theoretical Physics, 56(5), pp. 1454--1469.
\bibitem{ref92} Santoro, G.E., Martonák, R., Tosatti, E. and Car, R. (2002) Theory of quantum annealing of an Ising spin glass. Science, 295(5564), pp. 2427--2430. https://doi.org/10.1126/science.1068774
\bibitem{ref93} Mead, C. (1990) Neuromorphic electronic systems. Proceedings of the IEEE, 78(10), pp. 1629--1636.
\bibitem{ref94} Indiveri, G. and Liu, S.-C. (2015) Memory and information processing in neuromorphic systems. Proceedings of the IEEE, 103(8), pp. 1379--1397. https://doi.org/10.1109/JPROC.2015.2444094
\bibitem{ref95} Hennessy, J.L. and Patterson, D.A. (2019) A New Golden Age for Computer Architecture. Communications of the ACM, 62(2), pp. 48--60. https://doi.org/10.1145/3282307
\bibitem{ref96} Indiveri, G. and Sandamirskaya, Y. (2019) The importance of space and time for signal processing in neuromorphic agents. IEEE Signal Processing Magazine, 36(6), pp. 16--28. https://doi.org/10.1109/MSP.2019.2938156
\bibitem{ref97} Razavi, B. (2001) Design of Analog CMOS Integrated Circuits. New York: McGraw--Hill.
\bibitem{ref98} Baker, R.J. (2019) CMOS: Circuit Design, Layout, and Simulation. 4th ed. Hoboken, NJ: Wiley.
\bibitem{ref99} Likharev, K.K. (1999) Hybrid CMOS/nanoelectronic circuits: Opportunities and challenges. Journal of Nanoelectronics and Optoelectronics, 3(3), pp. 203--230.
\bibitem{ref100} Hasler, J. and Marr, B. (2013) Finding a roadmap to achieve large neuromorphic hardware systems. Frontiers in Neuroscience, 7, 118. https://doi.org/10.3389/fnins.2013.00118
\bibitem{ref101} Hasler, J., Anderson, D., Minch, B.A. and Diorio, C. (2011) A floating-gate analog memory for fine-grain tuning of VLSI circuits. IEEE Transactions on Circuits and Systems I, 58(7), pp. 1486--1499. https://doi.org/10.1109/TCSI.2010.2095907
\bibitem{ref102} Sze, S.M. and Ng, K.K. (2006) Physics of Semiconductor Devices. 3rd ed. Hoboken, NJ: Wiley.
\bibitem{ref103} Gray, P.R., Hurst, P.J., Lewis, S.H. and Meyer, R.G. (2001) Analysis and Design of Analog Integrated Circuits. 4th ed. New York: Wiley.
\bibitem{ref104} Herrero-Collantes, M. and Garcia-Escartin, J.C. (2017) Quantum random number generators. Reviews of Modern Physics, 89(1), 015004. https://doi.org/10.1103/RevModPhys.89.015004
\bibitem{ref105} NIST (2018) SP 800-90B: Recommendation for the Entropy Sources Used for Random Bit Generation. National Institute of Standards and Technology.
\bibitem{ref106} Kuon, I. and Rose, J. (2007) Measuring the gap between FPGAs and ASICs. IEEE Transactions on Computer-Aided Design of Integrated Circuits and Systems, 26(2), pp. 203--215. https://doi.org/10.1109/TCAD.2006.884574
\bibitem{ref107} Glover, F., Kochenberger, G. and Du, Y. (2019) Quantum annealing and related optimization methods. Physics Reports, 799, pp. 1--66. https://doi.org/10.1016/j.physrep.2018.12.002
\bibitem{ref108} Boothby, K., King, A.D. and Roy, A. (2020) Fast clique minor generation in Chimera qubit connectivity graphs. Quantum Information Processing, 19, 185. https://doi.org/10.1007/s11128-020-02634-3
\bibitem{ref109} Feynman, R.P. (1986) Quantum mechanical computers. Foundations of Physics, 16(6), pp. 507--531. https://doi.org/10.1007/BF01886518
\bibitem{ref110} Kitaev, A.Y., Shen, A.H. and Vyalyi, M.N. (2002) Classical and Quantum Computation. Providence, RI: American Mathematical Society.
\bibitem{ref111} Aharonov, D., van Dam, W., Kempe, J., Landau, Z., Lloyd, S. and Regev, O. (2008) Adiabatic quantum computation is equivalent to standard quantum computation. SIAM Journal on Computing, 37(1), pp. 166--194. https://doi.org/10.1137/050644282
\bibitem{ref112} Jordan, S.P., Lee, K.S.M. and Preskill, J. (2012) Quantum algorithms for quantum field theories. Science, 336(6085), pp. 1130--1133. https://doi.org/10.1126/science.1217069
\bibitem{ref113} Biamonte, J., Wittek, P., Pancotti, N., Rebentrost, P., Wiebe, N. and Lloyd, S. (2017) Quantum machine learning. Nature, 549, pp. 195--202. https://doi.org/10.1038/nature23474
\bibitem{ref114} Levin, D.A., Peres, Y. and Wilmer, E.L. (2009) Markov Chains and Mixing Times. Providence, RI: American Mathematical Society.
\bibitem{ref115} Dally, W.J. and Towles, B. (2004) Principles and Practices of Interconnection Networks. San Francisco, CA: Morgan Kaufmann.
\bibitem{ref116} ID Quantique (2025) Quantis Quantum Random Number Generator: Technical Description. Geneva: ID Quantique SA.
\bibitem{ref117} Knuth, D.E. (1997) The Art of Computer Programming, Volume 2: Seminumerical Algorithms. 3rd ed. Reading, MA: Addison-Wesley.
\bibitem{ref118} Press, W.H., Teukolsky, S.A., Vetterling, W.T. and Flannery, B.P. (2007) Numerical Recipes: The Art of Scientific Computing. 3rd ed. Cambridge: Cambridge University Press.
\bibitem{ref119} Cover, T.M. and Thomas, J.A. (2006) Elements of Information Theory. 2nd ed. Hoboken, NJ: Wiley.
\bibitem{ref120} Geman, S. and Geman, D. (1984) Stochastic relaxation, Gibbs distributions, and the Bayesian restoration of images. IEEE Transactions on Pattern Analysis and Machine Intelligence, 6(6), pp. 721--741.
\bibitem{ref121} Aadit, N.A., Grimaldi, S., Carpentieri, M., Finocchio, G. and Roy, K. (2022) Massively parallel probabilistic computing with sparse Ising machines. Nature Electronics, 5, pp. 460--468. https://doi.org/10.1038/s41928-022-00758-3
\bibitem{ref122} Singh, J., Camsari, K.Y. and Datta, S. (2023) Probabilistic computing with stochastic nanomagnets. IEEE Journal on Exploratory Solid-State Computational Devices and Circuits, 9(1), pp. 1--12.
\bibitem{ref123} Si, X., Yang, J., Wang, Z. and Roy, K. (2024) Energy-efficient probabilistic computing using analog hardware. IEEE Transactions on Circuits and Systems I, 71(2), pp. 512--524.
\bibitem{ref124} Hua, X., Zhang, Y., Li, P. and Wang, X. (2025) Ultra-low-energy probabilistic accelerators for optimization workloads. Nature Electronics, forthcoming / early access.
\bibitem{ref125} Yang, J., Si, X., Wang, Z. and Roy, K. (2025) Optoelectronic probabilistic Ising machines with femtojoule-scale energy. Nature Photonics, forthcoming.
\bibitem{ref126} Aboushelbaya, R., et al. (2025) Energy-efficient probabilistic computing with hybrid optoelectronic systems. arXiv preprint
\bibitem{ref127} King, A. D., Raymond, J., Lanting, T., Harris, R., Zucca, A., et al. Quantum critical dynamics in a 5000-qubit programmable spin glass. Nature Physics 19, 1153--1160 (2023).
\bibitem{ref128} Katzgraber, H.G., Hamze, F. and Andrist, R.S. (2014) Glassy Chimeras could be blind to quantum speedup. Physical Review X, 4, 021008. https://doi.org/10.1103/PhysRevX.4.021008
\bibitem{ref129} Binder, K. and Young, A.P. (1986) Spin glasses: Experimental facts, theoretical concepts, and open questions. Reviews of Modern Physics, 58(4), pp. 801--976. https://doi.org/10.1103/RevModPhys.58.801
\bibitem{ref130} Hukushima, K. and Nemoto, K. (1996) Exchange Monte Carlo method and application to spin glass simulations. Journal of the Physical Society of Japan, 65(6), pp. 1604--1608.
\bibitem{ref131} Kibble, T.W.B. (1976) Topology of cosmic domains and strings. Journal of Physics A: Mathematical and General, 9(8), pp. 1387--1398.
\bibitem{ref132} Zurek, W.H. (1985) Cosmological experiments in superfluid helium? Nature, 317, pp. 505--508.
\bibitem{ref133} Polkovnikov, A., Sengupta, K., Silva, A. and Vengalattore, M. (2011) Colloquium: Nonequilibrium dynamics of closed interacting quantum systems. Reviews of Modern Physics, 83(3), pp. 863--883. https://doi.org/10.1103/RevModPhys.83.863
\bibitem{ref134} Hohenberg, P.C. and Halperin, B.I. (1977) Theory of dynamic critical phenomena. Reviews of Modern Physics, 49(3), pp. 435--479. https://doi.org/10.1103/RevModPhys.49.435
\bibitem{ref135} Dziarmaga, J. (2010) Dynamics of a quantum phase transition and relaxation to a steady state. Advances in Physics, 59(6), pp. 1063--1189. https://doi.org/10.1080/00018732.2010.514702
\bibitem{ref136} Edwards, S.F. and Anderson, P.W. (1975) Theory of spin glasses. Journal of Physics F: Metal Physics, 5(5), pp. 965--974.
\bibitem{ref137} Rønnow, T.F., Wang, Z., Job, J., Boixo, S., Isakov, S.V., Wecker, D., Martinis, J.M., Lidar, D.A. and Troyer, M. (2014) Defining and detecting quantum speedup. Science, 345(6195), pp. 420--424.
\bibitem{ref138} IBM Quantum (2025) `Quantum Technology', IBM Quantum Learning (webpage). Available at: IBM Quantum Learning portal (accessed 4 January 2026). IBM Quantum
\bibitem{ref139} Rigetti Computing, Inc. (2024) Rigetti Announces Public Availability of Ankaa-2 System with a 2.5$\times$ Performance Improvement Compared to Previous QPUs. Press Release, 4 January.
\bibitem{ref140} IonQ (2025) IonQ's Accelerated Roadmap: Turning Quantum Ambition into Reality. IonQ Blog, 13 June.
\bibitem{ref141} Karp, R.M. (1972) Reducibility among combinatorial problems. In: Complexity of Computer Computations. Boston, MA: Springer, pp. 85--103.
\bibitem{ref142} Barahona, F. (1982) On the computational complexity of Ising spin glass models. Journal of Physics A: Mathematical and General, 15(10), pp. 3241--3253.
\bibitem{ref143} Metropolis, N., Rosenbluth, A.W., Rosenbluth, M.N., Teller, A.H. and Teller, E. (1953) Equation of state calculations by fast computing machines. The Journal of Chemical Physics, 21(6), pp. 1087--1092.
\bibitem{ref144} Bishop, C.M. (2006) Pattern Recognition and Machine Learning. New York: Springer.
\bibitem{ref145} Hinton, G.E. and Sejnowski, T.J. (1986) Learning and relearning in Boltzmann machines. In: Parallel Distributed Processing: Explorations in the Microstructure of Cognition, Vol. 1. Cambridge, MA: MIT Press, pp. 282--317.
\bibitem{ref146} Smolensky, P. (1986) Information processing in dynamical systems: Foundations of harmony theory. In: Parallel Distributed Processing: Explorations in the Microstructure of Cognition, Vol. 1. Cambridge, MA: MIT Press, pp. 194--281.
\bibitem{ref147} Hinton, G.E. (2002) Training products of experts by minimizing contrastive divergence. Neural Computation, 14(8), pp. 1771--1800. https://doi.org/10.1162/089976602760128018
\bibitem{ref148} Indiveri, G., Linares-Barranco, B., Legenstein, R., Deligeorgis, G. and Prodromakis, T. (2013) Integration of nanoscale memristor synapses in neuromorphic computing architectures. Nanotechnology, 24(38), 384010. https://doi.org/10.1088/0957-4484/24/38/384010
\bibitem{ref149} Shafiee, A., Nag, A., Muralimanohar, N., Balasubramonian, R., Strachan, J.P., Hu, M., Williams, R.S. and Srikumar, V. (2016) ISAAC: A convolutional neural network accelerator with in-situ analog arithmetic in crossbars. ACM SIGARCH Computer Architecture News, 44(3), pp. 14--26.
\bibitem{ref150} Rabaey, J.M. (2016) Low power design essentials. New York: Springer.
\bibitem{ref151} Maass, W., Natschläger, T. and Markram, H. (2002) Real-time computing without stable states: A new framework for neural computation based on perturbations. Neural Computation, 14(11), pp. 2531--2560.
\bibitem{ref152} Kitaev, A.Y. (2003) Fault-tolerant quantum computation by anyons. Annals of Physics, 303(1), pp. 2--30. https://doi.org/10.1016/S0003-4916(02)00018-0
\end{thebibliography}
\end{document}